\documentclass[a4paper,preprintnumbers,nofootinbib,noshowpacs,prd]{revtex4}
\usepackage{graphicx,epsfig}
\usepackage{feynmp}
\usepackage{amsmath}
\usepackage{amsfonts}
\usepackage{amssymb}
\usepackage{bbold,bm,bbm}
\usepackage{url}
\usepackage{rotating}
\usepackage{axodraw}
\usepackage{color}
\usepackage{hhline}
\usepackage{fancyhdr}
\usepackage{subfigure}

\usepackage[normalem]{ulem}   % For crossline
\usepackage[makeroom]{cancel}  % For crossline
\usepackage[pass]{geometry}  % Make paper size consistent
\usepackage[usenames,dvipsnames]{xcolor}

\numberwithin{equation}{section}

\graphicspath{ {Plots/} }

\makeatletter       % Macros for arabic section numbering

\renewcommand{\p@subsection}{}

\makeatother

\newcommand{\s}{\\[-2mm]}

\begin{document}

\begin{flushright}
%preprint numbers
PITT-PACC-1502 \\
\end{flushright}

\vskip 0.5cm

\title{Spin and Chirality Effects in Antler-Topology Processes at High
       Energy \boldmath{$e^+e^-$} Colliders}

\author{S.Y. Choi$^{1,2}$, N.D. Christensen$^{3}$, D. Salmon$^{2}$, and
        X. Wang$^{2}$\\[-3mm]
        \mbox{ }\\
   $^1$ {\it Department of Physics, Chonbuk National University, Jeonbuk 561-756,
             Korea} \\
   $^2$ {\it Pittsburgh Particle physics, Astrophysics, and Cosmology Center,
             Department of Physics and Astronomy, University of Pittsburgh,
             Pittsburgh, PA 15260, USA}\\
   $^3$ {\it Department of Physics, Illinois State University, Normal, IL  61790, USA}
}

\date{\today}

\begin{abstract}
{\noindent
We perform a model-independent investigation of spin and chirality correlation effects in the
antler-topology processes $e^+e^-\to\mathcal{P}^+\mathcal{P}^-\to (\ell^+ \mathcal{D}^0)
(\ell^-\mathcal{\bar{D}}^0)$ at high energy $e^+e^-$ colliders with polarized beams.
Generally the production process $e^+e^-\to\mathcal{P}^+\mathcal{P}^-$ can occur not only
through the $s$-channel exchange of vector bosons, $\mathcal{V}^0$, including the neutral
Standard Model (SM) gauge bosons, $\gamma$ and $Z$, but also through the $s$- and $t$-channel
exchanges of new neutral states, $\mathcal{S}^0$ and $\mathcal{T}^0$, and  the $u$-channel
exchange of new doubly-charged states, $\mathcal{U}^{--}$. The general set of (non-chiral)
three-point couplings of the new particles and leptons allowed in a renormalizable
quantum field theory is considered. The general spin and chirality analysis is based
on the threshold behavior of the excitation curves for $\mathcal{P}^+\mathcal{P}^-$ pair
production in $e^+e^-$ collisions with longitudinal and transverse polarized beams, the
angular distributions in the production process and also the production-decay angular
correlations. In the first step, we present
the observables in the helicity formalism. Subsequently, we show how a set of observables
can be designed for determining the spins and chiral structures of the new particles
without any model assumptions. Finally, taking into account a typical set of approximately
chiral invariant scenarios, we demonstrate how the spin and chirality effects can be
probed experimentally at a high energy $e^+e^-$ collider.
}
\end{abstract}

\maketitle

% \addtocounter{section}{1}

\section{Introduction}
\label{sec:introduction}

The monumental discovery \cite{Aad:2012tfa,Chatrchyan:2012ufa} of the Higgs boson at the CERN
Large Hadron Collider (LHC) has filled in the only missing piece of the SM of electroweak and
strong interactions, completing its gauge symmetry structure and electroweak symmetry breaking
(EWSB) through the so-called Brout-Englert-Higgs (BEH) mechanism \cite{Englert:1964et,
Higgs:1964ia,Higgs:1964pj,Higgs:1966ev,Kibble:1967sv}.
Nevertheless, there are several compelling indications that the SM needs to be extended by
including new particles and/or new types of interactions. Once any new particle indicating new
physics beyond the SM is discovered at the LHC or high energy $e^+e^-$ colliders, one of the
first crucial steps is to experimentally determine its spin as well as its mass because spin
is one of the canonical characteristics of all particles required for defining a new
theoretical framework as a Lorentz-invariant quantum field theory \cite{Wigner:1939cj}.\s

Many models beyond the SM \cite{Wess:1974tw,Nilles:1983ge,Haber:1984rc,Chung:2003fi,
Weinstein:1973gj,Weinberg:1979bn,Susskind:1978ms,ArkaniHamed:1998nn,Randall:1999ee,
Appelquist:2000nn,ArkaniHamed:2001nc,Csaki:2003dt,Csaki:2003zu} have been proposed and studied
not only to resolve several conceptual issues like the gauge hierarchy problem but also to explain
the dark matter (DM) composition of the Universe with new stable weakly interacting massive
particles \cite{Griest:2000kj,Bertone:2004pz,Ade:2013zuv}. For this purpose, a (discrete)
symmetry such as $R$ paity in supersymmetric (SUSY) models and Kaluza-Klein (KK) parity
in universal extra-dimension (UED) models is generally introduced to guarantee the stability
of the particles and thus to explain the DM relic density quantitatively. As a consequence,
the new particles can be produced only in pairs at high energy hadron or lepton colliders,
leading to challenging signatures with at least two invisible final-state particles.\s

\begin{figure}[hbt]
\centering
\includegraphics[scale=.65]{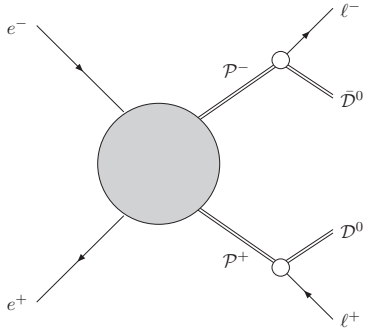}
\caption{\label{fig:antler_diagram}
        {The correlated process $e^+e^-\rightarrow \mathcal{P}^+\mathcal{P}^-
             \rightarrow (\ell^+\mathcal{D}^0) (\ell ^- \mathcal{\bar{D}}^0)$ characterized
             by the antler-topology diagram. Here, the invisible final-state particle
             $\mathcal{D}^0$ might be charge self-conjugate, i.e.
             $(\mathcal{D}^0)^c= \mathcal{D}^0$.}
        }
\end{figure}

At hadron colliders like the LHC such a signal with invisible particles is usually
insufficiently constrained for full kinematic reconstructions, rendering the unambiguous
and precise determination of the masses, spins and couplings of
(new) particles  produced in the intermediate or final stages challenging, even if
conceptually possible, as demonstrated in many previous works on mass
measurements \cite{Lester:1999tx,Barr:2003rg,Cho:2007qv,Barr:2007hy,Cho:2007dh,Tovey:2008ui,
Cheng:2008hk,Barr:2009jv,Matchev:2009ad,Polesello:2009rn,Konar:2009wn,Cohen:2010wv,
Alwall:2009sv,Artoisenet:2010cn,Alwall:2010cq,Han:2009ss,Han:2012nm,Han:2012nr,Swain:2014dha} and on spin
determination \cite{Barr:2004ze,Smillie:2005ar,Datta:2005zs,Barr:2005dz,Meade:2006dw,
Alves:2006df,Athanasiou:2006ef,Wang:2006hk,Smillie:2006cd,Choi:2006mt,Kilic:2007zk,
Alves:2007xt,Csaki:2007xm,Wang:2008sw,Burns:2008cp,Cho:2008tj,Gedalia:2009ym,
Ehrenfeld:2009rt,Edelhauser:2010gb,Horton:2010bg,Cheng:2010yy,Buckley:2010jv,Chen:2010ek,
Chen:2011cya,Nojiri:2011qn,MoortgatPick:2011ix}.\s

In contrast to hadron colliders, an $e^+e^-$ collider \cite{Behnke:2013xla,Baer:2013cma,
Behnke:2013lya,Koratzinos:2013chw,Gomez-Ceballos:2013zzn,Accomando:2004sz,Linssen:2012hp}
has a fixed center-of-mass (c.m.) energy and c.m. frame and the collider can
be equipped with longitudinally
and/or transversely polarized beams. These characteristic features allow us to exploit several
complementary techniques at $e^+e^-$ colliders for unambiguously determining the spins as well
as the masses of new pairwise-produced particles, the invisible particles
from the decays of the parent particles and the particles exchanged as intermediate
states, with good precision. In the present work we focus on the following production-decay
correlated processes
\begin{eqnarray}
e^+e^-  \,\to\, \mathcal{P}^+\mathcal{P}^-
        \,\to\, (\ell^+ \mathcal{D}^0) (\ell^-\mathcal{\bar{D}}^0)
\label{eq:antler_process}
\end{eqnarray}
dubbed antler-topology events \cite{Han:2009ss}, which contain the production of an
electrically charged pair $\mathcal{P}^+\mathcal{P}^-$ in $e^+e^-$ collisions followed
by the two-body decays, $\mathcal{P}^+\to\ell^+\mathcal{D}^0$ and $\mathcal{P}^-\to\ell^-\mathcal{\bar{D}}^0$, giving rise to a charged lepton pair
$\ell^\pm(= e^\pm,\mu^\pm)$ and an invisible pair $\mathcal{D}^0\mathcal{\bar{D}}^0$
(See Fig.$\,$\ref{fig:antler_diagram}).\s

The invisible particle $\mathcal{D}^0$ may be charge self-conjugate, i.e.
$\mathcal{\bar{D}}^0=\mathcal{D}^0$. Nevertheless, it is expected to be insubstantial
quantitatively whether the particle is self-conjugate or not, unless the width of the parent
particle $\mathcal{P}^\pm$ is very large and there exist large chirality mixing
contributions \cite{Hagiwara:2005ym}. So, any interference effects due to the charge
self-conjugateness of the invisible particle will be ignored in the present
work.\footnote{An indirect but powerful way of checking the charge self-conjugateness
of the particle $\mathcal{D}^0$ is to study the process $e^-e^-\to\mathcal{P}^-\mathcal{P}^-$
to which the self-conjugate particle $\mathcal{D}^0$ can contribute through its $t$-channel
exchange. The $e^-e^-$ mode is under consideration as a satellite mode
at the ILC. }\s

If the parent particle $\mathcal{P}^-$ carries an electron number $L_e(\mathcal{P}^-)=+1$
or a muon number $L_\mu(\mathcal{P}^-)=+1$, then the final-state leptons must be $e^-e^+$ or
$\mu^-\mu^+$, respectively, if electron and muon numbers are conserved individually and the
invisible particles, $\mathcal{D}^0$ and $\mathcal{\bar{D}}^0$, carry no lepton numbers.
On the other hand, if the parent particle carries no lepton number, the final-state leptons
can be any of the four combinations, $\{e^-e^+, e^-\mu^+, \mu^- e^+, \mu^-\mu^+\}$, and
the invisible particles, $\mathcal{D}^0$ and $\mathcal{\bar{D}}^0$, must carry the same
lepton number as $\ell^\mp= e^\mp, \mu^\mp$, respectively. \s

Once the masses of new particles are determined by (pure) kinematic
effects \cite{Christensen:2014yya}, a sequence of techniques increasing in complexity can
be applied to determine the spins and chirality properties of particles in the correlated
antler-topology process at $e^+e^-$ colliders \cite{Battaglia:2005zf,Choi:2006mr,
Buckley:2007th,Buckley:2008eb,Boudjema:2009fz,Christensen:2013sea}:
\begin{itemize}
\item[{(a)}] Rise of the excitation curve near threshold with polarized
             electron and positron beams;
\item[{(b)}] Angular distribution of the production process;
\item[{(c)}] Angular distributions of the decays of polarized particles;
\item[{(d)}] Angular correlations between decay products of two particles.
\end{itemize}
While the first and second steps (a) and (b) are already sufficient in the case with
a spin-0 scalar $\mathcal{P}^\pm=S^\pm_p$ as will be demonstrated in detail, the
production-decay correlations need to be considered for the case with a spin-1/2 fermion
$\mathcal{P}^\pm=F^\pm_p$ and a spin-1 $\mathcal{P}^\pm=V^\pm_p$ to determine the
$\mathcal{P}$ spin unambiguously; in principle a proper combination of these complementary
techniques enables us to determine the spins of the invisible particles, $\mathcal{D}^0$
and $\mathcal{\bar{D}}^0$, and all the intermediate particles exchanged in $s$-, $t$- or
$u$-channel diagrams participating in the production process. For our numerical analysis
we follow the standard procedure. We show through detailed simulations how the
theoretically predicted distributions can be reconstructed after including initial state
QED radiation (ISR), beamstrahlung and  width effects as well as typical kinematic
cuts.\s

The paper is organized as follows. In Sect.$\,$\ref{sec:setup} we describe a general
theoretical framework for the spin and chiral effects in antler-topology
processes at high energy $e^+e^-$ colliders. In Sect.$\,$\ref{sec:production} we present
the complete amplitudes and polarized cross sections for the production process
$e^+e^-\to\mathcal{P}^+\mathcal{P}^-$ in the $e^+e^-$ center-of-mass (c.m.) frame with
the general set of couplings listed in Appendix \ref{sec:appendix_a_feynman_rules}.
The technical framework we have employed is the helicity formalism \cite{Jacob:1959at}.
Then, we present in Sect.$\,$\ref{sec:two_body_decays} the complete helicity amplitudes
of the two-body decays $\mathcal{P}^+ \to \ell^+ \mathcal{D}^0$ and
$\mathcal{P}^-\to\ell^-\mathcal{\bar{D}}^0$ with general couplings given in
Appendix \ref{sec:appendix_a_feynman_rules}.
Sect.$\,$\ref{sec:full_angular_correlations} describes how to obtain the fully-correlated
six-dimensional production-decay angular distributions by combining the production helicity
amplitudes and the two two-body decay helicity amplitudes and by implementing arbitrary
electron and positron polarizations \cite{Hikasa:1985qi,Hagiwara:1985yu,MoortgatPick:2005cw,
Choi:2006vh,Ananthanarayan:2008dr}.
Sect.$\,$\ref{sec:observables}, the main
part of the present work, is devoted to various observables: the threshold-excitation
patterns, the production angle distributions equipped with polarized beams, the lepton
decay polar-angle distributions and the lepton angular-correlations of the two two-decay
modes. They provide us with powerful tests of the spin and chirality effects in the
production-decay correlated process. While all the analytic results are maintained to
be general, the numerical analyses are given for the theories with (approximate)
electron chirality conservation such as SUSY and UED models and a subsection will
be devoted to a brief discussion of the possible influence from electron chirality
violation effects. Finally, we summarize our findings and conclude in
Sect.$\,$\ref{sec:summary_conclusion}.
For completeness, we include three appendices in addition to
Appendix \ref{sec:appendix_a_feynman_rules}. In Appendix \ref{sec:appendix_b_d_functions},
we list all of the Wigner $d$-functions used in the main text \cite{d functions:1957rs}.
In Appendix \ref{sec:appendix_c_arbitrary_polarized_beams}, we describe how to obtain the
expression of the production matrix element-squared for arbitrary polarized electron and
positron beams. Finally, in Appendix \ref{sec:appendix_d_kinematics_of_the_antler_process}
we give an analytic proof of the presence of a twofold discrete
ambiguity in determining the $\mathcal{P}^\pm$ momenta in the process
$e^+e^-\to\mathcal{P}^+\mathcal{P}^-\to (\ell^+\mathcal{D}^0) (\ell^-\mathcal{\bar{D}}^0)$,
even if the masses of the particles, $\mathcal{P}^\pm$ and $\mathcal{D}^0$
($\mathcal{\bar{D}}^0$), are a priori known.\s

\section{Setup for model-independent spin determinations}
\label{sec:setup}

\begin{figure}[htb]
\centering
\includegraphics[scale=.90]{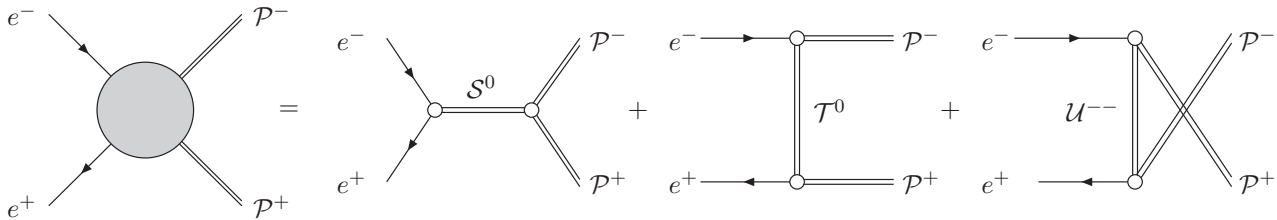}
\caption{\label{fig:production_diagram}
        {New $s$-channel $\mathcal{S}^0$-exchange diagrams (including the standard
             $\gamma$- and $Z$-exchange diagrams), new $t$-channel $\mathcal{T}^0$-exchange
             diagrams and new $u$-channel $\mathcal{U}^{--}$-exchange diagrams to the
             pair-production process $e^+e^-\to \mathcal{P}^+\mathcal{P}^-$.}
         }
\end{figure}

Generally, the production part $e^+e^-\to\mathcal{P}^+\mathcal{P}^-$ of the antler-topology
process (\ref{eq:antler_process}) can occur through $s$-, $t$- and/or $u$-channel diagrams in
renormalizable field theories, as shown in Fig.$\,$\ref{fig:production_diagram}. Which types of
diagrams are present and/or significant depend crucially on the nature of the new particles,
$\mathcal{P}^\pm$, $\mathcal{D}^0$ and $\mathcal{\bar{D}}^0$ as well as the SM leptons
$\ell^\pm$ and on the constraints from the discrete symmetries conserved in the theory. \s

We assume that the new particles, $\mathcal{P}^\pm$, $\mathcal{D}^0$ and $\mathcal{\bar{D}}^0$,
are produced on-shell in the antler-topology process (\ref{eq:antler_process}), and they are
uncolored under the SM strong-interaction group so that they are not strongly
interacting.\footnote{In addition, assuming the widths of the new particles to be
much smaller than their corresponding masses, we neglect their width effects for any
analytic expressions, although we consider them in numerical simulations in the
present work.}  Motivated mainly by the DM problem, the new particles are assumed to be odd
under a conserved discrete $Z_2$-parity symmetry. Therefore, they can only be
produced in pairs at high energy hadron and lepton colliders with an initial $Z_2$-parity
even environment such as LHC, ILC, TLEP and CLIC, etc. Furthermore, the invisible particle
$\mathcal{D}^0$ participating in the two-body decay $\mathcal{P}^+\to e^+ \mathcal{D}^0$,
if the decay mode is present, is included among the particles $\mathcal{T}^0$
exchanged in the $t$-channel diagram of the production process
$e^+e^-\to\mathcal{P}^+\mathcal{P}^-$. This implies that unavoidably at least one of the particles
$\mathcal{T}^0$ is lighter than the particle $\mathcal{P}^\pm$ in the antler-topology
process with $\ell^\pm =e^\pm$.   \s

As the $\mathcal{P}^-$ as well as the electron $e^-$ is singly electrically-charged, the $s$-
and $t$-channel processes are mediated by (potentially several) neutral particles, $\mathcal{S}^0$ and
$\mathcal{T}^0$, but any $u$-channel processes must be mediated by (potentially several)
doubly-charged particles, $\mathcal{U}^{--}$. In passing, we note that most of the popular
extensions of the SM such as supersymmetry (SUSY) and universal extra-dimension (UED) models
contain no doubly-charged particles so that there exist only $s$-channel and/or $t$-channel
exchange diagrams but no $u$-channel exchange diagrams contributing to the production process
$e^+e^-\to\mathcal{P}^+\mathcal{P}^-$. The $s$-channel scalar-exchange contributions may
be practically negligible as well because the electron-chirality violating couplings of any
scalar to the electron line are strongly suppressed in proportion to the tiny electron mass in
those SUSY and UED models.\s

Since the on-shell particles, $\mathcal{P}^\pm$, $\mathcal{D}^0$ and $\mathcal{\bar{D}}^0$
as well as the virtual intermediate particles, $\mathcal{S}^0,\, \mathcal{T}^0$ and
$\mathcal{U}^{\pm\pm}$, are not directly measured, their spins and couplings as well as
masses are not a priori known. The neutral state $\mathcal{S}^0$ can be a spin-0 scalar,
$S^0_s$, or a spin-1 vector boson, $V^0_s$, including the standard gauge bosons
$V^0_s=\gamma, Z$ as well. Each of the other intermediate particles can be a spin-0 scalar,
a spin-1/2 fermion or a spin-1 vector boson, assigned in relation to the spin of the
particle $\mathcal{P}^\pm$. In any Lorentz-invariant theories, there exist in total
twenty ($20=2+8+8+2$) different spin assignments for the production-decay correlated
antler-topology process (\ref{eq:antler_process}) as
\begin{eqnarray}
  \left( J_\mathcal{P},\!J_\mathcal{D}; J_\mathcal{S},\! J_\mathcal{T},\!
         J_\mathcal{U} \right)
\!= \!\bigg\{\! \left(0,\tfrac{1}{2}; 0\oplus 1, \tfrac{1}{2}, \tfrac{1}{2}\right),\!
          \left(\tfrac{1}{2},0; 0\oplus 1, 0\oplus 1, 0\oplus 1\right),\!
          \left(\tfrac{1}{2},1; 0\oplus 1, 0\oplus 1, 0\oplus 1\right),\!
          \left(1,\tfrac{1}{2}; 0\oplus 1, \tfrac{1}{2}, \tfrac{1}{2}\right)\!
  \bigg\}
\end{eqnarray}
with spins up to $1$ and couplings consistent with renormalizable interactions. The symbols
used for the particles in our analysis are listed in Tab.$\,$\ref{tab:list_of_symbols} along
with their charges, spins and $Z_2$ parities. Generically, the intermediate states,
$\mathcal{S}^0$, $\mathcal{T}^0$ and $\mathcal{U}^{--}$ may stand for several different
states, although typically the on-shell particle $\mathcal{P}^\pm$ or $\mathcal{D}^0$ stands
for a single state.  Note that, if the parent particle $\mathcal{P}^\pm$ turns out to be a
spin-0 or spin-1 particle, then the daughter particles, $\mathcal{D}^0$ and $\mathcal{\bar{D}}^0$,
and the $t$- and $u$-channel intermediate particles $\mathcal{T}^0$ and $\mathcal{U}^{\pm\pm}$
are guaranteed to be spin-1/2 particles. \s

\begin{table}[htb]
\begin{center}
\begin{tabular}{|l|ccc|r|c|}
\hline
\ \ Particle \ \  &             &  Spin         &               & \ \ Charge \ \
                                                                & \ \ $Z_2$ Parity \ \ \\
                  & \ \ $0$ \ \ & \ \ $1/2$ \ \  & \ \ $1$ \ \  &                & \\
\hline\hline
{ } \quad $\ell^- $              &             &  $\ell^-$      &          &  $-1$ \quad { }
                                                                           &  $+$ \\
\hline
{ } \quad $\mathcal{D}^0$        &  $S^0_d$    &  $F^0_d$     &  $V^0_d$   &  $0$  \quad { }
                                                                           &  $-$  \\
{ } \quad $\mathcal{P}^- $       &  $S^-_p $   &  $F^-_p$     &  $V^-_p $  &  $-1$ \quad { }
                                                                           &  $-$  \\
\hline\hline
{ } \quad $\mathcal{S}^0$        &  $S^0_s$    &  $F^0_s$     &  $V^0_s$   &  $0$  \quad { }
                                                                           &  $+$  \\
{ } \quad $\mathcal{T}^0$        &  $S^0_t$    &  $F^0_t$     &  $V^0_t$   &  $0$  \quad { }
                                                                           &  $-$  \\
{ } \quad $\mathcal{U}^{--}$     &  { } \! $S^{--}_u$   &  { } \! $F^{--}_u$
                                 &  { } \! $V^{--}_u$   &  $-2$ \quad { }  &  $-$  \\
\hline
\end{tabular}
\end{center}
\caption{\label{tab:list_of_symbols}
        {List of symbols used for the particles in our analysis with their electric
             charges, spins and $Z_2$ parities. The symbol $\ell^-$ denotes an electron $e^-$
             or a muon $\mu^-$. The last three lines are for the new particles exchanged
             in the $s$-, $t$- and $u$-channel diagrams including the neutral electroweak
             gauge bosons, $\gamma$ and $Z$, exchanged in the $s$-channel diagram in the production
             process, $e^+e^-\to \mathcal{P}^+\mathcal{P}^-$.}
        }
\end{table}

Among the elementary particles discovered so far, the electron is the lightest
electrically-charged particle in the SM. Its mass $m_e\simeq 0.51\, {\rm MeV}\,
\sim 2\times 10^{-6}\,v$ is much smaller than the
vacuum expectation value (vev) $v\simeq 246$ GeV of the SM Higgs field, the weak scale for
setting the masses of leptons and quarks, as well as the c.m. energies of future high-energy
$e^+e^-$ colliders. Any kinematic effects due to the electron mass are negligible so that
the electron will be regarded as a massless particle from the kinematic point of view in the
present work. The near masslessness of the electron is related to the approximate chiral
symmetry of the SM. Any new theory beyond the SM should guarantee the experimentally-established
smallness of the electron mass. This is a challenge in new theories beyond the SM since
they usually involve larger mass scale(s) than the weak scale. One simple and natural protection
mechanism is chiral symmetry.\footnote{Other possible solutions for getting
a massless fermion naturally is that the fermion is a Nambu-Goldstone fermion, the super-partner
of an unbroken gauge boson or the super-partner of a Goldstone boson.}\s

Nevertheless, we do not impose any type of chiral symmetry so as to maintain full generality
in our model-independent analysis of spin and chirality effects, emphasising the importance
of checking experimentally to what extent the underlying theory possesses chiral symmetry.
In each three-point vertex involving a fermion line, i.e. two spin-1/2 fermion states, we
allow for an arbitrary linear combination of right-handed and left-handed couplings. Only in our
numerical examples will every interaction vertex involving the initial $e^\pm$ line and
the final-state lepton $\ell^\pm$ ($=e^\pm, \mu^\pm$) be set to be purely chiral, as is
nearly valid in typical SUSY and UED models, apart from tiny contaminations proportional to
the electron or muon masses generated through the BEH mechanism of
EWSB \cite{Englert:1964et,Higgs:1964ia,Higgs:1964pj,Higgs:1966ev,Kibble:1967sv}.\s

\section{Pair Production Processes}
\label{sec:production}

In this section we present the analytic form of helicity amplitudes for the production process
\begin{eqnarray}
e^-(p_-,\sigma_-)\, + \, e^+(p_+,\sigma_+)\, \to\,
\mathcal{P}^-(q_-, \lambda_-)\, + \, \mathcal{P}^+(q_+,\lambda_+)
\end{eqnarray}
with the $s$-, $t$- and $u$-channel contributions as depicted in
Fig.$\,$\ref{fig:production_diagram} with the general three-point couplings listed
in Appendix A. Here, we discuss only the amplitudes for on-shell $\mathcal{P}$ pair
production. The technical framework for our analytic results is the standard
helicity formalism \cite{Jacob:1959at}.\s

The helicity of a massive particle is not a relativistically invariant quantity.
It is invariant only for rotations or boosts along the particle's momentum, as
long as the momentum does not change its sign. In the present work, we define
the helicities of the $\mathcal{P}^\pm$ in the $e^+e^-$ c.m. frame. Helicity
amplitudes contain full information on the production process and enable us to
take into account polarization of the initial $e^+e^-$ beams in a straightforward
way as described in Appendix \ref{sec:appendix_c_arbitrary_polarized_beams}.\s

Generically, ignoring the electron mass, we can cast the helicity amplitude into a
compact form composed of two parts - an electron-chirality conserving (ECC) part $Q^c$
and an electron-chirality violating (ECV) part $Q^v$ - as
\begin{eqnarray}
{\cal M}[e^-_{\sigma_-}e^+_{\sigma_+} \to
        \mathcal{P}^-_{\lambda_-}\mathcal{P}^+_{\lambda_+}]
= \sqrt{2} e^2
  \left[\, \delta_{\sigma_+,-\sigma_-} Q^c_{\sigma_-;\lambda_-,\lambda_+}
          +\delta_{\sigma_+,\sigma_-} Q^v_{\sigma_-;\lambda_-,\lambda_+}\right]\,
          d^{J_0}_{\Delta\sigma,\Delta\lambda}(\theta)
\label{eq:generic_production_amplitude}
\end{eqnarray}
where $J_0={\rm max}(|\Delta\sigma|,|\Delta\lambda|)$ with the difference of the $e^\mp$
helicities  $\Delta\sigma = J_e (\sigma_- - \sigma_+) = \pm 1, 0$ and that of the
$\mathcal{P}^\mp$ helicities $\Delta\lambda = J_{\mathcal{P}} (\lambda_- -\lambda_+)$.
Here, $J_e= 1/2$ and $J_{\mathcal{P}}$ are the spin of the electron and the particle
$\mathcal{P}$, respectively. No helicity indices are needed when the spin of the particle
$\mathcal{P}$ is zero, i.e. $J_{\mathcal{P}}=0$. After extracting the spin value of
the electron and $\mathcal{P}$, $\sigma_\pm$ takes two values of $\pm 1$ while $\lambda_\pm$
takes two values of $\pm 1$ or three values $\pm 1, 0$ for $J_{\mathcal{P}}=1/2$ or $1$,
respectively. Frequently, in the present work we adopt the conventions, $\sigma_{-,+} =\pm$
and $\lambda_{-,+}=\pm, 0$,  will be used to denote the sign of the re-scaled helicity values
for the sake of notational convenience.
The angle $\theta$ in Eq.$\,$(\ref{eq:generic_production_amplitude}) denotes the scattering
angle of $\mathcal{P}^-$ with respect to the $e^-$ direction in the $e^+e^-$ c.m. frame.
The explicit form of the $d$ functions needed here is reproduced in
Appendix \ref{sec:appendix_b_d_functions}. \s

The polarization-weighted polar-angle differential cross sections of the production process
can be cast into the form
\begin{eqnarray}
   \frac{d\sigma^{\mathcal{P}}_{\rm pol}}{d\cos\theta}
&=& \frac{\pi\alpha^2\beta}{4s}
    \bigg[ (1 - P^L_- P^L_+) (\mathcal{C}^{+}_{+} + \mathcal{C}^{-}_{-})
          +(P^L_- - P^L_+) (\mathcal{C}^{+}_{+} - \mathcal{C}^{-}_{-})     \nonumber\\
  && { }\hskip 0.3cm  \quad
          +(1 + P^L_- P^L_+) (\mathcal{V}^{+}_{+} + \mathcal{V}^{-}_{-})
          +(P^L_- + P^L_+) (\mathcal{V}^{+}_{+} - \mathcal{V}^{-}_{-})       \nonumber\\
  && { }\hskip 0.3cm  \quad
          +2 P^T_- P^T_+ \cos\delta\, {\rm Re}(\mathcal{V}^{-}_{+})
          +2 P^T_- P^T_+ \sin\delta\, {\rm Im}(\mathcal{V}^{-}_{+}) \bigg]
\label{eq:generic_differential_cross_section}
\end{eqnarray}
with $\delta$ the relative opening angle of the electron and positron transverse
polarizations and $\beta$ the speed of pair-produced particles, where $P_{\pm}^{L,T}$ is the degrees of longitudinal and transverse polarizations and $\delta$ is the relative opening angle of the $e^{\pm}$ transverse polarizations. The ECC and ECV production tensors $\mathcal{C}$ and $\mathcal{V}$
are defined in terms of the reduced production helicity amplitudes by
\begin{eqnarray}
    \mathcal{C}^{\sigma}_{\sigma}
&=& \sum_{\lambda_-,\lambda_+} \left|\, Q^c_{\sigma;\lambda_-,\lambda_+}\right|^2\,
    \left[d^{J_0}_{\sigma,\Delta\lambda}(\theta)\right]^2
    \label{eq:production_tensor_c} \\
    \mathcal{V}^{\sigma'}_{\sigma}
&=& \sum_{\lambda_-,\lambda_+} \left(Q^v_{\sigma;\lambda_-,\lambda_+}
                               Q^{v*}_{\sigma';\lambda_-,\lambda_+}\right)\,
    \left[d^{J_0}_{0,\Delta\lambda}(\theta)\right]^2\,
    \label{eq:production_tensor_v}
\end{eqnarray}
with $\sigma,\sigma' = \pm 1$ or simply $\pm$ for notational convenience.
(For more detailed derivation of the polarized cross sections, see
Appendix \ref{sec:appendix_c_arbitrary_polarized_beams}.)
The polarized total cross section $\sigma^{\mathcal{P}}_{\rm pol}$ can
then be obtained by integrating the differential cross section over the full
range of $\cos\theta$. \s

If all of the coupling coefficients are real and all the particle widths are neglected,
the following relations must hold for both the ECC and ECV parts of the production helicity
amplitudes:
\begin{eqnarray}
  Q^{c}_{\sigma;\lambda_-,\lambda_+}
= Q^{c*}_{\sigma; -\lambda_+,-\lambda_-} \quad \mbox{and}\quad
  Q^{v}_{\sigma;\lambda_-,\lambda_+}
= Q^{v*}_{-\sigma; -\lambda_+,-\lambda_-}
\label{eq:cpt_relation}
\end{eqnarray}
as a consequence of $CPT$ invariance in the absence of any absorptive parts. Therefore,
violation of this relation indicates the presence of re-scattering effects. On the
other hand, $CP$ invariance leads to the relation:
\begin{eqnarray}
  Q^{c}_{\sigma;\lambda_-,\lambda_+}
= Q^{c}_{\sigma; -\lambda_+,-\lambda_-} \quad \mbox{and}\quad
  Q^{v}_{\sigma;\lambda_-,\lambda_+}
= Q^{v}_{-\sigma; -\lambda_+,-\lambda_-}
\label{eq:cp_relation}
\end{eqnarray}
independently of the absorptive parts so that the relation can be directly used as
a test of {\it CP} conservation. Similarly, it is easy to see that {\it P} invariance
leads to the relation for both the ECC and ECV amplitudes:
\begin{eqnarray}
  Q^{c,v}_{\sigma;\lambda_-,\lambda_+}
= Q^{c,v}_{-\sigma; -\lambda_-,-\lambda_+}
\label{eq:p_relation}
\end{eqnarray}
which is violated usually through chiral interactions such as weak interactions in
the SM.\s

Applying the $P$ and $CP$ symmetry relations to the ECC and ECV production tensors,
(\ref{eq:production_tensor_c}) and (\ref{eq:production_tensor_v}), we can classify the six
polar-angle distributions in Eq.$\,$(\ref{eq:generic_differential_cross_section}) according
to their $P$ and $CP$ properties as shown in Tab.$\,$\ref{tab:symmetry_relations_on_production}.
We find that the two combinations, $\mathcal{C}^+_+  + \mathcal{C}^-_-$ and
$\mathcal{V}^+_+ + \mathcal{V}^-_-$, contributing
to the unpolarized part are both $P$- and $CP$-even  whereas the terms,
$\mathcal{C}^+_+ - \mathcal{C}^-_-$ and $\mathcal{V}^+_+ - \mathcal{V}^-_-$,
linear in the degrees of longitudinal polarization are
$P$-odd and $CP$-even. One of the two transverse-polarization dependent parts,
${\rm Re}(\mathcal{V}^-_+)$, is both $P$- and $CP$-even
and the other one, ${\rm Im}(\mathcal{V}^-_+)$, is  both $P$- and $CP$-odd.
Unlike the other five distributions, the distribution ${\rm Im}(\mathcal{V}^-_+)$
vanishes due to {\it CPT} invariance if all the couplings are real. \s

\begin{table}[htb]
\begin{center}
\begin{tabular}{|c||c|c|}
\hline
\ \ Polar-angle distributions \ \         &        $P$         &       $CP$        \\
\hline\hline
 $\mathcal{C}^+_+ + \mathcal{C}^-_- $     &  \ \   even  \ \   &  \ \  even  \ \   \\
 $\mathcal{C}^+_+ - \mathcal{C}^-_- $     &  \ \   odd   \ \   &  \ \  even  \ \   \\
\hline
 $\mathcal{V}^+_+ + \mathcal{V}^-_- $     &  \ \   even  \ \   &  \ \  even  \ \   \\
 $\mathcal{V}^+_+ - \mathcal{V}^-_- $     &  \ \   odd   \ \   &  \ \  even  \ \   \\
\hline
     ${\rm Re}(\mathcal{V}^-_+) $         &  \ \   even  \ \   &  \ \  even  \ \   \\
     ${\rm Im}(\mathcal{V}^-_+) $         &  \ \   odd   \ \   &  \ \  odd   \ \   \\
\hline
\end{tabular}
\end{center}
\caption{\label{tab:symmetry_relations_on_production}
        {$P$ and $CP$ properties of the production polar-angle distributions
        separable with initial beam polarizations.}
        }
\end{table}

As can be checked with the expression of the last line in
Eq.$\,$(\ref{eq:generic_differential_cross_section}), the transverse-polarization dependent
parts can be non-zero only in the presence of some non-trivial ECV contributions so that
they serve as a useful indicator for the ECV parts. If both the electron and positron
longitudinal polarizations are available, then we can obtain the ECC and ECV parts of
the unpolarized cross section separately. For the degrees $\xi_\pm$ of $e^\pm$
longitudinal polarization the ECC and ECV parts of the cross section are given by the
relations:
\begin{eqnarray}
   \frac{d\sigma^{\mathcal{P}c}_{\rm unpol}}{d\cos\theta}
\! &=&\! \frac{\pi\alpha^2\beta}{4s} (\mathcal{C}^+_+ + \mathcal{C}^-_-)
 = \frac{1}{8\xi_-\xi_+}
    \left[ (1+\xi_-\xi_+) \left(\frac{d\sigma^{\mathcal{P}}_{\uparrow\downarrow}}{d\cos\theta}
                               +\frac{d\sigma^{\mathcal{P}}_{\downarrow\uparrow}}{d\cos\theta}
                               \right)
          -(1-\xi_-\xi_+) \left(\frac{d\sigma^{\mathcal{P}}_{\uparrow\uparrow}}{d\cos\theta}
                               +\frac{d\sigma^{\mathcal{P}}_{\downarrow\downarrow}}{d\cos\theta}
                               \right)
         \right]
    \label{eq:ecc_part_extraction} \\
   \frac{d\sigma^{\mathcal{P}v}_{\rm unpol}}{d\cos\theta}
\! &=&\! \frac{\pi\alpha^2\beta}{4s} (\mathcal{V}^+_+ + \mathcal{V}^-_-)
 = \frac{1}{8\xi_-\xi_+}
    \left[ (1+\xi_-\xi_+) \left(\frac{d\sigma^{\mathcal{P}}_{\uparrow\uparrow}}{d\cos\theta}
                               +\frac{d\sigma^{\mathcal{P}}_{\downarrow\downarrow}}{d\cos\theta}
                               \right)
          -(1-\xi_-\xi_+) \left(\frac{d\sigma^{\mathcal{P}}_{\uparrow\downarrow}}{d\cos\theta}
                               +\frac{d\sigma^{\mathcal{P}}_{\downarrow\uparrow}}{d\cos\theta}
                               \right)
         \right]
     \label{eq:ecv_part_extraction}
\end{eqnarray}
where the upper arrow ($\uparrow$) or down arrow ($\downarrow$) indicates that the direction of
longitudinal polarization is parallel or anti-parallel to the particle momentum with the first
and second one for the electron and positron, respectively. Furthermore, we can construct two
$P$-odd $LR$-asymmetric quantities, of which one is ECC and the other is ECV, as
\begin{eqnarray}
   \mathcal{A}^{\mathcal{P}c}_{LR}
&\equiv& \frac{\pi\alpha^2\beta}{4s} (\mathcal{C}^+_+ - \mathcal{C}^-_-)
  = \frac{1}{8\,\xi_-\xi_+}
     \left[ (\xi_- + \xi_+) \left(\frac{d\sigma^{\mathcal{P}}_{\uparrow\downarrow}}{d\cos\theta}
                               -\frac{d\sigma^{\mathcal{P}}_{\downarrow\uparrow}}{d\cos\theta}
                               \right)
          -(\xi_- - \xi_+) \left(\frac{d\sigma^{\mathcal{P}}_{\uparrow\uparrow}}{d\cos\theta}
                               -\frac{d\sigma^{\mathcal{P}}_{\downarrow\downarrow}}{d\cos\theta}
                               \right)
         \right]
    \label{eq:p-odd_lr_asymmetry_c} \\
   \mathcal{A}^{\mathcal{P}v}_{LR}
&\equiv& \frac{\pi\alpha^2\beta}{4s} (\mathcal{V}^+_+ - \mathcal{V}^-_-)
  = \frac{1}{8\,\xi_-\xi_+}
    \left[ (\xi_- + \xi_+) \left(\frac{d\sigma^{\mathcal{P}}_{\uparrow\uparrow}}{d\cos\theta}
                               -\frac{d\sigma^{\mathcal{P}}_{\downarrow\downarrow}}{d\cos\theta}
                               \right)
          -(\xi_- - \xi_+) \left(\frac{d\sigma^{\mathcal{P}}_{\uparrow\downarrow}}{d\cos\theta}
                               -\frac{d\sigma^{\mathcal{P}}_{\downarrow\uparrow}}{d\cos\theta}
                               \right)
         \right]
     \label{eq:p-odd_lr_asymmetry_v}
\end{eqnarray}
These observables, $\mathcal{A}^{\mathcal{P}c}_{LR}$ and $\mathcal{A}^{\mathcal{P}v}_{LR}$,
are expected to play a crucial role in diagnosing the chiral structure of the ECC and ECV parts
of the production process, respectively. Furthermore, Eq.$\,$(\ref{eq:ecc_part_extraction}) and
Eq.$\,$(\ref{eq:p-odd_lr_asymmetry_c}) are powerful even when electron chirality invariance
is violated. As we will see, they enable us to extract the ECC parts separately so
that the analysis of observables discussed in Sect.$\,$\ref{sec:observables} can be
adopted without any further elaboration. \s

\subsection{Charged spin-0 scalar pair \boldmath{$S^+_p S^-_p$} production}
\label{subsec:charged_scalar_pair}

The production of an electrically charged spin-0 scalar pair $S^+_p S^-_p$ in $e^+e^-$
collisions
\begin{eqnarray}
 e^-(p_-,\sigma_-) + e^+(p_+,\sigma_+)
 \, \to\,
 S^-_p (q_-) + S^+_p (q_+)
\label{eq:ee_to_SS}
\end{eqnarray}
is generally mediated by the $s$-channel exchange of neutral spin-0 $S^0_s$ and spin-1
$V^0_s$ (including the standard $\gamma$ and $Z$ bosons), by the $t$-channel exchange of
neutral spin-1/2 fermions $F^0_t$, and also by the $u$-channel exchange of doubly-charged
spin-1/2 fermions $F^{--}_u$. The $t$- or $u$-channel diagrams can contribute to the process
only when the produced scalar $S^-_p$ has the same electron number as the electron or
positron in theories with conserved electron number. (Again, $\sigma_{-,+}= \pm 1$ are
twice the electron and positron helicities and the convention $\sigma_{-,+}=\pm$
is used.) \s

The amplitude of the scalar-pair production process in Eq.$\,$(\ref{eq:ee_to_SS}) can be
expressed in terms of four generalized ECC and ECV bilinear charges, $Q^c_\pm$ and $Q^v_\pm$,
in the $e^+e^-$ c.m. frame as
\begin{eqnarray}
  {\cal M} \left[\, e^-_{\sigma_-} e^+_{\sigma_+}\, \to\, S^-_p S^+_p \right]
=  \sqrt{2} e^2 \left[ \delta_{\sigma_+, -\sigma_-}\, Q^c_{\sigma_-}
                     +\delta_{\sigma_+,\sigma_-}\, Q^v_{\sigma_-} \right]\,
    d^{J_0}_{\Delta\sigma, 0}(\theta)
\end{eqnarray}
where $J_0=|\Delta\sigma|$ with $\Delta\sigma=(\sigma_- -\sigma_+)/2 =\pm 1, 0$
and $\theta$ is the scattering polar angle between $S^-_p$ with respect to the $e^-$ direction
in the $e^+e^-$ c.m. frame. Explicitly, the ECC and ECV reduced helicity amplitudes are given
in terms of all the relevant 3-point couplings listed in Appendix A by
\begin{eqnarray}
&& Q^c_\pm = \beta \left[s^V_{ee\pm} s_V^{SS} D_s(M^2_{V_s})
                      - |t^{eS}_{F\pm}|^2 D_t(M^2_{F_t}, M^2_{S_p})
                      + |u^{eS}_{F\pm}|^2 D_u(M^2_{F_u}, M^2_{S_p})\right]
                      \label{eq:scalar_reduced_amplitude_c} \\
&& Q^v_\pm = -\frac{1}{\sqrt{2}\gamma}
            \left[ s^S_{ee\pm} s^{SS}_S  D_s(M^2_{S_s})
                 -\frac{M_{F_t}}{M_{S_p}}\,
                        t^{eS}_{F\pm} t^{eS*}_{F\mp}\,
                        D_t(M^2_{F_t}, M^2_{S_p})
                 -\frac{M_{F_u}}{M_{S_p}}\,
                        u^{eS}_{F\pm} u^{eS*}_{F\mp}\,
                        D_u(M^2_{F_u}, M^2_{S_p})\right]
                     \label{eq:scalar_reduced_amplitude_v}
\end{eqnarray}
in terms of the boost factor $\gamma=\sqrt{s}/2 M_{S_p}$ and the re-scaled
angle-independent $s$-channel propagator $D_s(M^2_a)$ and the re-scaled
angle-dependent $t$-channel and $u$-channel propagators, $D_t(M^2_a, M^2_b)$
and $D_u(M^2_a, M^2_b)$ defined as
\begin{eqnarray}
   D_s(M^2_a)
&=& \frac{1}{1-M^2_a/s+i M_a \Gamma_a/s}  \\
   D_{t/u}(M^2_a, M^2_b)
&=& \frac{1}{\Delta_{ab}\mp \beta\cos\theta}
\label{eq:t/u propagators}
% \\
%   D_u(M^2_a, M^2_b)
% &=& \frac{1}{\Delta_{ab}+\beta\cos\theta}
\end{eqnarray}
with $\Delta_{ab}=1+2(M^2_a-M^2_b)/s$ and $\cos\theta$ in the $e^+e^-$ c.m. frame.
All of the propagators are constant, i.e. independent of the polar angle at threshold
with $\beta=0$, i.e. when the scalar pair $S^+_p S^-_p$ are produced
at rest. (The width $\Gamma_a$ appearing in
the $s$-channel propagator is supposed to be much smaller than $M_a$ and  the c.m. energy
so that their effects will be ignored in our later numerical analyses.)\s

Using the explicit form of $d$ functions (see Appendix~\ref{sec:appendix_b_d_functions}),
we obtain the polarization-weighted differential cross sections of the production of scalar
particles as
\begin{eqnarray}
    \frac{d\sigma^S_{\rm pol}}{d\cos\theta}
&=& \frac{\pi\alpha^2}{8s} \beta
    \left[ (1-P_-^L P_+^L) (|Q^c_+|^2 + |Q^c_-|^2)\,\sin^2\theta
          +(P_-^L-P_+^L) (|Q^c_+|^2-|Q^c_-|^2)\,\sin^2\theta \right. \nonumber \\
& & \left. \qquad  +2(1+P_-^L P_+^L) (|Q^v_+|^2 + |Q^v_-|^2)
                   +2(P_-^L+P_+^L) (|Q^v_+|^2-|Q^v_-|^2) \right. \nonumber\\
& & \left. \qquad  + 4 P_-^T P_+^T \cos\delta\, {\rm Re}(Q^v_+ Q^{v*}_-)
                   + 4 P_-^T P_+^T \sin\delta\, {\rm Im}(Q^v_+ Q^{v*}_-)
     \right]
\label{eq:scalar_differential_cross_section}
\end{eqnarray}
where $P_\mp^{L,T}$ and $\delta$ are the degrees of longitudinal and transverse $e^\mp$
polarizations and the relative opening angle of the $e^\mp$ transverse polarizations.
The polarized total cross section $\sigma^S_{\rm pol}$ can be then obtained by integrating
the differential cross section over the full range of $\cos\theta$. One noteworthy point
is that the transverse-polarization dependent parts on the last line in
Eq.$\,$(\ref{eq:scalar_differential_cross_section}) survive even after the integration
if there exist any non-trivial ECV amplitudes.\s

Inspecting the polarization-weighted differential cross sections in
Eq.$\,$(\ref{eq:scalar_differential_cross_section}), we find the following aspects of the
scalar pair production:
\begin{itemize}
\item As previously demonstrated in detail for the production of scalar smuon or selectron
      pairs in SUSY models, the ECC part of the production cross section of an
      electrically-charged scalar pair in $e^+e^-$ collisions, originated from the
      $J=1$ $e^+e^-$ system, has two characteristic features. Firstly, the cross section
      rises slowly in $P$-waves near the threshold, i.e. $\sim\, \beta^3$ as the ECC
      amplitudes $Q^c_\pm$ are proportional to $\beta$. Secondly, as the total spin
      angular momentum of the final system of two spinless scalar particles is zero,
      angular momentum conservation generates the $\sin^2\theta$ dependence of the ECC part
      of the differential cross section, leading to the angular distribution
      $\sim\sin^2\theta$ near the threshold.
\item However, the two salient features of the ECC parts are spoiled by any
      non-trivial ECV contributions originated from $s$-channel scalar exchanges
      or $t$- and $u$-channel spin-1/2 fermion exchanges with both left-handed and
      right-handed couplings. Near the threshold the ECV amplitudes become constant.
      Therefore, in contrast to the ECC part the ECV part of the total cross section
      rises sharply in $S$-waves $\sim \beta$ and the ECV part of the differential
      cross section is isotropic.
\item As mentioned before, even in the presence of both the ECC and ECV contributions,
      the electron and positron beam polarizations can provide powerful diagnostic handles
      for differentiating the ECC and ECV parts. On one hand, the presence of the ECV
      contributions, if not suppressed, can be confirmed by transverse $e^\pm$
      polarizations.\footnote{As is well known, transversely-polarized electron and
      positron beams can be produced at $e^+e^-$ circular colliders by the guiding
      magnetic field of storage rings through its coupling to the magnetic moment
      of electrons and positrons.}
      On the other hand, longitudinal electron and positron polarizations enable
      us to extract out the ECC parts and to check the chiral structure of the three-point
      $ee S_s$, $e F_t S_p$ and $e F_u S_p$ couplings.
\item Then, the polar-angle distribution can be used for confirming the presence of $t$- or
      $u$-channel exchanges, as the distribution is peaked near the forward and/or backward
      directions for the $t$- and/or $u$-channel contributions.
\item If there exist only $s$-channel contributions, then the ECC and ECV part of the angular
      distribution is proportional to $\sin^2\theta$ and to a constant in the scalar-pair
      production in $e^+e^-$ collisions, respectively.
\end{itemize}
To find which of the these aspects are unique to the spin-0 case we need to compare them
with the spin-1/2 and spin-1 case. \s

Asymptotically the ECV amplitudes become vanishing $\sim M^2_{S_p}/s$ and the ECC ones
remain finite as can be checked with Eqs.$\,$(\ref{eq:scalar_reduced_amplitude_c}) and
(\ref{eq:scalar_reduced_amplitude_v}). As the c.m. energy increases, the ECV contributions
diminish and the ECC part of the unpolarized cross section of a scalar-pair production
scales as
\begin{eqnarray}
\sigma^{Sc}_{\rm unpol} \ \ \rightarrow \ \
\frac{\pi\alpha^2}{6s} \left(| s^V_{ee+} s^{SS}_V|^2
                                      + |s^V_{ee-} s^{SS}_V|^2\right)
\ \ \mbox{and} \ \ s\,\to\,\infty
\end{eqnarray}
in the absence of both $t$- and $u$-channel contributions, following the simple scaling
law $\propto 1/s$, and the cross section scales in the presence of the $t$-channel and
$u$-channel contributions as
\begin{eqnarray}
\sigma^{Sc}_{\rm unpol}
 & \rightarrow &
\frac{\pi\alpha^2}{4s}
   \left\{\left[\left(|t^{eS}_{F+}|^2\right)^2
               +\left(|t^{eS}_{F-}|^2\right)^2\right]
         \log \frac{s}{M^2_{F_t}}
        +\left[\left(|u^{eS}_{F+}|^2\right)^2
              +\left(|u^{eS}_{F-}|^2\right)^2\right]
         \log \frac{s}{M^2_{F_u}}\right\}
 \ \  \mbox{as} \ \ s\,\to\, \infty \nonumber\\
 & \rightarrow &
 \frac{\pi\alpha^2}{4s}
      \left[\left(|t^{eS}_{F+}|^2\right)^2
           +\left(|t^{eS}_{F-}|^2\right)^2
           +\left(|u^{eS}_{F+}|^2\right)^2
           +\left(|u^{eS}_{F-}|^2\right)^2\right]\,
           \log \frac{s}{M^2_{S_p}}
 \ \  \mbox{as} \ \  s\,\to\, \infty
\label{eq:scalar_pair_asymptotic_limit}
\end{eqnarray}
as expected from the near-forward and near-backward enhancements of the $t$- and $u$-channel
exchanges. (The expression on the last line in Eq.$\,$(\ref{eq:scalar_pair_asymptotic_limit})
is obtained by replacing all the intermediate masses by the scalar mass $M_{S_p}$ as
a typical mass scale.) As the ECC part of the $S^\pm_p$-pair production cross section is
zero in strict forward and backward direction $\theta=0,\pi$ due to angular momentum
conservation, the cross section remains scale-invariant apart from the logarithmic
coefficients.\s

\subsection{Charged spin-1/2 fermion pair \boldmath{$F^+_p F^-_p$} production}
\label{subsec:charged_fermion_pair}

The analysis presented in Subsect.$\,$\ref{subsec:charged_scalar_pair} for the scalar pair
production repeats itself rather closely for new spin-1/2 fermion states, $F^\pm_p$.
In addition to the standard $\gamma$ and $Z$ exchanges, there may exist the $s$-, $t$- and
$u$-channel exchanges of new spin-0 scalar states, $S^0_s, S^0_t$ and $S^{--}_u$, and new
spin-1 vector states, $V^0_s, V^0_t$ and $V^{--}_u$. Despite the complicated superposition
of scalar and vector interactions, the helicity amplitudes of the production
of an electrically-charged fermion pair, $F^+_p F^-_p$, can be decomposed into the ECC and
ECV parts as in Eq.$\,$(\ref{eq:generic_production_amplitude}) with $\Delta\sigma=
(\sigma_- -\sigma_+)/2=\pm 1, 0$, $\Delta\lambda=(\lambda_- - \lambda_+)/2 =\pm 1, 0$,
and $J_0={\rm max}(|\Delta\sigma|, |\Delta\lambda|)=1,0$. Explicitly, employing
the general couplings listed in Appendix A, we obtain for the ECC helicity amplitudes
$Q^{c}_{\sigma_-;\lambda_-, \lambda_+}$ for which $J_0=1$:
\begin{eqnarray}
    Q^c_{\pm;\lambda,\lambda}
&=& -\frac{1}{2\gamma} s^V_{ee\pm} \left(s^{FF}_{V+}+s^{FF}_{V-}\right)
     D_s (M^2_{V_s}) \nonumber\\
  &&  + \frac{1}{2\gamma}\left[ |t^{eF}_{S\pm}|^2
                                D_t (M^2_{S_t},M^2_{F_p})
                               -|u^{eF}_{S\pm}|^2
                                D_u (M^2_{S_u},M^2_{F_p}) \right] \nonumber\\
  && + \frac{1}{2\gamma}\left[ \left(2+\frac{M^2_{F_P}}{M^2_{V_t}}\right)
                                |t^{eF}_{V\pm}|^2
                                D_t (M^2_{V_t},M^2_{F_p})
                              -\left(2+\frac{M^2_{F_p}}{M^2_{V_u}}\right)
                                |u^{eF}_{V\pm}|^2
                                D_u (M^2_{V_u},M^2_{F_p})\right]
\label{eq:fermion_reduced_amplitude_c_v}
\end{eqnarray}
for the same $F^\mp_p$ helicities, $\lambda_- = \lambda_+ = \lambda = \pm $, and
\begin{eqnarray}
    Q^c_{\pm;\lambda,-\lambda}
&=& -\frac{1}{\sqrt{2}} s^V_{ee\pm}
     \left[(s^{FF}_{V+}+s^{FF}_{V-})+\lambda\beta\, (s^{FF}_{V+}-s^{FF}_{V-})\right]
           D_s (M^2_{V_s}) \nonumber\\
   && +\frac{1}{\sqrt{2}}\left[\left(1\mp\lambda\beta\right)
                                |t^{eF}_{S\pm}|^2
                                D_t (M^2_{S_t},M^2_{F_p})
                            -\left(1\pm\lambda\beta\right)
                                |u^{eF}_{S\pm}|^2
                                D_u (M^2_{S_u},M^2_{F_p})\right]\nonumber\\
   && +\frac{1}{\sqrt{2}} \left[ 2(1\pm\lambda\beta)
                                    +\frac{M^2_{F_p}}{M^2_{V_t}}(1\mp\lambda\beta)\right]
                                   |t^{eF}_{V\pm}|^2
                                   D_t (M^2_{V_t},M^2_{F_p})\nonumber\\
   && -\frac{1}{\sqrt{2}} \left[2(1\mp\lambda\beta)
                                    +\frac{M^2_{F_p}}{M^2_{V_u}}(1\pm\lambda\beta)\right]
                                   |u^{eF}_{V\pm}|^2
                                   D_u (M^2_{V_u},M^2_{F_p})
\label{eq:fermion_reduced_amplitude_c_c}
\end{eqnarray}
for the opposite $F^\mp_p$ helicities, $\lambda_- = - \lambda_+ = \lambda = \pm$ with
the boost factors, $\gamma=\sqrt{s}/2 M_{F_p}$ and $\beta=\sqrt{1-4 M^2_{F_p}/s}$.
On the other hand, the ECV reduced helicity amplitudes $Q^{v}_{\sigma_-;\lambda_-,\lambda_+}$
read
\begin{eqnarray}
    Q^v_{\pm;\lambda,\lambda}
&=& \frac{1}{2\sqrt{2}}\, s^S_{ee\pm}
                        \left[\lambda (s^{FF}_{S+}-s^{FF}_{S-})
                             -(s^{FF}_{S+}+s^{FF}_{S-})\,\beta\right]
                             D_s (M^2_{S_s}) \nonumber\\
&& -\frac{1}{2\sqrt{2}}\left(\beta \mp\lambda\right)
                       \left(1\pm\lambda\cos\theta\right)
                        t^{eF}_{S\pm}t^{eF*}_{S\mp}\,
                        D_t (M^2_{S_t},M^2_{F_p}) \nonumber\\
&& -\frac{1}{2\sqrt{2}}\left(\beta \mp\lambda\right)
                       \left(1\mp\lambda\cos\theta\right)
                        u^{eF}_{S\pm}u^{eF*}_{S\mp}\,
                        D_u (M^2_{S_u},M^2_{F_p}) \nonumber\\
&& +\frac{1}{\sqrt{2}} \left[2 (\beta\pm\lambda)
                            -\frac{M^2_{F_p}}{M^2_{V_t}}
                             (\beta \mp\lambda)
                              (1\pm\lambda\cos\theta)\right]
                             t^{eF}_{V\pm}t^{eF*}_{V\mp}\,
                             D_t (M^2_{V_t},M^2_{F_p}) \nonumber\\
&& +\frac{1}{\sqrt{2}} \left[2 (\beta\pm\lambda)
                            -\frac{M^2_{F_p}}{M^2_{V_u}}
                             (\beta\mp\lambda)
                             (1\mp\lambda\cos\theta)\right]
                             u^{eF}_{V\pm}u^{eF*}_{V\mp}\,
                             D_u (M^2_{V_u},M^2_{F_p})
\label{eq:fermion_reduced_amplitude_v_v}
\end{eqnarray}
for the same $F^\mp_p$ helicities, $\lambda_- = \lambda_+ = \lambda = \pm $, and
\begin{eqnarray}
   Q^v_{\pm;\lambda,-\lambda}
&=& \frac{1}{2\gamma} \left[ t^{eF}_{S\pm}t^{eF*}_{S\mp}\, D_t (M^2_{S_t},M^2_{F_p})
                            -u^{eF}_{S\pm}u^{eF*}_{S\mp}\, D_u (M^2_{S_u},M^2_{F_p})\right]
                           \nonumber\\
&& +\frac{1}{2\gamma}\left[ \frac{M^2_{F_p}}{M^2_{V_t}}\,
                            t^{eF}_{V\pm}t^{eF*}_{V\mp}\,
                            D_t (M^2_{V_t},M^2_{F_p})
                           -\frac{M^2_{F_p}}{M^2_{V_u}}\,
                            u^{eF}_{V\pm}u^{eF*}_{V\mp}\,
                            D_u (M^2_{V_u},M^2_{F_p}) \right]
\label{eq:fermion_reduced_amplitude_v_c}
\end{eqnarray}
for the opposite $F^\mp_p$ helicities, $\lambda_- = - \lambda_+ = \lambda = \pm$.
From these ECC and ECV reduced amplitudes, one can get the polarized differential cross
section by using Eq.$\,$(\ref{eq:generic_differential_cross_section}). \s

Inspecting the explicit form of the ECC and ECV reduced helicity amplitudes leads to the
following features of the amplitudes:
\begin{itemize}
\item Near threshold, the ECC reduced amplitudes become independent
      of the $F^\pm_p$ helicities, leading to the relation
      $Q^c_{\pm;\lambda,-\lambda}=\sqrt{2}\, Q^c_{\pm;\lambda,\lambda}$.
      This implies that the ECC part of the unpolarized differential cross section
      behaves like
\begin{eqnarray}
 \frac{d\sigma^{Fc}_{\rm unpol}}{d\cos\theta}
\,\sim\, \left[1+\beta^2\cos^2\theta\right]\, \mathcal{G}(\beta\cos\theta)
          + \cdots
\quad   \to\quad \mbox{flat near the threshold}
\end{eqnarray}
\item Because not only the $e^\pm$ but also the particle $F^\pm_p$ are electrically charged,
      there exists at least an $s$-channel $\gamma$ exchange contribution to the
      production process with pure vector-current couplings as $s^\gamma_{ee\pm}
      = s^{FF}_{\gamma\pm} = +1$. This contribution generates a non-zero significant
      amplitude at threshold with $\beta=0$ as can be proved with
      Eq.$\,$(\ref{eq:fermion_reduced_amplitude_c_c}). Therefore, the rise of the
      excitation curve of the unpolarized production cross section must be of
      an $S$-wave type, i.e. $\sigma^{Fc}_{\rm unpol} \sim \beta$ near the threshold.
      Note that this threshold pattern is not spoiled by the ECV contributions.
\item If there are neither $t$-channel nor $u$-channel exchange diagrams, the
      ECV reduced helicity amplitudes  $Q^v_{\pm;\lambda,-\lambda}$ are vanishing and
      all the other non-vanishing ECV reduced amplitudes are constant. Therefore, the
      ECV part of the polar-angle distribution is isotropic. On the other hand, in this case,
      the production cross section rises in $P$-waves or $S$-waves when the $S_s F_p F_p$
      coupling is of a pure scalar type ($s^{FF}_{S+}=s^{FF}_{S-}$) or of a pure
      pseudoscalar type ($s^{FF}_{S+}=-s^{FF}_{S-}$).
\item The ECV ECC $t$-channel and/or $u$-channel contributions arise from non-chiral $e S_tF_p,
      eS_u F_p$ scalar and/or $e V_t F_p, e V_u F_p$ vector couplings. They develop a
      non-trivial angular dependence near the threshold
\begin{eqnarray}
                \frac{d\sigma^{Fv}_{\rm unpol}}{d\cos\theta}
\quad \to \quad \left[ a_v + b_v \cos^2\theta\right] + \cdots \ \
                \mbox{with $a_v>0$ and $b_v\neq 0$ near the threshold}
\end{eqnarray}
The sign of the coefficient $b_v$ depends on the relative size of the scalar and vector
contributions in the $t$- and $u$-channel diagrams.
\end{itemize}
Compared with the spin-0 case, we can claim that the spin-1/2 case has distinct
characteristics in the threshold behavior and the polar-angle distribution.\s

As the c.m. energy increases, the ECC amplitudes with the same $F^\pm_p$ helicities and the
ECV amplitudes with the opposite $F^\pm_p$ helicities vanish $\sim M^2_{F_p}/s$.
However, the ECC amplitudes with the opposite $F^\pm_p$ helicities and the ECV amplitudes
with the same $F^\pm_p$ helicities are finite in the asymptotic high-energy limit as can
be checked with Eqs.$\,$(\ref{eq:fermion_reduced_amplitude_c_v}),
(\ref{eq:fermion_reduced_amplitude_c_c}),(\ref{eq:fermion_reduced_amplitude_v_v}) and
(\ref{eq:fermion_reduced_amplitude_v_c}). Therefore, unlike the spin-0 case,
both the ECC and ECV parts of the unpolarized cross section of the fermion-pair production
scale asymptotically as
\begin{eqnarray}
\sigma^{Fc}_{\rm unpol} \ \
& \rightarrow & \ \
\frac{\pi\alpha^2}{3s}
\left(|s^V_{ee+} s^{FF}_{V+}|^2 + |s^V_{ee-} s^{FF}_{V-}|^2
     +|s^V_{ee+} s^{FF}_{V-}|^2 + |s^V_{ee-} s^{FF}_{V-}|^2 \right)
\ \ \mbox{as} \ \ s\,\to\,\infty \\
\sigma^{Fv}_{\rm unpol} \ \
& \rightarrow & \ \
\frac{\pi\alpha^2}{4s}
\left(|s^S_{ee+} s^{FF}_{S+}|^2 + |s^S_{ee-} s^{FF}_{S-}|^2
     +|s^V_{ee+} s^{FF}_{S-}|^2 + |s^V_{ee-} s^{FF}_{S-}|^2 \right)
\ \ \mbox{as} \ \ s\,\to\,\infty
\end{eqnarray}
in the absence of both $t$- and $u$-channel contributions, following the simple scaling
law $\propto 1/s$, and both the ECC and ECV parts of the cross section scale
in the presence of the $t$-channel and $u$-channel contributions as
\begin{eqnarray}
\sigma^{Fc}_{\rm unpol}
 & \rightarrow &
     \pi\alpha^2
   \left[\frac{1}{M^2_{V_t}} \left((|t^{eF}_{V+}|^2)^2 + (|t^{eF}_{V-}|^2)^2\right)
        +\frac{1}{M^2_{V_u}} \left((|u^{eF}_{V+}|^2)^2 + (|u^{eF}_{V-}|^2)^2\right)
        \right]
 \quad  \mbox{as} \quad  s\,\to\, \infty \\
\sigma^{Fv}_{\rm unpol}
 & \rightarrow &
      \pi\alpha^2
      \left[\frac{M^4_{F_p}}{M^6_{V_t}}
            |t^{eF}_{V+} t^{eF*}_{V-}|^2
           +\frac{M^4_{F_p}}{M^6_{V_u}}
            |u^{eF}_{V+} u^{eF*}_{V-}|^2 \right]
 \quad  \mbox{as} \quad  s\,\to\, \infty
\label{eq:vector_pair_asymptotic_limit}
\end{eqnarray}
as expected from the forward and backward enhancements of the $t$- and $u$-channel
exchanges, which is a remnant of the Rutherford pole damped by the Yukawa mass cut-off
in the exchange of heavy particles. The size of the cross section is set by the
Compton wave-lengths of the particles exchanged in the $t$-channel and/or $u$-channel.\s

\subsection{Charged spin-1 vector-boson pair \boldmath{$V^+_p V^-_p$} production}
\label{subsec:charged_vector-boson_pair}

Similarly to the production of an electrically-charged spin-0 scalar pair, the production of
an electrically-charged spin-1 vector-boson pair $V^+_p V^-_p$ in $e^+e^-$ collisions
\begin{eqnarray}
 e^-(p_-,\sigma_-) + e^+(p_+,\sigma_+)
 \, \to\,
 V^-_p (q_-,\lambda_-) + V^+_p (q_+,\lambda_+)
\label{eq:ee_to_VV}
\end{eqnarray}
is generally mediated by the $s$-channel exchange of neutral spin-0 particles $S^0_s$ and
spin-1 particles $V^0_s$ (including the standard $\gamma$ and $Z$ bosons), by the $t$-channel
exchange of neutral spin-1/2 fermions $F^0_t$, and also by the $u$-channel exchange of
doubly-charged spin-1/2 fermions $F^{--}_u$, if the produced scalar $V^-_p$ has the same
lepton number as the positron, when electron number conservation is imposed on the theory.
Here, $\sigma_-, \sigma_+$ are twice the electron and positron helicities and
$\lambda_-, \lambda_+=\pm 1, 0$ are the $V^\mp_p$ helicities, respectively.\s

The amplitude describing the production process in Eq.$\,$(\ref{eq:ee_to_VV}) can be expressed
in terms of the scattering angle $\theta$ between the $e^-$ and $V^-_p$ momentum directions in
the $e^+e^-$ c.m. frame as in Eq.$\,$(\ref{eq:generic_production_amplitude})
with $\Delta\sigma=(\sigma_- - \sigma_+)/2 = 0,\pm 1$,
$\Delta\lambda=\lambda_- - \lambda_+  =0, \pm 1, \pm 2$ and
$J_0={\rm max}(|\Delta\sigma|,|\Delta\lambda|)$.
Explicitly, the ECC reduced helicity amplitudes $Q^c_{\sigma_-;\lambda_-, \lambda_+}$ are
given by
\begin{eqnarray}
    Q^c_{\sigma_-;\pm,\pm}
&=& -\beta s^V_{ee\sigma_-}s^{VV}_V\,
           D_s (M^2_{V_s})
       + (\beta-\cos\theta) | t^{eV}_{F\sigma_-}|^2\,
          D_t (M^2_{F_t},M^2_{V_p})
       \!-\! (\beta+\cos\theta) | u^{eV}_{F\sigma_-}|^2\,
          D_u (M^2_{F_u},M^2_{V_p}) \\
    Q^c_{\sigma_-;\,0,\,\, 0}
&=& (2\gamma^2+1)\, Q^c_{\sigma_-;\pm,\pm}
    +\cos\theta \left[|t^{eV}_{F\sigma_-}|^2\,
                       D_t (M^2_{F_t},M^2_{V_p})
                     +|u^{eV}_{F\sigma_-}|^2\,
                       D_u (M^2_{F_u},M^2_{V_p})\right] \\
    Q^c_{\sigma_-;\pm,\, 0}
&=& Q^c_{\sigma_-;0,\,\mp}
 = 2\gamma\, Q^c_{\sigma_-;\pm,\pm}
    \pm\frac{\sigma_-}{\gamma}
    \left[ |t^{eV}_{F\sigma_-}|^2\, D_t (M^2_{F_t},M^2_{V_p})
          +|u^{eV}_{F\sigma_-}|^2\, D_u (M^2_{F_u},M^2_{V_p})\right] \\
   Q^c_{\sigma_-;\pm,\mp}
&=& -\sqrt{2} \left[ |t^{eV}_{F\sigma_-}|^2\, D_t (M^2_{F_t},M^2_{V_p})
                    +|u^{eV}_{F\sigma_-}|^2\, D_u (M^2_{F_u},M^2_{V_p}) \right]
\label{eq:vector_reduced_amplitude_c}
\end{eqnarray}
and the ECV reduced helicity amplitudes by
\begin{eqnarray}
   Q^v_{\sigma_-;\pm,\pm}
&=& -\frac{1}{\sqrt{2}\gamma}
            s^S_{ee\sigma_-} s^{VV}_S
            D_s (M^2_{S_s}) \nonumber\\
    && +\frac{1}{\sqrt{2}\gamma} \frac{M_{F_t}}{M_{V_p}}
             (1\pm\sigma_-\cos\theta)\,
             t^{eV}_{F\sigma_-} t^{eV*}_{F-\sigma_-}
             D_t (M^2_{F_t},M^2_{V_p}) \nonumber\\
    && +\frac{1}{\sqrt{2}\gamma} \frac{M_{F_u}}{M_{V_p}}
             (1\mp\sigma_-\cos\theta)\,
             u^{eV}_{F\sigma_-} u^{eV*}_{F-\sigma_-}
             D_u (M^2_{F_u},M^2_{V_p})
\end{eqnarray}
for both transversely polarized vector bosons with the same helicity, and
\begin{eqnarray}
   Q^v_{\sigma_-;\,0,\, 0}
&=& \sqrt{2}\gamma(1-\gamma^{-2}/2)\, s^S_{ee\sigma_-}s^{VV}_S D_s (M^2_{S_s}) \nonumber\\
 && -\sqrt{2}\gamma \frac{M_{F_t}}{M_{V_p}}
                         (1-\gamma^{-2}/2-\beta\cos\theta)\,
                         t^{eV}_{F\sigma_-} t^{eV*}_{F-\sigma_-}
                         D_t (M^2_{F_t},M^2_{V_p}) \nonumber\\
 && -\sqrt{2}\gamma \frac{M_{F_u}}{M_{V_p}}
                         (1-\gamma^{-2}/2+\beta\cos\theta)\,
                         u^{eV}_{F\sigma_-} u^{eV*}_{F-\sigma_-}
                         D_u (M^2_{F_u},M^2_{V_p})
\end{eqnarray}
for both longitudinally polarized vector bosons, respectively. For
$|\Delta\lambda|= 1$ with one transversely polarized and one longitudinally polarized
vector bosons and for $|\Delta\lambda|=2$, we have
\begin{eqnarray}
    Q^v_{\sigma_-;\pm,\, 0}
&=& Q^v_{\sigma_-;0,\,\pm}
   = \frac{(\beta\pm\sigma_-)}{\sqrt{2}}
     \left[ \frac{M_{F_t}}{M_{V_p}}\,
           t^{eV}_{F\sigma_-} t^{eV*}_{F-\sigma_-}\,
           D_t (M^2_{F_t},M^2_{V_p})
           -\frac{M_{F_u}}{M_{V_p}}\,
           u^{eV}_{F\sigma_-} u^{eV*}_{F-\sigma_-}\,
           D_u (M^2_{F_u},M^2_{V_p}) \right] \\
     Q^v_{\sigma_-;\pm,\mp}
&=& 0
\label{eq:vector_reduced_amplitude_v}
\end{eqnarray}
Here, the boost factors are $\gamma=\sqrt{s}/2 M_{V_p}$ and $\beta=\sqrt{1-4 M^2_{V_p}/s}$.
The ECC diagrams with $s$-channel $V_s^0$-exchange such as the
standard $s$-channel $\gamma$ and $Z$ exchange have only a $J=1$ partial wave because of
angular momentum conservation, contributing to only the seven final helicity
combinations with $J_0=1$. On the other hand, the diagrams with $t$-channel and $u$-channel
fermion exchanges have all the partial waves with $J\geq J_0$.\s

In the case with $J_0=2$ only the $F^0_t$ and $F^{--}_u$ exchange diagrams can contribute
to this final-state configuration. Moreover, because $|\Delta\lambda|=2$, the final vector
bosons are both transverse $[(\lambda,\bar{\lambda})=(\pm,\mp)]$.
Thus these amplitudes do not have any bad high-energy behavior. \s

The other seven ECC final helicity combinations give $J_0=1$. Five of them have at least
one longitudinal $V_p$, which could give a divergent behavior at high energies. Some parts
of the amplitudes $Q^{c}_{\sigma_-;\pm, 0}=Q^{c}_{\sigma_-;0,\pm}$ and $Q^c_{\sigma_-;0,0}$
are proportional to the ECC amplitude $Q^{c}_{\sigma_-;\pm,\pm}$ with the proportionality
coefficients, $\gamma$ or $\gamma^2$, respectively, as expected from longitudinal
$V^\pm_p$ counting. To avoid the bad high-energy behavior, it is necessary to satisfy the
two relations\footnote{If the electron mass is not ignored, additional
divergent parts proportional to the mass  appear in the ECV parts with longitudinally
polarized $V^\pm_p$. They can be cancelled by the $s$-channel scalar exchanges with
their couplings proportional to the electron mass as in the SM.} among the couplings as
provided by gauge symmetry in the SM \cite{Cornwall:1973tb,Cornwall:1974km,
Llewellyn Smith:1973ey}:
\begin{eqnarray}
     s^V_{ee\pm} s^{VV}_V
&=& |t^{eV}_{F\pm}|^2 - |u^{eV}_{F\pm}|^2
\label{eq:cancellation_condition_c}\\
    s^S_{ee\pm} s^{VV}_S
&=& \frac{M_{F_t}}{M_{V_p}} t^{eV}_{F\pm} t^{eV*}_{F\mp}
   +\frac{M_{F_u}}{M_{V_p}} u^{eV}_{F\pm} u^{eV*}_{F\mp}
\label{eq:cancellation_condition_v}
\end{eqnarray}
for each electron helicity $\sigma_-=\pm$, leading to an effective cancellation among
the $s$-channel, $t$-channel and $u$-channel contributions so that the ECC
amplitudes $Q^c_{\sigma_-;\pm,\pm}$ and $Q^c_{\sigma_-;0,0}$ and the ECV ampltitudes $Q^v_{\sigma_-;0,0}$
vanish asymptotically as the c.m.
energy increases.\footnote{The cancellation conditions enforce the condition that
$s^V_{ee\pm} s^{VV}_V$ are real and $s^S_{ee\pm } s^{VV}_S$ are complex conjugate to
each other.} \s

If the ECC cancellation condition (\ref{eq:cancellation_condition_c}) for the ECC part
is satisfied, the ECC amplitudes $Q^c_{\sigma_-;\pm, 0}=Q^c_{\sigma_-;0,\pm}$ for one
longitudinal and one transverse $V_p$ pair decrease as $\gamma^{-1}$ at high energies, while
the ECC amplitudes $Q^c_{\sigma_-;\pm,\pm}$ are suppressed by $(1-\beta)\sim\gamma^{-2}$ since $D_{t/u}\sim \left(1\mp\beta\cos\theta\right)^{-1}$ at high energies.
Therefore, only three of the nine ECC helicity combinations, $(+,-), (-,+)$ and $(0,0)$,
survive at high energies.  On the other hand, if the ECV cancellation condition
(\ref{eq:cancellation_condition_v}) is satisfied, the $J_0=0$ ECV amplitudes,
$Q^v_{\sigma_-;\pm,\pm}$ and $Q^v_{\sigma_-;0,0}$ are suppressed by $\gamma^{-1}$ while
the $J_0=1$ ECV amplitudes, $Q^v_{\sigma_-;\pm, 0}$ and $Q^v_{\sigma_-;0,\pm}$, survive
at high energies.\s

The three ECC amplitudes surviving at high energies do not contribute to the ECC cross
section equally. The $J_0=2$ ECC amplitudes with the $(\pm,\mp)$ helicity combinations
dominate over the other $(0,0)$ ECC amplitude at high energies because of the $t$-channel
and/or $u$-channel polar factors $1/(\Delta \mp\beta \cos\theta)$ which peaks at
$\cos\theta=\pm 1$ with a $\left(1-\beta\right)^{-1}\sim\gamma^2$ enhancement. (In practice the peaks appear below
$|\cos\theta|=1$ because the relevant $d^2_{\sigma_-;\pm 2}$ functions with $|\sigma_-|=1$
are proportional to $\sin\theta$ and vanish at $|\cos\theta|=1$.)
As there must exist the $t$-channel and/or $u$-channel contributions for preserving the
good high-energy behavior of the cross section by compensating the $s$-channel $\gamma, Z$
contributions for both of the $e^+e^-$ helicity combinations $(\sigma_-,\sigma_+)=(\pm,\mp)$,
the ECC unpolarized cross section scales asymptotically as
\begin{eqnarray}
           \sigma^{Vc}_{\rm unpol} \ \
\to \ \ \frac{4\pi\alpha^2}{s}
     \left\{\left[(|t^{eV}_{F+}|^2)^2+(|t^{eV}_{F-}|^2)^2\right]\,
            \log \frac{s}{M^2_{F_t}}
            +\left[(|u^{eV}_{F+}|^2)^2+(|u^{eV}_{F-}|^2)^2\right]\,
            \log \frac{s}{M^2_{F_u}} \right\} \quad \mbox{as} \ \
            s\, \to\, \infty
\end{eqnarray}
which follows the typical scaling law $\propto 1/s$ apart from the logarithmic parts.\s

In contrast, the $J_0=2$ ECV amplitudes are zero and, with the ECV cancellation condition
(\ref{eq:cancellation_condition_v}), only the $J_0=1$ ECV amplitudes $Q^v_{\pm;\pm, 0}$
and $Q^v_{\pm;0,\pm}$ survive asymptotically, leading to the form of the ECV cross section:
\begin{eqnarray}
 \sigma^{Vv}_{\rm unpol} \ \
\to \ \ \frac{2\pi\alpha^2}{s}
     \left[ \frac{M^2_{F_t}}{M^2_{V_p}}\,
            |t^{eV}_{F+} t^{eV*}_{F-}|^2\, \log \frac{s}{M^2_{F_t}}
           +\frac{M^2_{F_u}}{M^2_{V_p}}\,
            |u^{eV}_{F+} u^{eV*}_{F-}|^2\, \log \frac{s}{M^2_{F_u}} \right]
           \quad \mbox{as} \ \  s\, \to\, \infty
\end{eqnarray}
which follows the scaling law $\propto 1/s$ apart from the logarithmic parts
with the mass-squared of the intermediate particles indicating the chiral-flipping
phenomena.\s

At threshold of the spin-1 vector pair production, the total spin becomes equal to
the total angular momentum so that it takes only the three values, $J=0,1, 2$,
because no orbital angular momentum is developed between the final $V^\pm_p$.
Among the three possible angular momenta, $J=0$ is forbidden for the ECC parts
because the initial $e^+e^-$ state can have only $J\geq 1$ if the electron mass is
neglected. The ECC part of the cross section needs to have a $J=2$ contribution from
$t$-channel or $u$-channel spin-1/2 fermion exchanges or a $J=1$ contribution from
new $s$-channel spin-1 vector-boson exchanges, partly as a means for erasing the bad
high-energy behavior. In the presence of the $t$- or $u$-channel contributions as in
the SM, the ECC part of the total cross section rises sharply in $S$-waves near
threshold as
\begin{eqnarray}
    \sigma^{Vc}_{\rm unpol}
\sim  \frac{4\pi\alpha^2}{M^2_{V_p}}\,
        \left\{ \bigg[\frac{|t^{eV}_{F_+}|^2}{1+M^2_{F_t}/M^2_{V_p}}
                     +\frac{|u^{eV}_{F_+}|^2}{1+M^2_{F_u}/M^2_{V_p}}\bigg]^2
               +\bigg[\frac{|t^{eV}_{F_-}|^2}{1+M^2_{F_t}/M^2_{V_p}}
                     +\frac{|u^{eV}_{F_-}|^2}{1+M^2_{F_u}/M^2_{V_p}}\bigg]^2
              \right\}\, \beta
\end{eqnarray}
while the ECC part of the angular distribution
\begin{eqnarray}
   \frac{1}{\sigma^{Vc}_{\rm unpol}}\frac{d\sigma^{Vc}_{\rm unpol}}{d\cos\theta}
\ \ \sim \ \ \frac{1}{2} + O(\beta) \cos\theta
\end{eqnarray}
is essentially flat in the threshold region and the flat behavior is modified
linearly in $\beta$ above the threshold, unless the theory is $P$-invariant.\s

If there exist any ECV contributions in the $s$-, $t$- and/or $u$-channel diagrams
due to non-chiral couplings, the ECV amplitudes for the spin-1 vector-boson pair
production are finite at threshold so that the ECV part of the cross
section rises sharply in $S$-waves near threshold as
\begin{eqnarray}
   \sigma^{Vv}_{\rm unpol}
\ \ \sim \ \
   \frac{8\pi\alpha^2}{M^2_{V_p}}
   \bigg[\, 3\, \mathcal{B}_1
          + 2\, \mathcal{B}_2\, \bigg]\, \beta
\end{eqnarray}
with the non-negative functions defined as $\mathcal{B}_1$ and $\mathcal{B}_2$
\begin{eqnarray}
    \mathcal{B}_1
&=& \bigg|\, 2 s^{S}_{ee+} s^{VV}_S \frac{M^2_{V_p}}{4M^2_{V_p}-M^2_{S_s}}
           - t^{eV}_{F+} t^{eV*}_{F-} \frac{M_{V_p} M_{F_t}}{M^2_{V_p}+M^2_{F_t}}
           - u^{eV}_{F+} u^{eV*}_{F-} \frac{M_{V_p} M_{F_u}}{M^2_{V_p}+M^2_{F_u}}
           \bigg|^2 \\
    \mathcal{B}_2
&=& \bigg|\, t^{eV}_{F+} t^{eV*}_{F-} \frac{M_{V_p} M_{F_t}}{M^2_{V_p}+M^2_{F_t}}
         -u^{eV}_{F+} u^{eV*}_{F-} \frac{M_{V_p} M_{F_u}}{M^2_{V_p}+M^2_{F_u}}
           \bigg|^2
\end{eqnarray}
and, similarly to the ECC part,  the ECV part of the angular distribution is essentially
flat in the threshold region.\s

Comparing the predictions for the excitations of the spin-1 electrically-charged
vector bosons with those of the spin-1/2 electrically-charged fermions leads us to the
conclusion that the onset of the excitation curves alone does not
discriminate one from the other. Therefore, the analyses of the final-state two-body
decay processes and/or production-decay angular correlations are required for
discriminating the spin-1 vector bosons from the spin-1/2 fermions.\s

\begin{figure}[htb]
\centering
\includegraphics[scale=1.0]{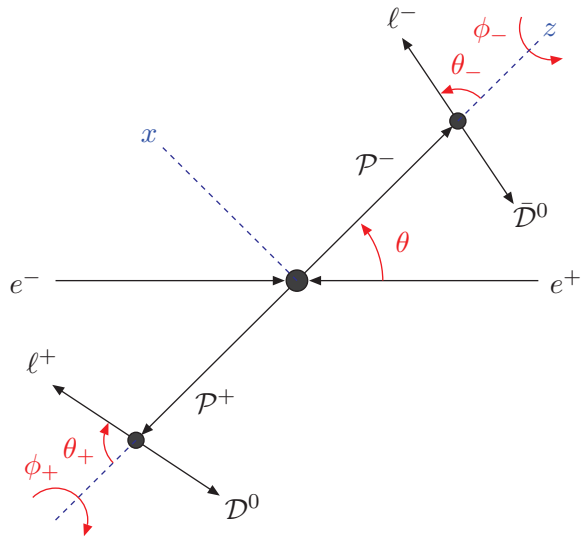}
\caption{\label{fig:coordinate_system_diagram}
        {The coordinate system in the colliding $e^+e^-$ c.m. frame. The $y$-axis is chosen
             along the $\vec{p}_-(e^-)\times \vec{q}_- (\mathcal{P}^-)$  direction and it is
             pointing towards the observer. The coordinate systems in the $\mathcal{P}^-$ and
             $\mathcal{P}^+$ rest frames are obtained from it by boosts along the $z$-axis.
             The angles $\theta_\pm$ and $\phi_\pm$ are the polar and azimuthal angles of the
             lepton $\ell^\pm$ associated with the two-body decay of the $\mathcal{P}^\pm$
             particles in their respective rest frames.}
         }
\end{figure}

\section{Two-body Decays}
\label{sec:two_body_decays}

The decay amplitudes ${\cal D}_-$ and ${\cal D}_+$ of the two-body decays,
$\mathcal{P}^- \to \ell^-\mathcal{\bar{D}}^0$ and $\mathcal{P}^+\to\ell^+\mathcal{D}^0$,
are most simply expressed in the $\mathcal{P}^-$ and $\mathcal{P}^+$ rest frames, respectively.
We define each of these frames by a boost of the $e^+e^-$ c.m. frame along the $z$-axis as
shown in Fig.$\,$\ref{fig:coordinate_system_diagram}. In the $\mathcal{P}^\mp$ rest frame,
we parameterize the $\ell^\mp$ four-momenta, $p_1$ and $p_3$, as
\begin{eqnarray}
p_1^\mu = p^\mu_{\ell^-} &=& \frac{M^2_{\mathcal{P}}-M^2_{\mathcal{D}}}{2 M_{\mathcal{P}}}
               \left(1, \sin\theta_- \cos\phi_-,\sin\theta_-\sin\phi_-, \cos\theta_-\right)\\
p_3^\mu = p^\mu_{\ell^+} &=& \frac{M^2_{\mathcal{P}}-M^2_{\mathcal{D}}}{2 M_{\mathcal{P}}}
               \left(1, \sin\theta_+ \cos\phi_+,\sin\theta_+\sin\phi_+, -\cos\theta_+\right)
\end{eqnarray}
In this convention of the coordinate systems the angles of the charged lepton are chosen
as $(\theta_-,\phi_-)$ in the $\mathcal{P}^-$ decays and $(\pi-\theta_+, \phi_+)$ in the
$\mathcal{P}^+$ decays.\s

It is a straightforward exercise to evaluate the helicity amplitudes of the decays
$\mathcal{P}^- \to\ell^-\mathcal{\bar{D}}^0$ and $\mathcal{P}^+\to\ell^+ \mathcal{D}^0$ with
the general couplings listed in Appendix A in the $\mathcal{P}^\mp$ rest frames described before.
Generically, when the charged lepton masses are ignored, the decay amplitudes can be written as
\begin{eqnarray}
    {\cal D}_-[\mathcal{P}^-_{\lambda_-}
               \to \ell^-_{\sigma_1}\mathcal{\bar{D}}^0_{\sigma_2}]
&=& e\, K_{\mathcal{PD}}\, D_-[\mathcal{P}^-\to\ell^-\mathcal{\bar{D}}^0]_{\sigma_1 \sigma_2}\,
    d^{J_{\mathcal{P}}}_{\lambda_-,\sigma_1-\sigma_2}(\theta_-)\,
    {\rm e}^{i(\lambda_- - \sigma_1+\sigma_2)\,\phi_-}
\label{eq:generic_decay_amplitude_12}
     \\
    {\cal D}_+[\mathcal{P}^+_{\lambda_+}
               \to \ell^+_{\sigma_3} \mathcal{D}^0_{\sigma_4}]
&=& e\, K_{\mathcal{PD}}\, D_+[\mathcal{P}^+\to\ell^+ \mathcal{D}^0]_{\sigma_3\sigma_4}\,
    d^{J_{\mathcal{P}}}_{\lambda_+,\sigma_3-\sigma_4}(\theta_+)\,
    {\rm e}^{-i(\lambda_+ - \sigma_3+\sigma_4)\,\phi_+}
\label{eq:generic_decay_amplitude_34}
\end{eqnarray}
with $K_{\mathcal{PD}}=\sqrt{M^2_{\mathcal{P}}-M^2_{\mathcal{D}}}$ and
$\lambda_-(\lambda_+) , \sigma_1 (\sigma_3)$ and $\sigma_2 (\sigma_4)$ the helicities of
the particles $\mathcal{P}^- (\mathcal{P}^+), \ell^- (\ell^+)$ and $\mathcal{\bar{D}}^0
(\mathcal{D}^0)$. We obtain for all the decay combinations with $\mathcal{P}^\pm
=S^\pm_p ,F^\pm_p ,V^\pm_p$ and $\mathcal{D}^0=F^0_d, S^0_d, V^0_d$
the reduced decay helicity amplitudes:
\begin{eqnarray}
    D_-[S^-_p \to \ell^- \bar{F}^0_d ]_{\sigma_1\sigma_2}
&=& \delta_{\sigma_2\sigma_1}\, d^{\ell F}_{S\sigma_1}
    \label{eq:decay_amplitude_sf} \\
    D_-[F^-_p \to \ell^-\, \bar{S}^0_d]_{\sigma_1}\ \
&=& d^{\ell S}_{F\sigma_1}
    \label{eq:decay_amplitude_fs} \\
    D_-[F^-_p \to \ell^- \bar{V}^0_d]_{\sigma_1\sigma_2}
&=& - \left[\sqrt{2}\,\delta_{\sigma_2\sigma_1}
         +\delta_{\sigma_2 0} (M_{F_p}/M_{V_d}) \right]
         d^{\ell V}_{F\sigma_1}
    \label{eq:decay_amplitude_fv}  \\
    D_-[V^-_p \to \ell^- \bar{F}^0_d]_{\sigma_1\sigma_2}
&=& \left[\delta_{\sigma_2 \sigma_1} (M_{F_d}/M_{V_p})
        +\sqrt{2}\,\delta_{\sigma_2,-\sigma_1}\right]
         d^{\ell F}_{V\sigma_1}
    \label{eq:decay_amplitude_vf}
\end{eqnarray}
and the reduced decay amplitudes for the charge-conjugated decays $\mathcal{P}^+\to\ell^+
\mathcal{D}^0$ are given by the relation
\begin{eqnarray}
  D_+[\mathcal{P}^+\to \ell^+ \mathcal{D}^0]_{\sigma_3 \sigma_4}
= \mp D_-[\mathcal{P}^- \to \ell^- \mathcal{\bar{D}}^0]^*_{-\sigma_3,-\sigma_4}
\end{eqnarray}
up to an overall sign. The sign $+$ is for $\mathcal{P}^\pm = V^\pm_p$ and the sign
$-$ for $\mathcal{P}^\pm= S^\pm_p, F^\pm_p$.\s

\section{Full angular-correlations of the final-state leptons}
\label{sec:full_angular_correlations}

In this section we present the most general angular distribution of the decay products in
the correlated production-decay process, following the formalism in
Ref.$\,$\cite{Hagiwara:1986vm}
\begin{eqnarray}
         e^-(p_-,\sigma_-) + e^+(p_+,\sigma_+)\,
&\to& \, \mathcal{P}^-(q_-,\lambda_-) + \mathcal{P}^+(q_+,\lambda_+) \nonumber\\
         \mathcal{P}^-(q_-,\lambda_-)\,
&\to& \, \ell^-(p_1,\sigma_1) + \mathcal{\bar{D}}^0 (p_2,\sigma_2) \nonumber\\
         \mathcal{P}^+(q_+,\lambda_+)\,
&\to& \, \ell^+(p_3,\sigma_3) + \mathcal{D}^0(p_4,\sigma_4)
\end{eqnarray}
with two visible massless charged leptons $\ell^\pm$ and two invisible
neutral particles $\mathcal{D}^0$ and $\mathcal{\bar{D}}^0$ in the final state. Combining
the production process and two decay processes, we can extract explicitly the dependence
of the correlated cross section on final charged lepton angles as well as
the production angles and beam polarizations.\s

\begin{figure}[htb]
\centering
\includegraphics[scale=.75]{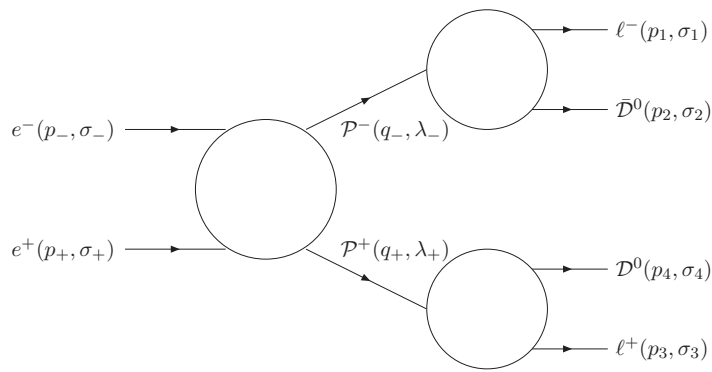}
\caption{\label{fig:helicity_diagram}
        {A schematic view of the process $e^+e^-\,\to\, \mathcal{P}^+\mathcal{P}^- \,
              \to\, (\ell^+\mathcal{D}^0) (\ell^- \mathcal{\bar{D}}^0)$. Shown in the
              parentheses are the four-momenta and re-scaled helicities of the corresponding
              particles.}
         }
\end{figure}

\subsection{Derivation of the correlated distributions}
\label{subsec:cross_sections}

The fully production-decay correlated amplitudes can be expressed in terms of the production
and decay helicity amplitudes as follows:
\begin{eqnarray}
{\cal M}(p_-,\sigma_-; p_+,\sigma_+; p_{1,2}, \sigma_{1,2}; p_{3,4},\sigma_{3,4})
&=& \Pi_{\mathcal{P}^-}(q^2_-)\, \Pi_{\mathcal{P}^+}(q^2_+)\sum_{\lambda_-,\lambda_+}
  {\cal M}(p_-,\sigma_-;p_+,\sigma_+; q_-,\lambda_-; q_+,\lambda_+)\nonumber\\
  && \times
  {\cal D}_-(q_-,\lambda_-; p_1,\sigma_1; p_2,\sigma_2)\times
  {\cal D}_+(q_+,\lambda_+; p_3,\sigma_3; p_4,\sigma_4)
\end{eqnarray}
where the Breit-Wigner propagator factors $\Pi_{\mathcal{P}^\pm}$ for
the $\mathcal{P}^\pm$ particles are
\begin{eqnarray}
   \Pi_{\mathcal{P}^\pm}(q^2_\pm)
= (q^2_\pm-M^2_{\mathcal{P}}+i M_{\mathcal{P}} \Gamma_{\mathcal{P}})^{-1}
\end{eqnarray}
Here we take the summations over intermediate $\mathcal{P}^\pm$ polarizations in the
helicity basis, i.e. helicities, which are most convenient for theoretical considerations.\s

In the c.m. frame of the colliding $e^+e^-$ beams, we choose the $\mathcal{P}^-$ momentum
direction as the $z$-axis and the $\vec{p}_-(e^-)\times \vec{q}_-(\mathcal{P}^-)$ direction as
the $y$-axis so that the scattering $e^+e^-\to \mathcal{P}^+\mathcal{P}^-$ takes place in
the $x$-$z$ plane (see Fig.$\,$\ref{fig:coordinate_system_diagram}).\footnote{The dependence
of the distribution on the production azimuthal-angle $\varphi$ can be encoded in terms
dependent on the transverse beam polarizations  as shown in Appendix C.} The production amplitude
${\cal M}$ is then a function of the scattering angle $\theta$ between $e^-$ and $\mathcal{P}^-$
momentum directions, as explicitly shown in the previous section. The explicit form of the
production amplitude and two decay amplitudes in the $e^+e^-$ c.m. frame can be
derived by the relations:
\begin{eqnarray}
    {\cal M}(p_-,\sigma_-;p_+,\sigma_+; q_-,\lambda_-; q_+,\lambda_+)
&=& {\cal M} [e^-_{\sigma_-}e^+_{\sigma_+}
                 \to \mathcal{P}^-_{\lambda_-}\mathcal{P}^+_{\lambda_+}]
    \\
    {\cal D}_-(q_-,\lambda_-; p_1,\sigma_1; p_2,\sigma_2)
&=& {\cal D}_- [\mathcal{P}^-_{\lambda_-}
                    \to \ell^-_{\sigma_1}\mathcal{\bar{D}}^0_{\sigma_2}]
    \\
    {\cal D}_+(q_+,\lambda_+; p_3,\sigma_3; p_4,\sigma_4)
&=& {\cal D}_+ [\mathcal{P}^+_{\lambda_+}
                    \to \ell^+_{\sigma_3}\mathcal{D}^0_{\sigma_4}]
\end{eqnarray}
with the expressions defined in Eq.$\,$(\ref{eq:generic_production_amplitude}) for the
production amplitudes and Eqs.$\,$(\ref{eq:generic_decay_amplitude_12}) and
(\ref{eq:generic_decay_amplitude_34}) for the decay amplitudes, respectively. \s

\subsection{Polarization-weighted cross sections}
\label{subsec:cross_sections}

Generally, the full correlations of the production and two two-body decay processes can
contain maximally $2^4\times (2J_{\mathcal{P}}+1)^4$ independent observables expressed
in terms of the $e^+e^-$ c.m. energy $\sqrt{s}$ and six production and decay angles - two
angles $(\theta,\varphi)$ for the production process and four angles $(\theta_\pm, \phi_\pm)$
for two decay processes - for arbitrarily-polarized electron and positron beams.
(Here, $J_{\mathcal{P}}$ is the spin of the particle $\mathcal{P}^\pm$.) The factor $2^4=16$
comes from the production part and the other $(2J_{\mathcal{P}}+1)^4$ ($=1,\, 16$ and $81$ for
$J_{\mathcal{P}}=0, 1/2$ and $1$) from the production-decay correlations. \s

The polarization-weighted squared matrix elements can be cast into a decomposed form:
\begin{eqnarray}
    \overline{\sum} \left|{\cal M}\right|^2
= \left|\Pi_{\mathcal{P}^-}(q^2_-)\right|^2 \left|\Pi_{\mathcal{P}^+}(q^2_+)\right|^2\,
     {\cal P}^{\lambda_- \lambda_+}_{\lambda'_- \lambda'_+}\,
     {\cal D}^{\lambda_-}_{\lambda'_-}\,
     \overline{\cal D}^{\lambda_+}_{\lambda'_+}
\label{eq:generic_polarization_summed_amplitude_squared}
\end{eqnarray}
with the summation over repeated indices $(\lambda_-,\lambda'_-,\lambda_+,\lambda'_+)$
assumed here and in the following equations. The polarization-weighted production tensor
reads
\begin{eqnarray}
 {\cal P}^{\lambda_- \lambda_+}_{\lambda'_- \lambda'_+}
 = \sum_{\sigma_-,\sigma'_-}\sum_{\sigma_+,\sigma'_+}\,
   P^-_{\sigma_-\sigma'_-} P^+_{\sigma_+\sigma'_+}\,
   {\cal M} [e^-_{\sigma_-}e^+_{\sigma_+}
             \to \mathcal{P}^-_{\lambda_-}\mathcal{P}^+_{\lambda_+}]\,
   {\cal M}^* [e^-_{\sigma_-}e^+_{\sigma_+}
                 \to \mathcal{P}^-_{\lambda'_-}\mathcal{P}^+_{\lambda'_+}]
\end{eqnarray}
in terms of the production helicity amplitudes, where the electron and positron polarization
tensors $P^\mp$ are given in the $(+,-)$ helicity basis by \cite{Hagiwara:1985yu}
\begin{eqnarray}
  P^-_{\sigma_- \sigma'_-}
&=& \frac{1}{2}
    \left(\begin{array}{cc}
      1+P^L_-                  &  P^T_- {\rm e}^{-i\phi_-}  \\[2mm]
      P^T_- {\rm e}^{i\phi_-}  & 1-P^L_-
         \end{array}\right) \\
  P^+_{\sigma_+ \sigma'_+}
&=& \frac{1}{2}
    \left(\begin{array}{cc}
      1+P^L_+                  &  P^T_+ {\rm e}^{-i\phi_+}  \\[2mm]
      P^T_+ {\rm e}^{i\phi_+}  & 1-P^L_+
         \end{array}\right)
\end{eqnarray}
respectively, where $\phi_- = -\varphi$ and $\phi_+ = -\varphi+\delta$ with the azimuthal
angle $\varphi$ of the $\mathcal{P}^-$ flight direction as measured from the
electron transverse polarization direction and $\delta$ the relative opening
angle of the electron and positron transverse-polarization directions. Details of this
calculation for incorporating beam polarizations are given in Appendix C.  The decay
density matrices with the daughter particle polarizations summed in
Eq.$\,$(\ref{eq:generic_polarization_summed_amplitude_squared}) are given by
\begin{eqnarray}
  {\cal D}^{\lambda_-}_{\lambda'_-} (\theta_-,\phi_-)
&=& \sum_{\sigma_1,\sigma_2} {\cal D}_- [\mathcal{P}^-_{\lambda_-}
                    \to \ell^-_{\sigma_1}\mathcal{\bar{D}}^0_{\sigma_2}]\,
                    {\cal D}^*_- [\mathcal{P}^-_{\lambda'_-}
                    \to \ell^-_{\sigma_1}\mathcal{\bar{D}}^0_{\sigma_2}]\\
\overline{\cal D}^{\lambda_+}_{\lambda'_+} (\theta_+,\phi_+)
&=& \sum_{\sigma_3,\sigma_4} {\cal D}_+ [\mathcal{P}^+_{\lambda_+}
                    \to \ell^+_{\sigma_3}\mathcal{D}^0_{\sigma_4}]\,
                    {\cal D}^*_+ [\mathcal{P}^+_{\lambda'_+}
                    \to \ell^+_{\sigma_3}\mathcal{D}^0_{\sigma_4}]
\end{eqnarray}
After integration over the virtual $\mathcal{P}$ masses squared, $q^2_-$ and $q^2_+$,
the unpolarized differential cross section can be expressed in the narrow width
approximation as
\begin{eqnarray}
\frac{d\sigma}{d\cos\theta d\varphi\, d\cos\theta_- d\phi_- d\cos\theta_+ d\phi_+}
=  \frac{(2J_{\mathcal{P}}+1)^2\beta}{1024\pi^4 s}\,
    {\rm Br}[\mathcal{P}^-\to\ell^-\mathcal{\bar{D}}^0]\,
    {\rm Br}[\mathcal{P}^+\to\ell^+\mathcal{D}^0]\,\,
    {\cal P}^{\lambda_- \lambda_+}_{\lambda'_- \lambda'_+}\,
     \rho^{\lambda_-}_{\lambda'_-}\,
     \overline{\rho}^{\lambda_+}_{\lambda'_+}
\label{eq:fully_correlated_distributions}
\end{eqnarray}
with $\beta=(1-4M^2_{\mathcal{P}}/s)^{1/2}$. Here, $\rho$ and $\overline{\rho}$ are the
normalized decay density matrices defined as
\begin{eqnarray}
   \rho^{\lambda_-}_{\lambda'_-}
 = \frac{{\cal D}^{\lambda_-}_{\lambda'_-}}{
         {\rm Tr}({\cal D})}
    \quad\,  \mbox{and} \quad\,
   \overline{\rho}^{\lambda_+}_{\lambda'_+}
 = \frac{\overline{\cal D}^{\lambda_+}_{\lambda'_+}}{
         {\rm Tr}(\overline{\cal D})}
\end{eqnarray}
satisfying the normalization conditions ${\rm Tr}(\rho) = 1$ and
${\rm Tr}(\overline{\rho}) = 1$.
With this normalization condition the overall constant $\mathcal{K}$ is fixed in terms of
the branching fractions ${\rm Br}(\mathcal{P}^-\to\ell^-\mathcal{\bar{D}}^0)$ and
${\rm Br}(\mathcal{P}^+\to\ell^+\mathcal{D}^0)$. By integrating over $\mathcal{P}^+$
decays, we obtain the inclusive $\mathcal{P}^-\to\ell^-\mathcal{\bar{D}}^0$ decay
distribution
\begin{eqnarray}
\frac{d\sigma}{d\cos\theta d\varphi\, d\cos\theta_- d\phi_-}
=  \frac{(2J_{\mathcal{P}}+1)\beta}{256\pi^3 s}\,
    {\rm Br}[\mathcal{P}^-\to\ell^-\mathcal{\bar{D}}^0]\,\,
    {\cal P}^{\lambda_- \lambda_+}_{\lambda'_- \lambda_+}\,
     \rho^{\lambda_-}_{\lambda'_-}
\label{eq:semi_inclusive_correlation_-}
\end{eqnarray}
and alternatively we obtain the $\mathcal{P}^+\to\ell^+\mathcal{D}^0$ decay distribution
as
\begin{eqnarray}
\frac{d\sigma}{d\cos\theta d\varphi\, d\cos\theta_+ d\phi_+}
=  \frac{(2J_{\mathcal{P}}+1)\beta}{256\pi^3 s}\,
    {\rm Br}[\mathcal{P}^+\to\ell^+\mathcal{D}^0]\,\,
    {\cal P}^{\lambda_- \lambda_+}_{\lambda_- \lambda'_+}\,
     \overline{\rho}^{\lambda_+}_{\lambda'_+}
\label{eq:semi_inclusive_correlation_+}
\end{eqnarray}
By further integrating out all the decay lepton angles, we simply get the unpolarized
differential cross section for the production process $e^+e^-\to\mathcal{P}^+\mathcal{P}^-$:
\begin{eqnarray}
\frac{d\sigma}{d\cos\theta d\varphi}
=  \frac{\beta}{64\pi^2 s}\,
    {\cal P}^{\lambda_- \lambda_+}_{\lambda_- \lambda_+}
\label{eq:unpolarized_differential_cross_section}
\end{eqnarray}
whose explicit form for the process $e^+e^-\to\mathcal{P}^+\mathcal{P}^-$ can be found
in Eq.$\,$(\ref{eq:generic_differential_cross_section}).
By comparing Eqs.$\,$(\ref{eq:fully_correlated_distributions}),
(\ref{eq:semi_inclusive_correlation_-}) and (\ref{eq:semi_inclusive_correlation_+}) with
Eq.$\,$(\ref{eq:unpolarized_differential_cross_section}) we can get the additional
information on not only the $\mathcal{P}^+\mathcal{P}^-$ production amplitudes but also
the $\mathcal{P}^\pm$ decay amplitudes encoded in decay lepton angular distributions. \s

\subsection{Decay density matrices}
\label{subsec:decay_density_matrices}

The explicit form of the normalized decay density matrix for each spin combination
of the parent and daughter particles, $\mathcal{P}^-$ and $\mathcal{\bar{D}}^0$, in
the decay $\mathcal{P}^-\to\ell^- \mathcal{\bar{D}}^0$ can be derived with the explicit
form of each decay amplitude listed in Eqs.$\,$(\ref{eq:decay_amplitude_sf}),
(\ref{eq:decay_amplitude_fs}), (\ref{eq:decay_amplitude_fv}) and
(\ref{eq:decay_amplitude_vf}), respectively. For the spin-0 case with $\mathcal{P}^- = S^-_p$
and $\mathcal{\bar{D}}^0=F^0_d$, the decay matrix is a single number:
\begin{eqnarray}
\rho [S^-_p \to \ell^- \bar{F}^0_d]= 1
\end{eqnarray}
generating no production-decay correlations, independently of the chiral structure of the
couplings. On the other hand, for the two spin-1/2 cases, the $2\times 2$ decay density
matrices read
\begin{eqnarray}
    \rho [F^-_p\to \ell^- \bar{S}^0_d]
= \frac{1}{2}
    \left(\begin{array}{cc}
          1 +\xi_{\rm fs} \cos\theta_-  &  \xi_{\rm fs} \sin\theta_- {\rm e}^{i\phi_-} \\
           \xi_{\rm fs}\sin\theta_- {\rm e}^{-i\phi_-} & 1 -\xi_{\rm fs} \cos\theta_-
                     \end{array}\right)
                      \ \ \mbox{with}\ \
                     \xi_{\rm fs} = \frac{|d^{\ell S}_{F+}|^2-|d^{\ell S}_{F-}|^2}{
                                          |d^{\ell S}_{F+}|^2+|d^{\ell S}_{F-}|^2}
\label{eq:fermion_decay_density_matrix_fs}
\end{eqnarray}
for the spin-0 daughter particle $\bar{S}^0_d$ and
\begin{eqnarray}
    \rho [F^-_p \to \ell^- \bar{V}^0_d]
&=& \frac{1}{2}
    \left(\begin{array}{cc}
          1 +\xi_{\rm fv} \eta_{\rm fv} \cos\theta_-  &
          \xi_{\rm fv} \eta_{\rm fv} \sin\theta_- {\rm e}^{i\phi_-} \\
          \xi_{\rm fv} \eta_{\rm fv} \sin\theta_- {\rm e}^{-i\phi_-} &
          1 -\xi_{\rm fv} \eta_{\rm fv} \cos\theta_-
                     \end{array}\right) \nonumber\\
  &&  \mbox{with}\ \
          \eta_{\rm fv} = \frac{M^2_{F_p}-2 M^2_{V_d}}{M^2_{F_p}+ 2 M^2_{V_d}}
          \ \ \mbox{and}\ \
          \xi_{\rm fv} = \frac{|d^{\ell V}_{F+}|^2-|d^{\ell V}_{F-}|^2}{
                               |d^{\ell V}_{F+}|^2+|d^{\ell V}_{F-}|^2}
 \label{eq:fermion_decay_density_matrix_fv}
\end{eqnarray}
for the spin-1 daughter particle $\bar{V}^0_d$, and the $3\times 3$ decay density matrix
for the spin-1 parent particle $V^-_p$ reads:
\begin{eqnarray}
 \rho [V^-_p\to \ell^- \bar{F}^0_d]
&=&   (1-\eta_{\rm vf})\, \Bbb{1}_{3\times 3} + (3\eta_{\rm vf}-2)\, \rho_T
  + \xi_{\rm vf}\, \eta_{\rm vf}\, \delta_T \nonumber\\
 &&   \mbox{with}\ \
      \eta_{\rm vf} = \frac{2M^2_{V_p}}{2M^2_{V_p}+M^2_{F_d}} \ \ \mbox{and} \ \
      \xi_{\rm vf} = \frac{|d^{\ell F}_{V+}|^2-|d^{\ell F}_{V-}|^2}{
                           |d^{\ell F}_{V+}|^2+|d^{\ell F}_{V-}|^2}
\label{eq:vector_decay_density_matrix}
\end{eqnarray}
where $\Bbb{1}_{3\times 3}$ is the $3\times 3$ identity matrix, and the normalized
matrix $\rho_T$ and the traceless matrix $\delta_T$ are given by
\begin{eqnarray}
&& \rho_T = \frac{1}{4}\,
           \left(\begin{array}{ccc}
                 1+ c^2_-    & \sqrt{2} c_- s_- {\rm e}^{i\phi_-}
                                 & s^2_- {\rm e}^{2i\phi_-}  \\
                 \sqrt{2} c_- s_- {\rm e}^{-i\phi_-} & 2 s^2_-
                                 & -\sqrt{2} c_- s_- {\rm e}^{i\phi_-} \\
                 s^2_- {\rm e}^{-2i\phi_-} & -\sqrt{2} c_- s_- {\rm e}^{-i\phi_-}
                                 & 1+ c_-^2
                 \end{array}\right) \\
&& \delta_T = \frac{1}{4}\,
           \left(\begin{array}{ccc}
                 2 c_-    & \sqrt{2} s_- {\rm e}^{i\phi_-}
                                 & 0  \\
                 \sqrt{2} s_- {\rm e}^{-i\phi_-} & 0
                                 & \sqrt{2} s_- {\rm e}^{i\phi_-} \\
                 0 &  \sqrt{2} s_- {\rm e}^{-i\phi_-}
                                 & -2 c_-
                 \end{array}\right)
\end{eqnarray}
with the abbreviations $c_-=\cos\theta_-$ and $s_- =\sin\theta_-$. \s

The density matrices for the charge-conjugated decays $\mathcal{P}^+\to \ell^+ \mathcal{D}^0$
are related to those of the decays $\mathcal{P}^-\to \ell^- \mathcal{\bar{D}}^0$ as follows:
\begin{eqnarray}
  \overline{\rho}[\mathcal{P}^+\to\ell^+ \mathcal{D}^0]
= \rho[\mathcal{P}^-\to\ell^- \mathcal{\bar{D}}^0]\,
  \left(\theta_- \to \theta_+,\, \phi_- \to -\phi_+,\, \xi \to -\xi \right)
\end{eqnarray}
The two density matrices can be used for describing non-trivial final-state angular
correlations between two visible leptons through the connection linked by the production
process.\s

As shown clearly by the expressions in Eqs.$\,$(\ref{eq:fermion_decay_density_matrix_fs}),
(\ref{eq:fermion_decay_density_matrix_fv}) and (\ref{eq:vector_decay_density_matrix}), the
decay distributions are affected not just by the spins and masses of the particles but also
the chiralities of their couplings. We find:
\begin{itemize}
\item If the relative chirality $\xi_{\rm fs}$ is zero, i.e. the coupling is either
      pure vector-like or pure axial-vector-like, the decay density matrix becomes an
      identity matrix, washing out any correlation in the final-state leptons of the decays
      $F^-_p\to\ell^- \bar{S}^0_d$ and $F^+_p\to\ell^+ S^0_d$ completely.
      On the contrary, if the coupling is purely chiral with $\xi_{\rm fs}=\pm 1$,
      the decay distributions provide maximal information on the production-decay
      correlations.
\item In addition to the relative chirality $\xi_{\rm fv}$ there exists a kinematic
      factor $\eta_{\rm fv}$ determining the polarization analysis power in the decay
      $F^-_p\to\ell^-\bar{V}^0_d$. This purely mass-dependent factor vanishes for the
      special case with $M_{F_p}=\sqrt{2} M_{V_d}$ and takes its maximum value of unity
      only when $M_{V_d}=0$, i.e. the spin-1 daughter particle $V^0_d$ is massless.
      Nevertheless, if the coupling is purely chiral, then this decay mode with a spin-1
      daughter particle can be distinguished from the decay mode with a spin-0 daughter
      particle by measuring the polarization analysis power; in the latter case its
      magnitude is 1 and in the former case its magnitude is $\eta_{\rm fv} < 1$ for
      $M_{V_d}>0$.
\item In the spin-1 case, if the relative chirality $\xi_{\rm vf}$ is zero, the density
      matrix becomes an identity matrix only when the parent and daughter particles are
      degenerate, i.e. $M_{V_p}=M_{F_d}$. However, in this degenerate case, the decay is
      kinematically forbidden. Therefore, we can conclude that the spin-1 case can be
      distinguished from the spin-0 and spin-1/2 cases.
\end{itemize}
Before closing this subsection, we emphasize that, with all these spin- and
chirality-dependent characteristics of the decay density matrices, the decay angle correlations
of the final-state leptons become trivial unless the parent particles are polarized as will
be demonstrated below.\s

\section{Observables}
\label{sec:observables}

In the last section, we gave a detailed description of the angular distribution of the
final-state lepton-antilepton pairs arising from the decay of the $\mathcal{P}^+\mathcal{P}^-$
pair. Schematically, the 6-fold differential cross section has the form
\begin{eqnarray}
   d\sigma\ \ \sim \ \
   \sum^{N_{\rm tot}}\, \mathrm{P}_i(P^L_-, P^L_+; P^T_-, P^T_+, \delta;
                                             \theta, \varphi; \sqrt{s})\,\,
                        \mathrm{D}_i(\theta_-,\phi_-,\theta_+,\phi_+) \quad
                                 \mbox{with}\quad N_{\rm tot}=(2J_{\mathcal{P}}+1)^4
\end{eqnarray}
Here the functions $\mathrm{D}_i$ form a linearly independent set consisting
of low-energy spherical harmonics, which reflects the decay dynamics. The dynamics of the
production process is solely contained in the factors $\mathrm{P}_i$,
forming maximally 16 independent terms. These are given essentially by the density matrix
of the $\mathcal{P}^+\mathcal{P}^-$ pair and by beam polarizations. The fact that we can
in principle measure $16 \times (2J_{\mathcal{P}}+1)^4$ functions shows that it is possible
to extract an enormous amount of information on the production and decay mechanism.\s

However, unless we have a sufficient number of events, it is neither possible nor practical to
perform a fit with the large number of all independent angular and/or polarization
distributions. Rather it is meaningful to obtain from the experimental data a specific set of
observables depending on the c.m. energy, the beam polarizations, the production angles and the
decay angular distributions that are efficiently controllable and reconstructible and sensitive
to the spin and chirality effects. In the following numerical analysis we restrict ourselves
to five conventional kinematic variables --- the beam energy $\sqrt{s}$, the production
polar angle $\theta$, the two lepton polar angles, $\theta_-$ and $\theta_+$, in the decays,
$\mathcal{P}^-\to\ell^-\mathcal{\bar{D}}^0$ and $\mathcal{P}^+\to\ell^+\mathcal{D}^0$,
and the cosine of the azimuthal-angle difference $\phi$ between two decay planes.
The impact of beam polarizations on each observable is also diagnosed numerically.\s

In order to gauge the sensitivities of the observables mentioned in the previous subsection
to spin and chirality effects in the antler-topology processes, we investigate
their distributions for ten typical spin and  chirality assignments as shown with five examples
from the MSSM and five examples from the MUED listed in
Tab.$\,$\ref{tab:examples_antler_topology_process}. For the sake
of simplicity, when describing the specific examples, we impose electron
chirality invariance (which is valid to a very good precision in the popular models MSSM
and MUED), forcing us to neglect any $s$-channel scalar contributions and to set any
three-point $e \mathcal{T}\mathcal{P}$ and $\mathcal{P} \ell \mathcal{D}$ vertices with
$\ell=e, \mu$ to be purely chiral in the $t$-channel diagram and the two-body decay diagrams.
Furthermore, in the present numerical analysis we do not have any $u$-channel
exchange of doubly-charged particles, for which new higher representations of the SM gauge
group have to be introduced in the theories. In any case, note that in principle all the
$u$-channel contributions, if they exist, can be worked out through the analytic expressions
presented in Sect.$\,$\ref{sec:production}. For example, the major difference between a $u$-channel process and a $t$-channel process is that the production polar-angle 
distribution will be backward-peaked instead of forward-peaked, as can be seen from Eq.$\,$(\ref{eq:t/u propagators}).\s

\begin{table}[htb]
\begin{center}
\begin{tabular}{|c|c|c||c|c|c|c|}
\hline
\ \ Index                                          \ \  &
\ \ $[J_{\mathcal{P}}, J_{\mathcal{D}}]$           \ \  &
\ \ Chirality                                      \ \  &
\ \ Antler-topology process                        \ \  &
\ \ $s$-channel  $[J_{\mathcal{S}^0}]$             \ \  &
\ \ $t$-channel  $[J_{\mathcal{T}^0}]$             \ \  &
\ \ Model                                          \ \
\\
\hline\hline
\ \  $A_{L/R}$ \ \ &
    $[0, 1/2]$                                                    &
      $L/R$                                                       &
\ \  $e^+e^- \to \tilde{\mu}^+_{L/R}\tilde{\mu}^-_{L/R}
         \to (\mu^+\tilde{B}) (\mu^-\tilde{B})$ \ \ &
    $\gamma, Z$ \, $[J=1]$                                        &
    $-$                                                           &
    MSSM
\\
\ \ $B_{L/R}$ \ \ &
     $[0, 1/2]$                                                   &
     $L/R$                                                        &
\ \  $e^+e^- \to \tilde{e}^+_{L/R} \tilde{e}^-_{L/R}
         \to (e^+\tilde{B}) (e^-\tilde{B})$               \ \     &
    $\gamma, Z$ \, $[J=1]$                                        &
    $\tilde{B}, \tilde{W}^0$\, $[J=1/2]$                          &
    MSSM
\\
\hline
\ \ $C_{L/R}$ \ \ &
    $[1/2, 1]$                                                    &
      $L/R$                                                       &
\ \  $e^+e^- \to \mu^+_{L1/R1}\mu^-_{L1/R1}
     \to (\mu^+ B_1) (\mu^- B_1)$                     \ \ &
    $\gamma, Z$ \, $[J=1]$                                        &
    $-$                                                           &
    MUED
\\
\ \ $D_{L/R}$ \ \ &
    $[1/2, 1]$                                                    &
      $L/R$                                                       &
\ \  $e^+e^- \to e^+_{L1/R1}e^-_{L1/R1}
     \to (e^+ B_1) (e^- B_1)$                         \ \ &
    $\gamma, Z$ \, $[J=1]$                                        &
    $B_1, W^0_1$ \, $[J=1]$                                    &
    MUED
\\
\hline
\ \ $E_L$ \ \ &
    $[1/2, 0]$                                                    &
     $L$                                                          &
\ \  $e^+e^- \to \tilde{W}^+\tilde{W}^-
     \to (\ell^+\tilde{\nu}_\ell) (\ell^-\tilde{\nu}^*_\ell)$ \ \ &
    $\gamma, Z$ \, $[J=1]$                                        &
    $\tilde{\nu}_e$\, $[J=0]$                                     &
    MSSM
\\
\hline
\ \ $F_L$ \ \ &
    $[1, 1/2]$                                                    &
        $L$                                                       &
\ \  $e^+e^- \to W^+_1 W^-_1
     \to (\ell^+\nu_{\ell 1}) (\ell^-\bar{\nu}_{\ell 1})$     \ \ &
    $\gamma, Z$ \, $[J=1]$                                        &
    $\nu_{e1}$ \, $[J=1/2]$                                         &
    MUED
\\
\hline
\end{tabular}
\end{center}
\caption{\label{tab:examples_antler_topology_process}
        {Ten examples for the antler-topology processes - five in MSSM and five in MUED.
             Every ECV effect due to EWSB in these models is small so that the $e \mathcal{T}
             \mathcal{P}$ and $\mathcal{P}\ell \mathcal{D}$ couplings are purely chiral to
             very good approximation. The first index is introduced to specify each spin and
             chirality assignment. The chirality index, $R$ or $L$, in the third column stands
             for the chiral structure of the $e^- \mathcal{T}^0 \mathcal{P}^-$ vertex and the
             $\mathcal{P}^- \ell^- \mathcal{\bar{D}}^0$ vertex. We note that the chirality
             of each $t$-channel coupling is identical to the chirality of the vertex
             describing the decay $\mathcal{P}^-\to\ell^-\mathcal{\bar{D}}^0$ in every
             scenario.}
        }
\end{table}

In general several particles may contribute to the $s$-channel and/or $t$-channel diagrams
and the mass spectrum of the new particles depends strongly on the mass generation mechanism
unique to each model beyond the SM. Nevertheless, expecting no significant loss of generality,
we assume in our numerical analysis that only the SM neutral electroweak gauge bosons $\gamma$
and $Z$ contribute to the $s$-channel diagram and only one or two particles, named
$\mathcal{T}^0_1$ and $\mathcal{T}^0_2$ when two particles are involved, are exchanged
in the $t$-channel diagram. Then, we take the following simplified mass spectrum:
\begin{eqnarray}
M_{\mathcal{P}} = M_{\mathcal{T}_2} = 200\, {\rm GeV} \quad \mbox{and}\quad
M_{\mathcal{D}} = M_{\mathcal{T}_1} = 100\, {\rm GeV}
\label{eq:mass_spectrum}
\end{eqnarray}
We emphasize that the mass spectrum (\ref{eq:mass_spectrum}) is chosen only as a simple
illustrative example in the MSSM and MUED models with different spins but similar final
states and so the procedure for spin determination demonstrated in the present work can
be explored for any other BSM models as well as within the SM itself. The coupling of the
$Z$ boson as well as the photon $\gamma$ to the new spin-1/2 charged fermion pair $F^+_p F^-_p$
with $\mathcal{P}^\pm=F^\pm_p$ is taken to be
purely vector-like, as this is valid for the first Kaluza-Klein (KK) lepton states in
MUED with $F^\pm_p = \ell^\pm_{L1/R1}$ and for the pure charged wino or higgsino
states in the MSSM with $F^\pm_p =\tilde{W}^\pm, \tilde{H}$, valid to very good
approximation when the mixing between the gaugino and higgsino states due to EWSB
is ignored in the MSSM. It is also assumed that the lightest neutralino is a pure bino,
$\tilde{B}$, and the second lightest neutralino is a pure wino, $\tilde{W}^0$. In this
case, the lightest chargino is almost degenerate with the second lightest neutralino. \s

Applying all the assumptions mentioned above to the MSSM and MUED processes listed in
Tab.$\,$\ref{tab:examples_antler_topology_process}, we can obtain the full list of
non-zero ECC couplings for the processes \cite{Chung:2003fi,Datta:2010us}: for the
$s$-channel couplings
\begin{eqnarray}
&&   s^\gamma_{ee\pm}
   = s^{\tilde{\ell}_R\tilde{\ell}_R}_\gamma
   = s^{\tilde{\ell}_L\tilde{\ell}_L}_\gamma
   = s^{\ell_{R1}\ell_{R1}}_{\gamma \pm}
   = s^{\ell_{L1}\ell_{L1}}_{\gamma \pm}
   = s^{\tilde{W}\tilde{W}}_{\gamma \pm}
   = s^{W_1 W_1}_\gamma
   = 1
\label{eq:s-channel_coupling_1} \\
&&   s^Z_{ee+}
   = s^{\tilde{\ell}_R\tilde{\ell}_R}_Z
   = s^{\ell_{R1}\ell_{R1}}_{Z \pm}
   = -s_W/c_W
\label{eq:s-channel_coupling_2} \\
&&   s^Z_{ee-}
   = s^{\tilde{\ell}_L\tilde{\ell}_L}_Z
   = s^{\ell_{L1}\ell_{L1}}_{Z \pm}
   = (1/2-s^2_W)/c_Ws_W
\label{eq:s-channel_coupling_3} \\
&&   s^{\tilde{W}\tilde{W}}_{Z \pm}
   = s^{W_1 W_1}_Z
   = c_W/s_W
\label{eq:s-channel_coupling_4}
\end{eqnarray}
with $\ell^\pm=e^\pm, \mu^\pm$ and for the $t$-channel and decay couplings
\begin{eqnarray}
&&     t^{e\tilde{e}_R}_{\tilde{B}+}
   = d^{\ell\tilde{B}}_{\tilde{\ell}_R-}
   = -\sqrt{2}/c_W; \quad
     t^{e\tilde{e}_L}_{\tilde{B}-}
   = d^{\ell\tilde{B}}_{\tilde{\ell}_L+}
   = 1/\sqrt{2}c_W; \quad
     t^{e\tilde{e}_L}_{\tilde{W}-}
   = 1/\sqrt{2} s_W, \quad
     t^{e\tilde{W}}_{\tilde{\nu}_e-}
   = d^{\ell\tilde{\nu}_\ell}_{\tilde{W}+}
   = -1/s_W
\label{eq:t_d-channel_coupling_1} \\
&&     t^{e e_{R1}}_{B_1 +}
   = d^{\ell B_1}_{\ell_{R1} +}
   =  1/c_W; \quad
     t^{e e_{L1}}_{B_1 -}
   = d^{\ell B_1}_{\ell_{L1} -}
   = 1/2 c_W; \quad
     t^{e e_{L1}}_{W_1 -}
   = 1/2 s_W; \quad
     t^{e W_1}_{\nu_{e 1} -}
   = d^{\ell \nu_{\ell 1}}_{W_1 -}
   = -1/\sqrt{2} s_W
\label{eq:t_d-channel_coupling_2}
\end{eqnarray}
in the MSSM and in the MUED, respectively. All the other couplings are vanishing in the
ECC limit.\s

\subsection{Kinematics}
\label{subsec:kinematics}

Before presenting the detailed analytic and numerical analysis of spin and chirality effects
on each observable, we first describe how each kinematic observable can be constructed
for the antler-topology processes. The measurement of the cross section for $\mathcal{P}^+
\mathcal{P}^-$ pair production can be carried out by identifying acoplanar $\ell^+\ell^-$
pairs with respect to the $e^\pm$ beam axis accompanied by large missing energy carried by
the invisible $\mathcal{D}^0\mathcal{\bar{D}}^0$ pairs.\footnote{A detailed proof of the
twofold discrete ambiguity in reconstructing the full kinematics of the antler-topology
process production is given in Appendix D.} \s

For very high energy $\sqrt{s}\gg M_{\mathcal{P}}$ the flight direction of the parent
particle can be approximated by the flight direction of the daughter particles $\ell^\pm$
and the dilution due to the decay kinematics is small. However, at medium $e^+e^-$
energies the dilution increases, and the reconstruction of the $\mathcal{P}^\pm$ flight
direction provides more accurate results on the angular distribution of the $\mathcal{P}^\pm$
pairs. If all particle masses are known, the magnitude of the particle momenta is calculable
and the relative orientation of the momentum vectors of $\ell^\pm$ and $\mathcal{P}^\pm$
is fixed by the two-body decay kinematics:
\begin{eqnarray}
  M^2_{\mathcal{P}}-M^2_{\mathcal{D}}
= \sqrt{s} E_{\ell^\pm} \left(1- \beta\,  \hat{n}_{\mathcal{P}^\pm} \cdot\hat{n}_\pm\right)
            = \sqrt{s}\, E_{\ell^\pm} \left(1-\beta\cos\alpha_\pm\right)
\end{eqnarray}
where the unit vector $\hat{n}_{\mathcal{P}^\pm}$ stands for the $\mathcal{P}^\pm $ momentum
direction, the unit vectors $\hat{n}_\pm$ for the $\ell^\pm$ flight directions and the angles
$\alpha_\pm$ for the opening angles between the visible $\ell^\pm$ tracks and the parent
$\mathcal{P}^\pm$ momentum directions in the $e^+e^-$ c.m frame. The angles $\alpha_\pm$ can
be reconstructed event by event by measuring the lepton energies in the laboratory frame, i.e.
the $e^+e^-$ c.m. frame and they define two cones about the $\ell^+$ and $\ell^-$ axes
intersecting in two lines --- the true $\mathcal{P}^\pm$ flight direction and a false direction.
Thus the $\mathcal{P}^\pm$ flight direction can be reconstructed up to a two-fold discrete
ambiguity.\s

In contrast to the production angle, the decay polar angles $\theta_\pm$ in the
$\mathcal{P}^\pm$ rest frames can be unambiguously determined event by event independently
of the reconstruction of the $\mathcal{P}^\pm$ direction by the relation:
\begin{eqnarray}
  \cos\theta_\pm
= \frac{1}{\beta}
  \left( \frac{E_{\ell^\pm}}{E^*_{\ell^\pm}} - 1\right)
  \quad \mbox{with} \quad
E^*_{\ell^\pm} = \frac{M^2_{\mathcal{P}}-M^2_{\mathcal{D}}}{2M_{\mathcal{P}}}
\label{eq:theta_pm_energy_relation}
\end{eqnarray}
where $E^*_{\ell^\pm}$ is the fixed $\ell^\pm$ energy in the $\mathcal{P}^\pm$ rest frame.
Therefore, any decay polar-angle correlations between two leptons in the correlated process
can be reconstructed event by event by measuring the lepton energies in the laboratory
frame. \s

Another angular variable, which is reconstructible event by event in the antler-topology
processes, is the cosine of the difference $\phi=\phi_+-\phi_-$ of the azimuthal angles of
two leptons with respect to the production plane. Explicitly, it is related to the opening
angle of two visible leptons and two polar angles $\alpha_\pm$ in the laboratory frame as
\begin{eqnarray}
\cos\phi=\cos(\phi_+-\phi_-) = \frac{\hat{n}_+\cdot\hat{n}_- + \cos\alpha_+\cos\alpha_-}{
                            \sin\alpha_+ \sin\alpha_-}
\end{eqnarray}
Note that the $\cos 2\phi$ distribution also can be measured unambiguously as $\cos 2\phi
= 2\cos^2\phi-1$.\footnote{Actually, $\cos (n\theta)$ with any non-zero integer $n$ is a
polynomial of $\cos\phi$.} In contrast, the sign of the sine of the angular difference of two
azimuthal angles is not uniquely determined because of the intrinsic two-fold discrete ambiguity
in the determination of the $\mathcal{P}^\pm$ flight direction, although its magnitude is
determined. (For details, see Appendix \ref{sec:appendix_d_kinematics_of_the_antler_process}.) \s

There exist many other types of angular distributions which provide us with additional
information on the spin and chirality effects. Nevertheless, while postponing the complete
analysis based on the full set of energy and angular distributions, we will study the
four kinematic observables $\{\sqrt{s},\, \theta,\, \theta_-,\, \cos\phi\}$ supplemented
with beam polarizations.\s

\subsection{Beam energy dependence and threshold excitation pattern}
\label{subsec:threshold_excitation}

As described through a detailed analytical investigation before, the excitation curve
of the production cross section near threshold in the ECC scenario exhibits its characteristic
pattern according to the spin of the produced particle $\mathcal{P}^\pm$ and the chiral
patterns of the couplings among the on-shell particles and any intermediate particles exchanged
in the $s$-, $t$- and/or $u$-channel diagrams.\footnote{Very close
to the threshold the excitation curves may be distorted due to particle widths and Coulomb
attraction between two oppositely charged particles. However, the effects are
insubstantial for small widths so that they are ignored in the present work.}\s

The production cross section of a spin-0 scalar pair as in the scenario $A_{L/R}$ of
the $L$- or $R$-smuon pair production and the scenario $B_{L/R}$ of the $L$- or $R$-selectron
pair production shows a characteristic slow $P$-wave threshold excitation, i.e.
$\sigma\sim \beta^3$, despite the $t$-channel neutral bino and/or wino contributions
to the selectron pair production. In contrast, the production cross section of a spin-1/2
fermion pair as in the scenarios $C_{L/R}$ and $D_{L/R}$ for the $L$- or $R$-handed
first KK-muon and KK-electron pair production and as in the scenario $E_L$ of a wino
pair production always exhibits a sharp $S$-wave threshold excitation, i.e. $\sigma\sim \beta$,
(due to the unavoidable pure-vector coupling of a photon to the $e^+e^-$ and $F^+_pF^-_p$).
The excitation pattern in the scenario $F_L$ for the first KK $W$-boson pair production
is characterized dominantly by the presence of the $t$-channel contributions,
which should be present for preventing the cross section from developing a bad high energy
behavior as the $s$-channel $\gamma$ and $Z$ contributions cannot cancel each other at high
energies simultaneously for left- and right-chiral couplings. Note that the
polarized cross section with perfect right-handed electron polarization does not have the
$t$-channel spin-1/2 $\nu_{e1}$ contribution but only the $s$-channel $\gamma$ and $Z$
contributions leading to
complete asymptotic cancellation. In this case, the cross section exhibits a slow $P$-wave
behavior as in the scalar case. Otherwise, the cross section contains the non-zero $t$-channel
$\nu_{e1}$ contribution with the $J_0=2$ amplitude finite at threshold so that the cross
section rises in a sharp $S$-wave near threshold. These threshold patterns are summarized in
Tab.$\,$\ref{tab:threshold_excitation_pattern}.\s

\begin{table}[thb]
\begin{center}
\begin{tabular}{|c|c|c|c|}
\hline
\ \ Spin $J_{\mathcal{P}}$   \ \  &
\ \ Polarized cross section  \ \  &
\ \ Threshold excitation     \ \  &
\ \ Model                    \ \       \\
\hline
\ \     \raisebox{-1.8ex}{$0$}                  \ \  &
\ \ $\sigma_{L/R} [e^+e^-\to\tilde{\ell}^+_R\tilde{\ell}^-_R]$ \ \  &
\ \ $\beta^3$                \ \  &
\ \    MSSM                  \ \       \\[-1mm]
\ \     { }                  \ \  &
\ \ $\sigma_{L/R} [e^+e^-\to\tilde{\ell}^+_L\tilde{\ell}^-_L]$ \ \ &
\ \ $\beta^3$                \ \  &
\ \    MSSM                  \ \       \\
\hline
\ \      { }                 \ \  &
\ \ $\sigma_{L/R} [e^+e^-\to \ell^+_{R1} \ell^-_{R1}]$ \ \  &
\ \ $\beta$                  \ \  &
\ \   MUED                   \ \       \\
\ \     $1/2$                \ \  &
\ \ $\sigma_{L/R} [e^+e^-\to \ell^+_{L1} \ell^-_{L1}]$ \ \ &
\ \ $\beta$                  \ \  &
\ \   MUED                   \ \       \\
\ \     { }                  \ \  &
\ \ $\sigma_{L/R} [e^+e^-\to \tilde{W}^+ \tilde{W}^-]$ \ \  &
\ \ $\beta$                  \ \  &
\ \    MSSM                  \ \       \\
\hline
\ \     \raisebox{-1.8ex}{$1$}                  \ \  &
\ \ $\sigma_{L} [e^+e^-\to W^+_1 W^-_1]$  \ \  &
\ \ $\beta$                \ \  &
\ \   MUED                   \ \       \\[-1mm]
\ \     { }                  \ \  &
\ \ $\sigma_{R} [e^+e^-\to W^+_1 W^-_1]$  \ \  &
\ \ $\beta^3$                  \ \  &
\ \   MUED                   \ \       \\
\hline
\end{tabular}
\end{center}
\caption{\label{tab:threshold_excitation_pattern}
        {Threshold excitation of the polarization-weighted total cross sections for the
             ten MSSM and MUED processes with $\ell=e, \mu$.
             $\beta = (1-4 M^2_{\mathcal{P}}/s)^{1/2}$ is the speed of the particle
             $\mathcal{P}$ in the $e^+e^-$ c.m. frame. $\sigma_{L/R}$ stands for the
             polarization-weighted cross section with perfect left-handed or right-handed
             electron beam polarization.}
        }
\end{table}
\begin{figure}[htb]
\centering
\includegraphics[width=18.cm]{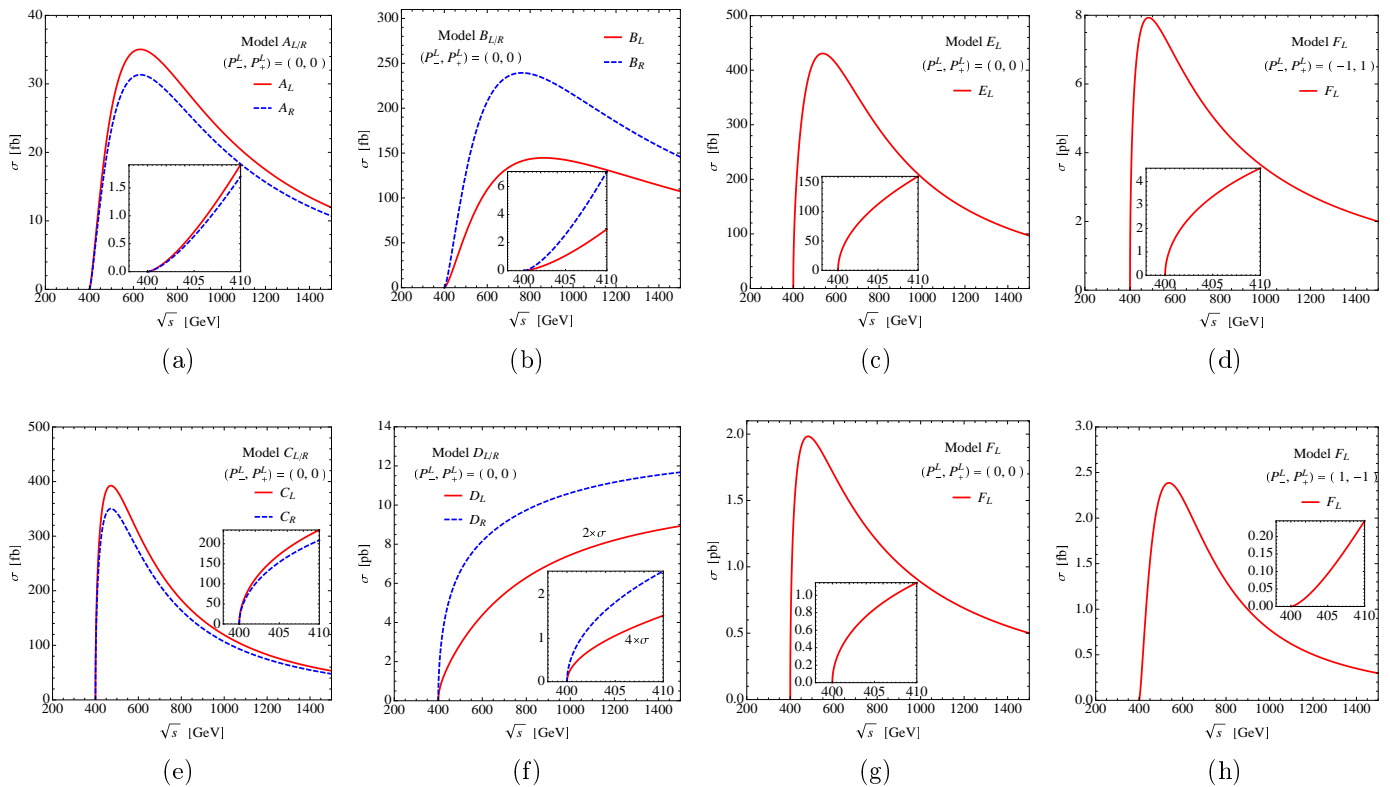}
\caption{\label{fig:threshold_excitation}
         Energy dependence of the total cross sections with the threshold
         excitation curves embedded for spin-0 scalar bosons ($A_{L/R}$ and $B_{L/R}$),
         spin-1/2 fermions ($C_{L/R}, D_{L/R}$ and $E_L$) and spin-1 vector bosons
         ($F_L$). The electron and positron beams are set to be unpolarized, except for
         the frames, (d) and (h); the upper (lower) frame is for purely
         left (right) polarized electron and purely right (left) polarized positron
         beams.
        }
\end{figure}

Based on the mass spectrum in Eq.$\,$(\ref{eq:mass_spectrum}) and the explicit form
of the couplings listed in Eqs.$\,$(\ref{eq:s-channel_coupling_1}),
(\ref{eq:s-channel_coupling_2}), (\ref{eq:s-channel_coupling_3}), and
(\ref{eq:s-channel_coupling_4}) and Eqs.$\,$(\ref{eq:t_d-channel_coupling_1}) and
(\ref{eq:t_d-channel_coupling_2}), we show in Figs.$\,$\ref{fig:threshold_excitation}
the energy dependence of total cross sections, with the threshold excitation curves
embedded, for spin-0 scalar bosons indexed with $A_{L/R}$ and $B_{L/R}$, for spin-1/2
fermions indexed with $C_{L/R}$, $D_{L/R}$ and $E_L$, and for spin-1 vector bosons
indexed with $F_L$. Here, the electron and positron beams are assumed to be unpolarized,
except for Figs.$\,$\ref{fig:threshold_excitation}(d) and (h).
In contrast to Figs.$\,$\ref{fig:threshold_excitation}(d),
the plot in Figs.$\,$\ref{fig:threshold_excitation}(h) clearly shows that
the cross section with purely right-handed electron and purely left-handed positron beams
killing the $t$-channel contributions while keeping only the $s$-channel spin-1 vector-boson
contributions exhibits a slow $P$-wave rise in the excitation curve. We note in passing
that it will be crucial to control beam polarization to very good precision in
extracting out the right-handed part as the right-handed cross section is more than
one thousand times smaller than the left-handed cross section. \s

{\it To summarize.} The threshold energy scan of the polarized
cross sections of the pair production process $e^+e^-\to\mathcal{P}^+\mathcal{P}^-$ can be
very powerful in identifying the spin of the new charged particles $\mathcal{P}^\pm$.
However, we note that this method may not be fully powerful enough for encompassing the most
general scenario including the case with simultaneous left-/right-chiral $t$- and/or
$u$-channel contributions and the case with neither of them. \s

\subsection{Polar-angle distribution in the production process}
\label{subsec:polar_angle_distribution}

As pointed out before and described in detail in Appendix D, there exists a twofold discrete
ambiguity in constructing the production polar angle $\theta$. For very high energy
$\sqrt{s} \gg M_{\mathcal{P}}$ the flight direction of the parent particle $\mathcal{P}^\pm$
can be approximated by the flight direction of daughter particle $\ell^\pm$ and the dilution
due to the decay kinematics is small. However, at medium energies the dilution increases and
so the reconstruction of the $\mathcal{P}^\pm$ flight direction provides more accurate results
on the angular distribution of the $\mathcal{P}^\pm$ pairs.\s

Analytically, the angle $\theta_{\rm ft}$ between the false and the true axis is related to
the azimuthal angle $\phi$ between two decay planes and to the boosts
$\gamma_\pm = \gamma (\cos\theta_\pm +\beta)$ of the leptons $\ell^\pm$ in the
laboratory frame as
\begin{eqnarray}
  \cos\theta_{\rm ft}
= 1-\frac{2\sin^2\phi}{\gamma^2_+ + \gamma^2_- + 2\gamma_+ \gamma_- \cos\phi + \sin^2\phi}
\end{eqnarray}
For high energies the maximum opening angle reduces effectively to $\theta_{\rm ft}\leq
O(1/\gamma)$ and approaches zero asymptotically when the two axes coincide. Quite generally,
as a result of the Jacobian root singularity in the relation between $\cos\theta_{\rm ft}$
and $\phi$, the false solutions tend to accumulate slightly near the true axis for all
energies \cite{Choi:2006mr}. \s

\begin{figure}[htb]
\centering
\includegraphics[width=18.cm]{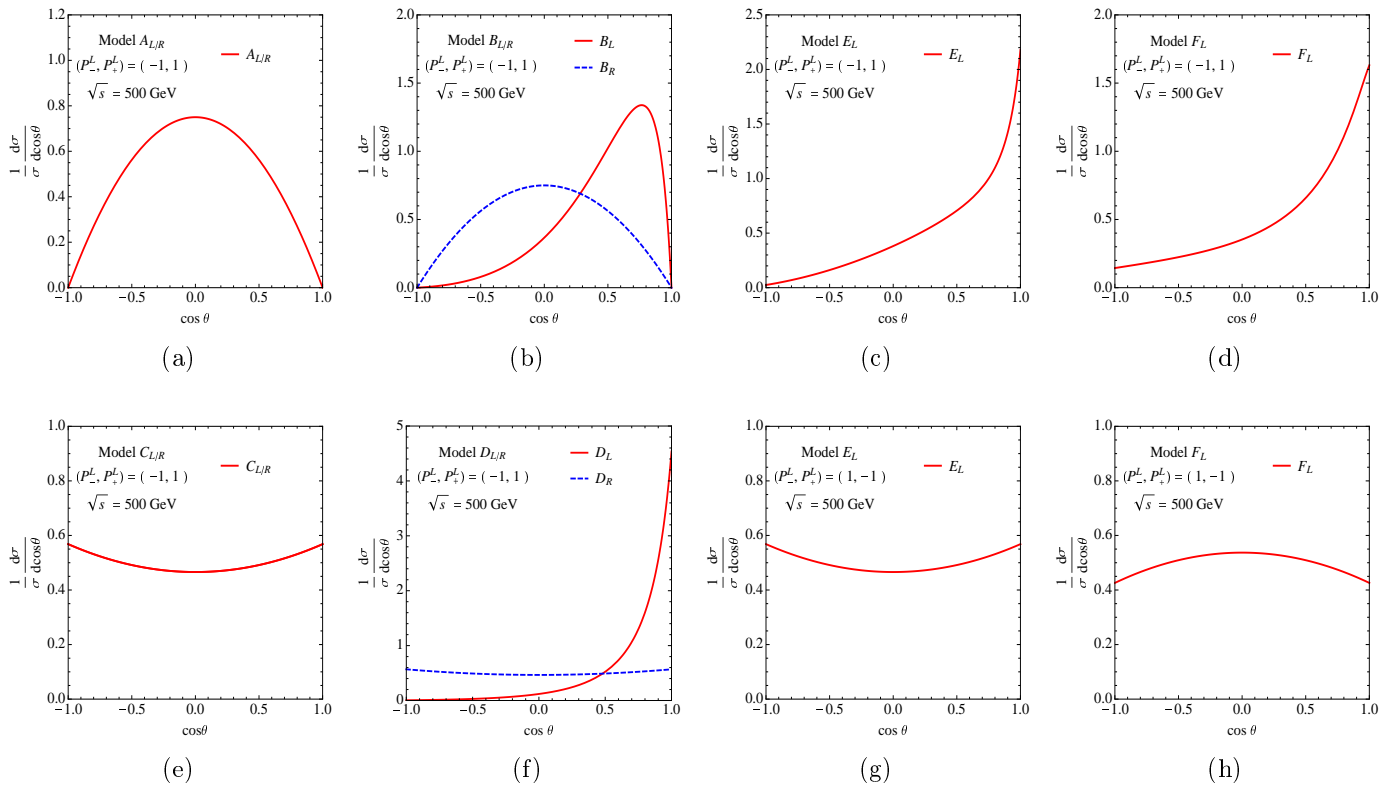}
\caption{\label{fig:production_polar-angle_distributions}
        {Production polar-angle distributions for the spin-0 scalar bosons indexed
             with $A_{L/R}$ and $B_{L/R}$, for the spin-1/2 fermions indexed with $C_{L/R},\,
             D_{L/R}$ and $E_L$, and for the spin-1 vector bosons indexed with $F_L$ in
             the MSSM and MUED models. The c.m. energy $\sqrt{s}$ is set to be 500 GeV.}
        }
\end{figure}

Experimentally, the absolute orientation in space is operationally obtained by rotating the
two $\mathcal{P}^\pm$ vectors around the $\ell^\pm$ axes against each other until they are
aligned back to back in opposite directions. The flattened false-axis distribution can be
extracted on the basis of Monte Carlo simulations.
Figs.$\,$\ref{fig:production_polar-angle_distributions} shows the normalized production
polar-angle distributions for the polarization-weighted differential cross sections,
$(1/\sigma_{L/R})\, d\sigma_{L/R}/d\cos\theta$, of the ten processes listed in
Tab.$\,$\ref{tab:examples_antler_topology_process}.
The plots in Figs.$\,$\ref{fig:production_polar-angle_distributions}(a) and (b) are for
the scalar-pair production processes, $A_{L/R}$ for smuon pairs and
$B_{L/R}$ for selectron pairs and the plots in
Figs.$\,$\ref{fig:production_polar-angle_distributions}(c), (e), (f) and (g) are
for the five fermion-pair
production processes,  $E_L$ for a wino pair, $C_{L/R}$ for the first KK-muon pairs and
$D_{L/R}$ for the first KK-electron pairs, respectively, while the
two plots in Figs.$\,$\ref{fig:production_polar-angle_distributions}(d) and (h)
are for a vector-boson-pair production process, $F_L$, for the first KK-$W$ pair.
\begin{itemize}
\item From Figs.$\,$\ref{fig:production_polar-angle_distributions}(a)
      and (b), we find that the cross
      sections vanish in the forward and back directions with $\cos\theta=\pm 1$ due to
      the overall angular factor proportional to $\sin^2\theta$, independently of the
      presence of $t$-channel contributions. If the $t$-channel fermion contributions are
      absent ($A_{L/R}$) or killed by beam polarization ($B_R$), the polar-angle distribution
      is forward and backward symmetric and simply $\sim \sin^2\theta$.
\item In contrast, the polar-angle distributions for spin-1/2 particles exhibit very distinct
      angular patterns. If the $t$-channel contributions are absent,
      as shown in Figs.$\,$\ref{fig:production_polar-angle_distributions}(c), or killed by
      right-handed  electron and left-handed positron beam polarizations, as in
      Figs.$\,$\ref{fig:production_polar-angle_distributions}(g), the
      differential cross sections having only the $s$-channel vector-boson contributions
      with pure vector-type couplings in the three cases have a typical angular
      distribution $1+\kappa_{1/2}\cos^2\theta$ with $\kappa_{1/2}= \beta^2/(2-\beta^2)
      =0$ at threshold and 1 at asymptotic high energies, leading to
      the characteristic distribution $1+\cos^2\theta$, reflecting the equal contributions
      of the dominant $(\lambda_-,\lambda_+) = (\pm,\mp)$ amplitudes. Once the $t$-channel
      contributions are included, the angular distribution is severely distorted.
      Nevertheless, as shown in  Figs.$\,$\ref{fig:production_polar-angle_distributions}(c)
      and (f), the cross sections are peaked at the forward direction.
\item Figs.$\,$\ref{fig:production_polar-angle_distributions}(d) and (h) show the angular
      distributions for spin-1 first KK $W$-boson pair production ($F_L$). If the $t$-channel
      contribution is absent as in Figs.$\,$\ref{fig:production_polar-angle_distributions}(h),
      the differential cross section has only $s$-channel spin-1
      vector-boson contributions with pure vector-type couplings ($F_L$) so that the
      $(\pm\mp)$ amplitudes with $J_0=2$ are zero and the $(0,0)$ amplitudes become dominant.
      As a result, the polar-angle distributions exhibit a characteristic energy-independent
      polar-angle distribution $\sim 1 - \kappa_1 \cos^2\theta$ with the energy-dependent
      coefficient $\kappa_1=3/19$  at threshold and 1 at asymptotically high energies,
      leading to the simple $\sin^2\theta$ distribution identical to the spin-0 case.
      This asymptotic behavior is a consequence of the so-called Goldstone boson equivalence
      theorem \cite{Cornwall:1974km}.
\end{itemize}
{\it To summarize.} The characteristic patterns of the polarized ECC polar-angle distributions
can be powerful in determining the spin of $\mathcal{P}^\pm$. Evidently it is crucial to
have the (longitudinal) polarization of electron and positron beams for the spin determination
through the angular distribution. However, we note that the polar-angle distributions alone
may not be powerful enough for covering the more general scenarios. \s

\subsection{Single lepton polar-angle distributions in the decays}
\label{subsec:lepton_polar_angle_distribution}

If the parent particle $\mathcal{P}^\pm$ is a spin-0 scalar boson $S^\pm_p$, there is no
production-decay angular correlation at all so that the (normalized) lepton
polar-angle distribution is flat, independently of any chirality assignments to the
couplings for the production and decay processes as well as of any initial
beam polarizations, i.e.
\begin{eqnarray}
   \frac{d\sigma^S\left[S^-_p\to\ell^- \bar{F}^0_d\right]}{d\cos\theta d\cos\theta_-}
 = \frac{d\sigma^S}{d\cos\theta}\cdot \frac{1}{2} \ \ \Rightarrow \ \
\frac{1}{{\cal C}_{\rm sf}}\frac{d {\cal C}_{\rm sf}}{d\cos\theta_-} = \frac{1}{2}
\end{eqnarray}
The linear relation in Eq.$\,$(\ref{eq:theta_pm_energy_relation}) between the polar angle
$\theta_\pm $ and the $\ell^-$ energy $E_{\ell^\pm }$ indicates that the lepton energy
distribution is flat with the energy between $E_{\rm min}=E^*_{\ell^\pm}(1-\beta)$
and $E_{\rm max}=E^*_{\ell^\pm}(1+\beta)$ with $\beta=\sqrt{1-4 M^2_{\mathcal{P}}/s}$.\s

When the parent particle $\mathcal{P}^\pm$ is a spin-1/2 fermion $F^\pm_p$, then we can
directly determine the differential or total cross section for fixed $F_p^\pm$ helicities
by measuring the polar angle distribution of the $F^\pm_p$ decay products. Depending on the
spin of the invisible particle $\mathcal{D}^0= S^0_d, V^0_d$ and the chirality
assignments to the $F_p S_d \ell$ and $F_p V_d \ell$ couplings,
the normalized and correlated polar-angle distributions can be expressed as
\begin{eqnarray}
    \frac{d\sigma^F [F^-_p \to \ell^- \bar{S}^0_d]}{d\cos\theta\, d\cos\theta_-}
&=& \frac{d\sigma^F}{d\cos\theta}\,\cdot \frac{1}{2}
     \left[ 1 + \xi_{\rm fs}\, \mathbb{P}_{F}\, \cos\theta_-\right]
     \hskip 5.mm \ \ \Rightarrow \ \
 \frac{1}{{\cal C}_{\rm fs}}\frac{d {\cal C}_{\rm fs}}{d\cos\theta_-}
 = \frac{1}{2}\left[1+ \xi_{\rm fs}\, \langle \mathbb{P}_{F}\rangle\, \cos\theta_-\right]
\label{eq:single_lepton_polar_angle_distribution_F_-} \\
    \frac{d\sigma^F [F^-_p \to \ell^- \bar{V}^0_d]}{d\cos\theta\, d\cos\theta_-}
&=& \frac{d\sigma}{d\cos\theta}\,\cdot \frac{1}{2}
    \left[ 1 + \xi_{\rm fv} \eta_{\rm fv}\, \mathbb{P}_{F}\, \cos\theta_- \right]
    \ \ \Rightarrow \ \
 \frac{1}{{\cal C}_{\rm fv}}\frac{d {\cal C}_{\rm fv}}{d\cos\theta_-}
 =\frac{1}{2}\left[1+ \xi_{\rm fv} \eta_{\rm fv}\,
                   \langle \mathbb{P}_{F}\rangle\, \cos\theta_-\right]
\label{eq:single_lepton_polar_angle_distribution_F_+}
\end{eqnarray}
where two relative chiralities $\xi_{\rm fs}$ and $\xi_{\rm fv}$ and one dilution factor
$\eta_{\rm fv}$ are defined by
\begin{eqnarray}
    \xi_{\rm fs}
&=& \frac{|d^{\ell S}_{F+}|^2-|d^{\ell S}_{F-}|^2}{
          |d^{\ell S}_{F+}|^2+|d^{\ell S}_{F-}|^2} \\
    \xi_{\rm fv}
&=& \frac{|d^{\ell V}_{F+}|^2-|d^{\ell V}_{F-}|^2}{
          |d^{\ell V}_{F+}|^2+|d^{\ell V}_{F-}|^2} \\
    \eta_{\rm fv}
&=& \frac{M^2_{F_p}-2M^2_{V_d}}{
          M^2_{F_p}+2M^2_{V_d}}
\label{eq:eta_fv}
\end{eqnarray}
in terms of the chiral coupling coefficients (which are introduced in
Appendix~\ref{sec:appendix_a_feynman_rules}) and the masses $M_{F_p}$ and $M_{V_d}$,
and the differential cross section and the polar-angle dependent polarization observable
are defined by
\begin{eqnarray}
    \frac{d\sigma^F}{d\cos\theta}
&=& \frac{d\sigma^F (\lambda_-=+)}{d\cos\theta}
    +\frac{d\sigma^F (\lambda_-=-)}{d\cos\theta} \\
    \mathbb{P}_{F}
&=& \left[\frac{d\sigma^F (\lambda_-=+)}{d\cos\theta}
         -\frac{d\sigma^F (\lambda_-=-)}{d\cos\theta}\right]\bigg{/}
          \frac{d\sigma^F}{d\cos\theta}
\end{eqnarray}
respectively. The average of the polarization observable over the production angle
$\theta$ are given by
\begin{eqnarray}
    \langle \mathbb{P}_{F}\rangle
= \frac{1}{\sigma^F}\, \int^1_{-1} \mathbb{P}_{F}\,
  \frac{d\sigma^F}{d\cos\theta}\, d\cos\theta
= \left(\mathbb{p}^{++}_{++}
       +\mathbb{p}^{+-}_{+-}\right)
 -\left(\mathbb{p}^{-+}_{-+}
       +\mathbb{p}^{--}_{--}\right)
\end{eqnarray}
satisfying the inequality condition $|\langle \mathbb{P}_F\rangle |\leq 1$ in terms of the
normalized production tensor $\mathbb{p}$ defined as an integral over the production polar
and azimuthal angles $\theta$ and $\varphi$ as
\begin{eqnarray}
   \mathbb{p}^{\lambda_- \lambda_+}_{\lambda'_- \lambda'_+}
= \int {\cal P}^{\lambda_- \lambda_+}_{\lambda'_- \lambda'_+}\,\,
   d\cos\theta\, d\varphi
  \bigg/
   \int \bigg(\sum_{\kappa_-,\kappa_+}
        {\cal P}^{\kappa_-\kappa_+}_{\kappa_-\kappa_+}\bigg)\,\,
        d\cos\theta\, d\varphi
\label{eq:normalized_integrated_production_tensor}
\end{eqnarray}
with the production tensor $\mathcal{P}$'s. The production tensor
$\mathbb{p}$ satisfies the normalization condition $\sum_{\lambda_-,\lambda_+}
\mathbb{p}^{\lambda_-\lambda_+}_{\lambda_-\lambda_+}=1$.\s

Any non-trivial $\ell^-$ polar-angle distribution can exist only when the parent particle
$F^\pm_p$ state has a non-zero degree of longitudinal polarization $\mathbb{P}_{F}$ which
may be generated by some parity-violating interactions or by electron (and positron) beam
polarizations. At the same time, the relative chiralities, $\xi_{\rm fs}$ and $\xi_{\rm fv}$,
and the polarization dilution factor $\eta_{\rm fv}$ should not be zero. \s

\begin{figure}[thb]
\centering
\includegraphics[width=16.cm]{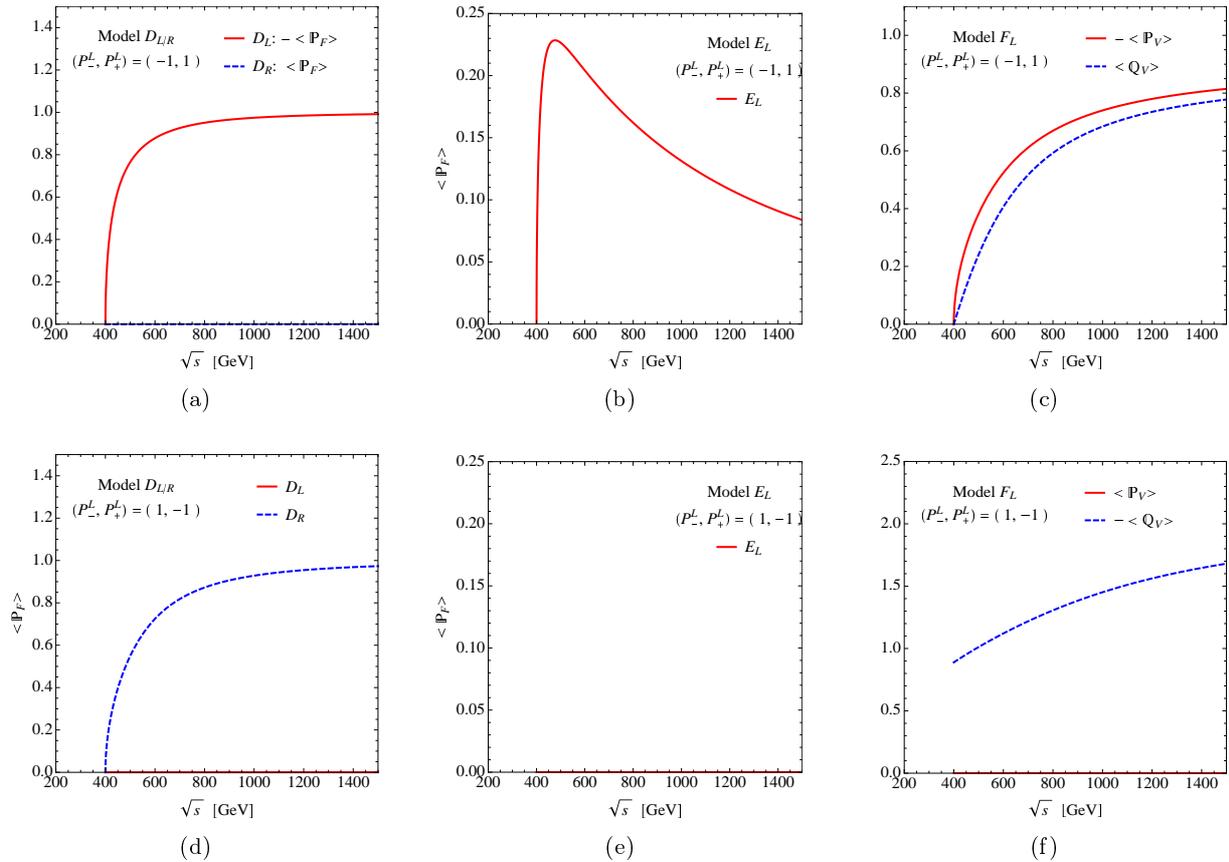}
\caption{\label{fig:energy_dependence_polarization_observables_1}
        {Energy dependence of the coefficients ${\langle \mathbb{P}_{F,V} \rangle}$ and
             ${\langle \mathbb{Q}_V \rangle}$ for a first KK electron ($D_{L/R}$), for a
             spin-1/2 charged wino ($E_L$), and for a spin-1 charged first KK $W$-boson
             ($F_L$). The upper (lower) frames are for left-handed (right-handed) electron
             and right-handed (left-handed) positron beams, respectively.
        }
        }
\end{figure}

It is evident from Eqs.$\,$(\ref{eq:single_lepton_polar_angle_distribution_F_-}) and
(\ref{eq:single_lepton_polar_angle_distribution_F_+}) that the single polar-angle
distributions are isotropic as in the scalar case if the relative chiralities,
$\xi_{\rm fs}$ and $\xi_{\rm fv}$, are zero, i.e. the couplings for the decays,
$F^-_p\to\ell^- \bar{S}^0_d$ and $F^-_p\to\ell^- \bar{V}^0_d$, are pure
scalar-type and pure vector-type. In the latter decay mode, not only the relative chirality
but also the dilution factor $\eta_{\rm fv}$ must not be zero, i.e, $M_{F_p}\neq
\sqrt{2} M_{V_d}$. Furthermore, as mentioned before, the $P$-odd polarization observable
$\langle \mathbb{P}_F\rangle$ needs to be non-zero in both of the decay modes, for
any non-trivial single decay polar-angle distributions.  \s

Before presenting the single decay polar-angle distributions at a fixed c.m. energy
$\sqrt{s}=500$\, GeV, we investigate the energy and polarization dependence of the
$P$-odd polarization observable $\langle \mathbb{P}_F\rangle$ in the $C_{L/R},\, D_{L/R}$
and $E_L$ scenarios of spin-1/2 particles.
\begin{itemize}
\item Firstly, we note that the polarization observable is identically zero, independently
      of beam polarization, in the $C_{L/R}$ scenario for the production of a first KK-muon
      pair $\mu^+_1\mu^-_1$, because the coupling of the $Z$ as well as $\gamma$ to the
      first KK muon pair in the $s$-channel exchange diagram is of a pure vector-type,
      generating no $P$-violating effects, so that the single decay polar-angle distribution
      is isotropic as in the spin-0 scalar-pair production. Therefore, the single decay
      polar-angle distribution cannot be exploited for distinguishing the spin-1/2 case of
      a first KK muon pair from the spin-0 case of a smuon pair.
\item In contrast, as the production of a first KK electron pair occurs through the $t$-channel
      spin-1 vector-boson contributions with pure left-chiral ($D_L$) or right-chiral ($D_R$)
      $e e_1 V^0_t$ couplings with the first KK vector boson $V^0_t= B_1,\, W^0_1$ as well as
      the $s$-channel $\gamma$ and $Z$-boson contributions with pure vector couplings with
      $F^\pm_p$, the $P$-odd polarization observable depends strongly on the c.m. energy
      and beam polarizations. As the c.m. energy increases, the $t$-channel contributions
      with maximally $P$-violating couplings become dominant rapidly due to the exchange
      of spin-1 neutral vector bosons $B_1$ and $W^0_1$ so that the $P$-odd observable
      approaches its maximum value of unity in magnitude in the $D_L$ ($D_R$) scenario
      for left-handed (right-handed) electron and  right-handed (left-handed) positron
      polarizations. In the former and
      latter cases ($D_L$ and $D_R$), the observable is negative and positive, respectively.
      On the other hand, for the opposite combination of beam polarizations the observable
      is zero because the $t$-channel contributions are killed. These features are clearly
      demonstrated in Figs.$\,$\ref{fig:energy_dependence_polarization_observables_1}(a)
      and (d).
\item In the charged wino case ($E_L$), the $t$-channel diagram is mediated by a
      spin-0 electron sneutrino $\tilde{\nu}_e$, killing the amplitude effectively
      in the forward direction due to chirality flipping. As a result, the $P$-odd
      observable decreases in size as the c.m. energy increases. Moreover, as the
      $e \tilde{\nu}_e \tilde{W}$ coupling is purely left-chiral, the $P$-odd observable
      is zero for right-handed electron and left-handed positron polarizations.
      These features can be verified with the plots in
      Figs.$\,$\ref{fig:energy_dependence_polarization_observables_1}(b) and (e).
\end{itemize}
It is necessary to compare these features of the spin-1/2 $F^\pm_p$ cases to those
for the spin-1 $V^\pm_p$ cases.\s

When the parent particle $\mathcal{P}^\pm$ is a spin-1 vector boson $V^\pm_p$, the correlated
polar-angle distributions and the normalized lepton polar-angle distribution are given
in terms of the $V^\pm_p$ helicity-dependent production cross sections by
\begin{eqnarray}
    \frac{d\sigma^V[V^-_p\to \ell^- \bar{F}^0_d]}{d\cos\theta\, d\cos\theta_-}
&=& \frac{d\sigma^V}{d\cos\theta}\,\cdot \frac{1}{2}
    \left[1 + \frac{3}{2}\,\xi_{\rm vf}\, \eta_{\rm vf}\, \mathbb{P}_{V}\, \cos\theta_-
            + \frac{1}{2} (3\eta_{\rm vf}-2)\, \mathbb{Q}_{V}\,
              \frac{(3\cos^2\theta_- -1)}{2}
             \right] \\
    \frac{1}{{\cal C}_{\rm vf}} \frac{d {\cal C}_{\rm vf}}{d\cos\theta_-}
&=& \frac{1}{2} \left[1 + \frac{3}{2}\,\xi_{\rm vf}\, \eta_{\rm vf}\,
                          \langle \mathbb{P}_{V}\rangle \, \cos\theta_-
            + \frac{1}{2} (3\eta_{\rm vf}-2)\, \langle \mathbb{Q}_{V}\rangle\,
              \frac{(3\cos^2\theta_- -1)}{2}
             \right]
\end{eqnarray}
with a relative chirality $\xi_{\rm vf}$ and a dilution factor
$\eta_{\rm vf}$ defined by
\begin{eqnarray}
   \xi_{\rm vf}
&=& \frac{|d^{\ell V}_{F+}|^2-|d^{\ell V}_{F-}|^2}{
          |d^{\ell V}_{F+}|^2+|d^{\ell V}_{F-}|^2} \\
   \eta_{\rm vf}
&=& \frac{2M^2_{V_p}}{2M^2_{V_p}+M^2_{F_d}}
\end{eqnarray}
with its minimum value of $\eta^{\rm min}_{\rm vf}=2/3$ for $M_{V_p}=M_{F_d}$,
where the differential cross section and two polarization observables $\mathbb{P}_V$
and $\mathbb{Q}_V$ are defined by
\begin{eqnarray}
   \frac{d\sigma^V}{d\cos\theta}
&=& \frac{d\sigma^V(\lambda_-=+)}{d\cos\theta}
   +\frac{d\sigma^V(\lambda_-=-)}{d\cos\theta}
   +\frac{d\sigma^V(\lambda_-=0)}{d\cos\theta} \\
   \mathbb{P}_{V}
&=& \left[\frac{d\sigma^V(\lambda_-=+)}{d\cos\theta}
         -\frac{d\sigma^V(\lambda_-=-)}{d\cos\theta}\right]\bigg{/}
                   \frac{d\sigma^V}{d\cos\theta} \\
   \mathbb{Q}_{V}
&=& \left[\frac{d\sigma^V(\lambda_-=+)}{d\cos\theta}
         +\frac{d\sigma^V(\lambda_-=-)}{d\cos\theta}
         -2\frac{d\sigma^V(\lambda_-=0)}{d\cos\theta}\right]\bigg{/}
                   \frac{d\sigma^V}{d\cos\theta}
\end{eqnarray}
and the averages of two polarization observables over the polar-angle distribution are
given by
\begin{eqnarray}
  \langle \mathbb{P}_V \rangle
&=& \frac{1}{\sigma} \int^1_{-1} \mathbb{P}_V \frac{d\sigma^V}{d\cos\theta}\,
     d\cos\theta
 =  \left(\mathbb{p}^{++}_{++}+\mathbb{p}^{+0}_{+0}
         +\mathbb{p}^{+-}_{+-}\right)
   -\left(\mathbb{p}^{-+}_{-+}
         +\mathbb{p}^{-0}_{-0}
         +\mathbb{p}^{--}_{--}\right)\\
     \langle \mathbb{Q}_V \rangle
&=& \frac{1}{\sigma} \int^1_{-1} \mathbb{Q}_V \frac{d\sigma^V}{d\cos\theta}\,
     d\cos\theta
 =  \left(\mathbb{p}^{++}_{++}
         +\mathbb{p}^{+0}_{+0}
         +\mathbb{p}^{+-}_{+-}\right)
   +\left(\mathbb{p}^{-+}_{-+}
         +\mathbb{p}^{-0}_{-0}
         +\mathbb{p}^{--}_{--}\right)
   -2\left(\mathbb{p}^{0+}_{0+}
          +\mathbb{p}^{00}_{00}
          +\mathbb{p}^{0-}_{0-}\right)
\end{eqnarray}
satisfying the inequality conditions $|\langle \mathbb{P}_V \rangle |\leq 1$ and
$|\langle \mathbb{Q}_V\rangle |\leq 2$ in terms of the $3\times 3$ normalized production
tensor matrix $\mathbb{p}$ defined similarly to
the equation (\ref{eq:normalized_integrated_production_tensor}).\s

Clearly, only if the vector boson $V^-_p$ is unpolarized, i.e. the production cross
section for each $V^-_p$ is identical with $d\sigma^V(\lambda_-=+) = d\sigma^V (\lambda_-=-)
=d\sigma^V (\lambda_-=0)$, will the decay polar-angle distribution be
isotropic. Note that, even if there are no parity-violating effects, i.e. $d\sigma^V(\lambda_-=+) =
d\sigma^V(\lambda_-=-)$, in the production process, there can exist a non-trivial lepton
polar-angle distribution proportional to $3\cos^2\theta_- -1$, unless the averaged degree of
longitudinal polarization $P_L(V^-_p) = \sigma^V(\lambda_-=0)/\sigma^V$ of the particle
$V^-_p$ is identical to $1/3$. These properties are demonstrated by the plots
in Figs.$\,$\ref{fig:energy_dependence_polarization_observables_1}(c) and (f)
for the production of a charged first KK $W$-boson pair with $s$-channel $\gamma, Z$ exchanges
with pure vector-type couplings and $t$-channel spin-1/2 first KK neutrino exchange with a
pure left-chiral coupling. Firstly, as the right-handed electron and left-handed positron
polarizations kill the $t$-channel contributions, the $P$-odd observable
$\langle \mathbb{P}_V\rangle$ is vanishing so that there is no term linear in $\cos\theta_-$.
Even in this case the $P$-even polarization observable $\langle \mathbb{Q}_V\rangle$ survives
and increases in size as the c.m. energy increases as shown in
Figs.$\,$\ref{fig:energy_dependence_polarization_observables_1}(f). Secondly, for the left-handed
electron and right-handed positron polarizations, the $P$-violating $t$-channel contribution
survives and both the $P$-even and $P$-odd observables increase in size as the c.m. energy
increases as shown in
Figs.$\,$\ref{fig:energy_dependence_polarization_observables_1}(c).\s

\begin{figure}[htb]
\centering
\includegraphics[width=18.cm]{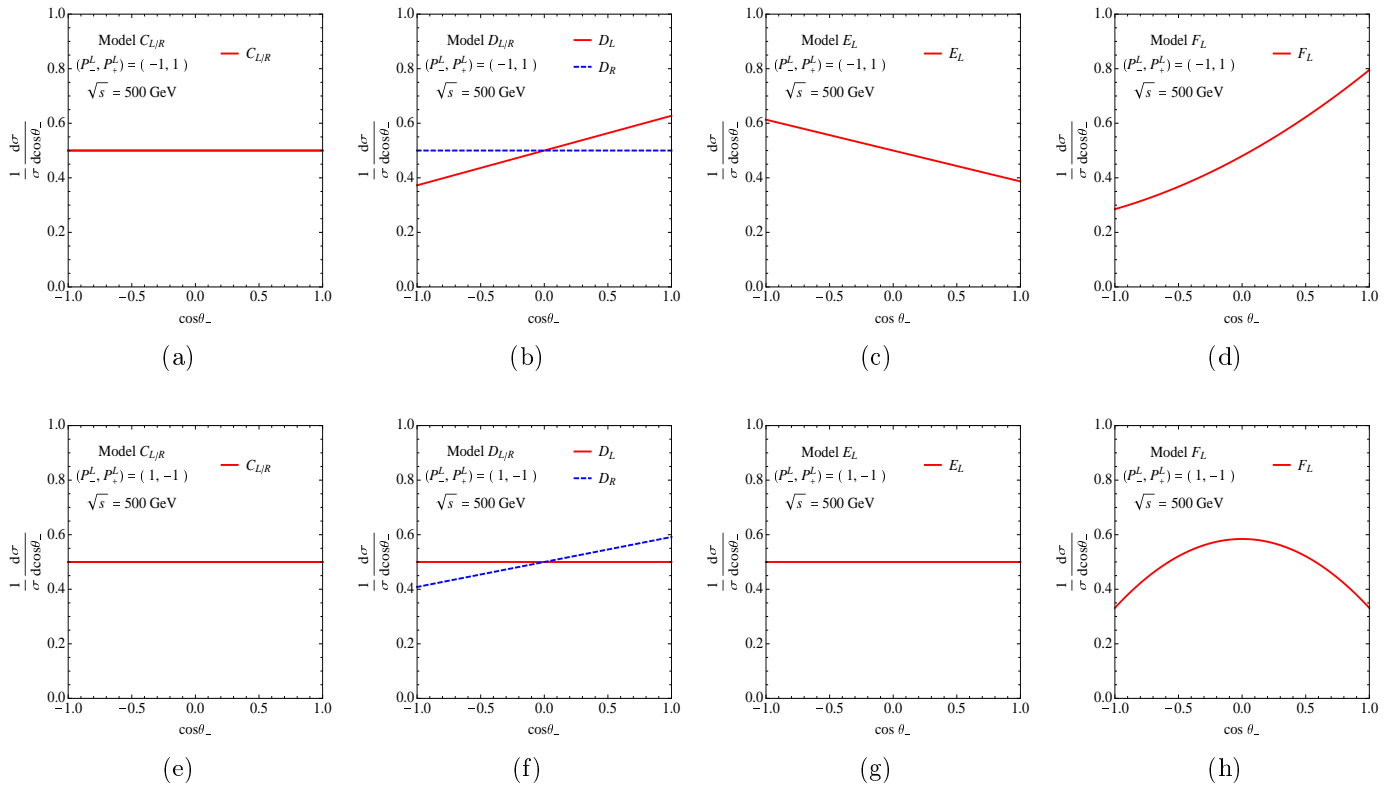}
\caption{\label{fig:decay_polar-angle_distributions}
        {Normalized single decay polar-angle distributions for a spin-1/2 charged first
             KK muon $\mu^-_{1L/1R}$ ($C_{L/R}$) and first KK electron $e^-_{1L/1R}$ ($D_{L/R}$),
             for a spin-1/2 charged wino $\tilde{W}^-$ ($E_L$) and for a spin-1 charged first
             KK $W$-boson $W^-_1$ ($F_L$), pair produced with its anti-particle in $e^+e^-$
             collisions at a fixed c.m. energy of 500 GeV. The upper (lower) frames are for
             left-handed (right-handed) electron and right-handed (left-handed) positron
             beams.}
        }
\end{figure}

Figure~\ref{fig:decay_polar-angle_distributions} shows the normalized single decay polar-angle
distributions for a spin-1/2 negatively charged first KK muon $\mu^-_{1L/1R}$ ($C_{L/R}$) and
first KK electron $e^-_{1L/1R}$ ($D_{L/R}$), for a spin-1/2 negatively charged wino
$\tilde{W}^-$ ($E_L$) and for a spin-1 negatively charged first KK $W$-boson $W^-_1$ ($F_L$),
pair produced with its anti-particle in $e^+e^-$ collisions at a fixed c.m. energy of 500 GeV.
\begin{itemize}
\item As shown in Figs.$\,$\ref{fig:decay_polar-angle_distributions}(a) and (e),
      the distribution for the $\mu^-_1$ decay
      is flat because the couplings of both $\gamma$ and $Z$ to the $\mu^+_1\mu^-_1$ pair
      are pure vector-type, preserving parity ($P$).
\item Similarly the flat distributions appear for the left-handed (right-handed) KK electron
      with right-handed (left-handed) electron and left-handed (right-handed) positron
      polarizations as shown by a (blue) dashed line in
      Figs.$\,$\ref{fig:decay_polar-angle_distributions}(b) and a (red)
      solid line in Figs.$\,$\ref{fig:decay_polar-angle_distributions}(f),
      as in both cases the $P$-violating $t$-channel
      contributions are killed. The same flat distribution in the $\tilde{W}$ decay occurs
      for right-handed electron and left-handed positron beams, killing the $t$-channel
      sneutrino contribution, as shown in Figs.$\,$\ref{fig:decay_polar-angle_distributions}(g).
\item There exist non-trivial decay polar-angle distributions with a positive
      slope in the $e^-_{1L/1R}$ decay for left-handed/right-handed electron and
      right-handed/left-handed positron beams as shown by the red solid line in
      Figs.$\,$\ref{fig:decay_polar-angle_distributions}(b) and by the blue dashed line
      in Figs.$\,$\ref{fig:decay_polar-angle_distributions}(f). This is due to
      the fact that both the $P$-odd polarization observable $\langle\mathbb{P}_F\rangle$ and
      the relative chirality factor $\xi_{\rm fv}$ is negative and positive for the $e^-_{1L}$
      and $e^-_{1R}$ decay, respectively, so that the product of two quantities is positive
      in both cases. In contrast, in the $E_L$ case with left-handed electron and right-handed
      positron beams, the $P$-odd polarization observable is positive but the relative
      chirality $\xi_{\rm fs}$ is negative so that the slope determined by the product of
      two quantities is negative as shown in
      Figs.$\,$\ref{fig:decay_polar-angle_distributions}(c).
\item Finally, in the $F_L$ case for a spin-1 negatively charged first KK $W$-boson $W^-_1$
      decay, the lines are clearly curved instead of being straight, as shown in
      Figs.$\,$\ref{fig:decay_polar-angle_distributions}(d) and (h). In particular,
      even though the coupling of $\gamma$ and $Z$ to a $W^\pm_1$ pair is $P$-conserving
      so that the $P$-odd observable $\langle \mathbb{P}_V\rangle$ vanishes for right-handed
      electron and left-handed positron beams, the single decay polar-angle distribution
      takes a non-trivial quadratic curve shape due to non-vanishing $P$-even polarization
      observable $\langle \mathbb{Q}_V\rangle$.
\end{itemize}
{\it To summarize.} It is necessary to have $P$-violating decays for any non-trivial single
decay polar-angle distribution. Moreover, in the spin-1/2 case, the production process
must have $P$-violating contributions due to the presence of $P$-violating interactions
which can be greatly enhanced by initial beam polarizations. In the spin-1 case,
in addition to the $P$-odd polarization observable, there can exist a $P$-even
polarization observable leading to non-trivial decay polar-angle distribution, the shape
of which is quadratic in $\cos\theta_\mp$.\s

\subsection{Angular correlations of two charged leptons}
\label{subsec:angular_correlation}

As can be checked with Eqs.$\,$(\ref{eq:single_lepton_polar_angle_distribution_F_-})
and (\ref{eq:single_lepton_polar_angle_distribution_F_+}), the lepton polar-angle
distribution of the process $e^-e^+\to F^-_p F^+_p$ followed by the decay
$F^-_p\to\ell^- S^0_d$ or $\ell^- V^0_d$ is isotropic if the integration of
the polarization observable $\mathbb{P}_{F}$ over the polar-angle $\theta$
is vanishing as in the KK muon-pair production due to the pure vector coupling
of the photon and $Z$ boson to the KK muon pair. Therefore, a single lepton angle
distribution cannot be exploited to distinguish the spin-1/2 case from the spin-0 case.
In this situation, we can exploit the angular correlations of two charged
leptons. \s

\subsubsection{Polar-angle correlations}
\label{subsubsec:polar-angle_correlation}

As the spin-1 case can usually be distinguished from the spin-0 and spin-1/2
cases through the coefficient proportional to $(3\cos^2\theta_--1)$ even when either
the $P$-odd observable $\langle \mathbb{P}_F\rangle $ or the $P$-odd relative
chirality is vanishing. On the contrary, in the spin-1/2 case there can exist a
non-trivial single lepton polar-angle distribution only when
both the $P$-odd coefficients and the $P$-odd integral are non-vanishing. Otherwise,
the spin-1/2 case cannot be distinguished from the spin-0 case by the single
lepton angular distribution. In this $P$-invariant case, we can consider the
polar-angle correlation of two final leptons, which is a $P$-even quantity.
In general, the polar-angle correlation in the spin-1/2 case can be decomposed
into four parts as
\begin{eqnarray}
  \frac{1}{{\cal C}_D }\frac{d{\cal C}_D}{d\cos\theta_- d\cos\theta_+}
\! =\! \frac{1}{4} \left\{ 1 \! +\! \xi_D \eta_D
        \left[(\cos\theta_- \! +\! \cos\theta_+)\, \Theta_{F1}
             +(\cos\theta_- \! -\! \cos\theta_+)\, \Theta_{F2}
             \right]
      + \xi^2_D \eta^2_D\, \cos\theta_- \cos\theta_+\, \Theta_{F3}\right\}
\label{eq:decay_polar_angle_correlations}
\end{eqnarray}
with $\xi_D=\xi_{\rm fs},\, \xi_{\rm fv}$ and $\eta_D=1,\, \eta_{\rm fv}$ for the decay modes,
$F^-_p\to\ell^- S^0_d,\, \ell^- V^0_d$, respectively. Here, the $P$-odd coefficients,
$\Theta_{F1,F2}$, and the $P$-even coefficient $\Theta_{F3}$, which are in general dependent on
the $e^+e^-$ c.m. energy and beam polarizations, are given by
\begin{eqnarray}
\Theta_{F1} &=& \mathbb{p}^{++}_{++}-\mathbb{p}^{--}_{--} \\
\Theta_{F2} &=& \mathbb{p}^{+-}_{+-}-\mathbb{p}^{-+}_{-+} \\
\Theta_{F3} &=& \left(\mathbb{p}^{+-}_{+-}+\mathbb{p}^{-+}_{-+}\right)
               -\left(\mathbb{p}^{++}_{++}+\mathbb{p}^{--}_{--}\right)
\label{eq:decay_polar_angle_correlation_coefficients}
\end{eqnarray}
We note in passing that the $P$-odd quantity $\langle \mathbb{P}_F\rangle$ appearing in the
single lepton polar-angle $\theta_-$ distributions is identical to the sum
$\Theta_{F1}+\Theta_{F2}$. An identical relation is valid also for the $P$-odd quantity
$\langle \mathbb{P}_V\rangle$ in the spin-1 case.\s

\begin{figure}[htb]
\centering
\includegraphics[width=16.cm]{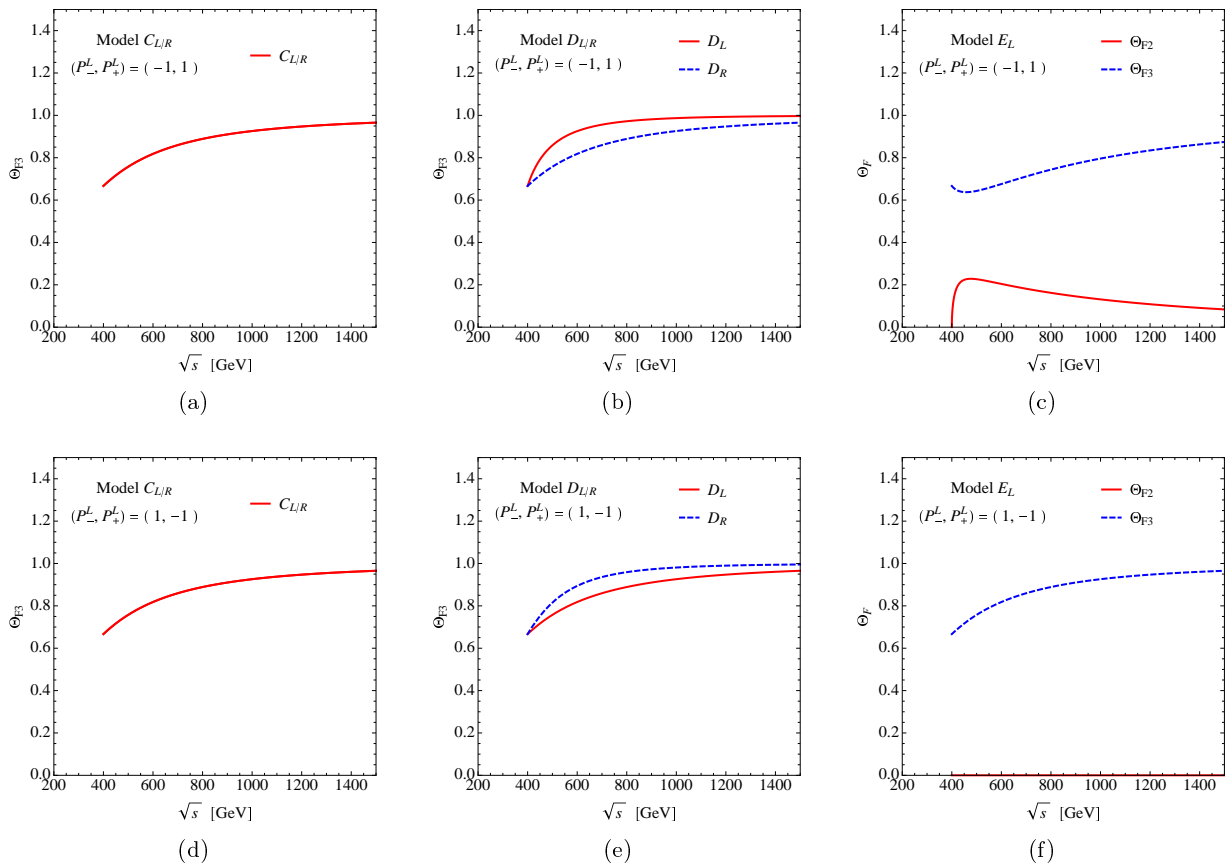}
\caption{\label{fig:decay_polar-angle_correlation_coefficients}
        {Energy dependence of the coefficients $\Theta_{F2}$ and/or $\Theta_{F3}$ of
             the correlated decay polar-angle distributions for a spin-1/2 charged first
             KK-muon ($C_{L/R}$) and first KK-electron ($D_{L/R}$) and for a spin-1/2 charged
             wino ($E_L$). The lines in the upper (lower) frames are for left-handed
             (right-handed) electron and right-handed (left-handed) positron beams,
             respectively.}
        }
\end{figure}

As indicated in the previous subsection, the $P$-odd quantities $\Theta_{F1,F2}$ are
vanishing\footnote{The quantity $\Theta_{F1}$ vanishes in the absence of any absorptive parts
as a consequence of $CPT$ invariance.}
in the production of a first KK-muon  $\mu^\pm_{1L/1R}$ pair, because the coupling of the spin-1
vector bosons $\gamma, Z$ to the first KK-muon pair is of a pure vector type. However, the
coefficient $\Theta_{F3}$ defining the $P$-conserving decay polar-angle correlation in
Eq.$\,$(\ref{eq:decay_polar_angle_correlations}) is $P$-even so that this quantity does not
have to be vanishing even in the $P$-conserving case. As shown numerically by the (red) solid
lines in Figs.$\,$\ref{fig:decay_polar-angle_correlation_coefficients}(a)
and (d), the $P$-even coefficient $\Theta_{F3}$ increases in size as the c.m. energy
increases.
As a consequence, it is evident that the spin-1/2 $\mu^\pm_1$ case can be distinguished
from the spin-0 $\tilde{\mu}^\pm_{L/R}$ case through the non-trivial polar-angle
correlation, which can be significant for pure right-chiral or left-chiral decays with
$\xi_{\rm fv}=\pm 1$ and a sizable dilution factor $\eta_{\rm fv}$. We note in passing
that the $P$-even coefficients in the other spin-1/2 cases ($D_{L/R}$ and $E_L$)
in Figs.$\,$\ref{fig:decay_polar-angle_correlation_coefficients}(b), (e), (c)and (f)
are also increasing in size with the c.m. energy and already sizable at the c.m. energy
of 500 GeV. \s

\subsubsection{Azimuthal-angle correlations}
\label{subsubsec:azimuthal_angle_correlation}

In this subsection, we study the fully-correlated azimuthal-angle distributions in the
production of a $\mathcal{P}^+\mathcal{P}^-$ pair in $e^+e^-$ collisions and both of their
sequential two-body decays $\mathcal{P}^-\to\ell^- \mathcal{\bar{D}}^0$ and
$\mathcal{P}^+\to \ell^+ \mathcal{D}^0$. The azimuthal-angle difference $\phi=\phi_- - \phi_+$
under consideration is the angle between the two decay planes, which is invariant under any
Lorentz boost along the $\mathcal{P}^\pm$ flight direction. These distributions
develop through quantum interference between the different helicity states in a coherent
sum, indicating that the effect is diluted as the $e^+e^-$ c.m. energy increases.
By extracting this angular dependence, we can determine which helicity states contribute
to the sum, and thus we can extract useful information on the spin of the pair-produced
particles in a model-independent way\cite{Buckley:2008eb}. \s

The general form of the azimuthal-angle correlation for the production and decays of
a spin $J$ particle pair is
\begin{eqnarray}
  \frac{1}{\cal C}\frac{d{\cal C}}{d\phi}
= \frac{1}{2\pi}
  \left[1+ \mathbb{A}_1 \cos(\phi) + \cdots + \mathbb{A}_{2J} \cos (2J\phi)\right]
\end{eqnarray}
We emphasize that the expression is still valid even in a $CP$-noninvariant theory as
all the sine terms are washed out by taking the average over two possible production
azimuthal angles, which is unavoidable due to a twofold ambiguity in reconstructing the
$\mathcal{P}^\pm$ momentum directions as shown in Eq.$\,$(\ref{eq:sin_phi_ambiguity}) in
Appendix \ref{sec:appendix_d_kinematics_of_the_antler_process}.
Each coefficient $\mathbb{A}_i$ ($i=1$-$2J$) can be worked out from the standard rules of
constructing matrix elements, as explicitly described below for the spin-1/2 and spin-1
cases. \s

Evidently, the correlated azimuthal-angle distribution for a spin-0 scalar-pair production
process is flat due to the absence of any production-decay spin correlations. In contrast,
the azimuthal-angle distribution for the spin-1/2 fermion-pair production process is given
by
\begin{eqnarray}
   \frac{1}{{\cal C}_D} \frac{d{\cal C}_D}{d\phi}
= \frac{1}{2\pi}\, \left[1- \xi^2_D \eta^2_D\,\Phi_F \cos\phi\right]
\end{eqnarray}
in terms of a $C$-even and $P$-even quantity $\Phi_F$ defined as
\begin{eqnarray}
\Phi_F = \frac{\pi^2}{16}\, {\rm Re} (\mathbb{p}^{++}_{--}+\mathbb{p}^{--}_{++})
\label{eq:azimuthal_angle_correlation_coefficient_f}
\end{eqnarray}
with $\xi_D= \xi_{\rm fs}, \xi_{\rm fv}$ and $\eta_D=1, \eta_{\rm fv}$ for
$\mathcal{D}^0=S^0_d$ and $V^0_d$, respectively, and the super/sub-scripts $\pm$ for the
helicities $\pm 1/2$ of the spin-1/2 fermion $F^\pm_p$. On the other hand, the correlated
azimuthal-angle distribution for a spin-1 vector boson pair $V^+_p V^-_p$ consists
of three parts as
\begin{eqnarray}
   \frac{1}{{\cal C}_{\rm vf}} \frac{d{\cal C}_{\rm vf}}{d\phi}
= \frac{1}{2\pi}\left[1- \xi^2_{\rm vf} \eta^2_{\rm vf}\, \Phi_{V1}\, \cos\phi
    +(3\eta_{\rm vf}-2)^2\, \Phi_{V2}\, \cos(2\phi)\right]
\end{eqnarray}
in terms of two $C$-even and $P$-even quantities $\Phi_{V1}$ and $\Phi_{V2}$
defined as\footnote{As the coefficients $\Phi_F$ and $\Phi_{V1, V2}$ are
$C$-even, no identification of the electric charges of two leptons
is required for reconstructing the azimuthal-angle correlations.}
\begin{eqnarray}
\Phi_{V1} &=& \frac{9\pi^2}{64}\,
              {\rm Re}(\mathbb{p}^{++}_{00}+\mathbb{p}^{00}_{--}
                      +\mathbb{p}^{+0}_{0-}+\mathbb{p}^{0+}_{-0})
\label{eq:azimuthal_angle_correlation_coefficient_v1} \\
\Phi_{V2} &=& \frac{1}{4}\,
              {\rm Re} (\mathbb{p}^{++}_{--}+\mathbb{p}^{--}_{++})
\label{eq:azimuthal_angle_correlation_coefficient_v2}
\end{eqnarray}
where the super/sub-scripts $\pm, 0$ stand for the helicities, $\pm 1, 0$ of the
spin-1 vector bosons $\mathcal{P}^\pm$. We note that the two-body decays do not
suppress the $\cos\phi$ terms, while the highest $\cos(2\phi)$ mode may be
suppressed if the polarization analyzing power $\eta_{\rm vf}$ is $2/3$, satisfied
only when the parent and daughter particles, $V^\pm_p$ and $F^0_d$, are
nearly degenerate. \s

\begin{figure}[tbh]
\centering
\includegraphics[width=18.cm]{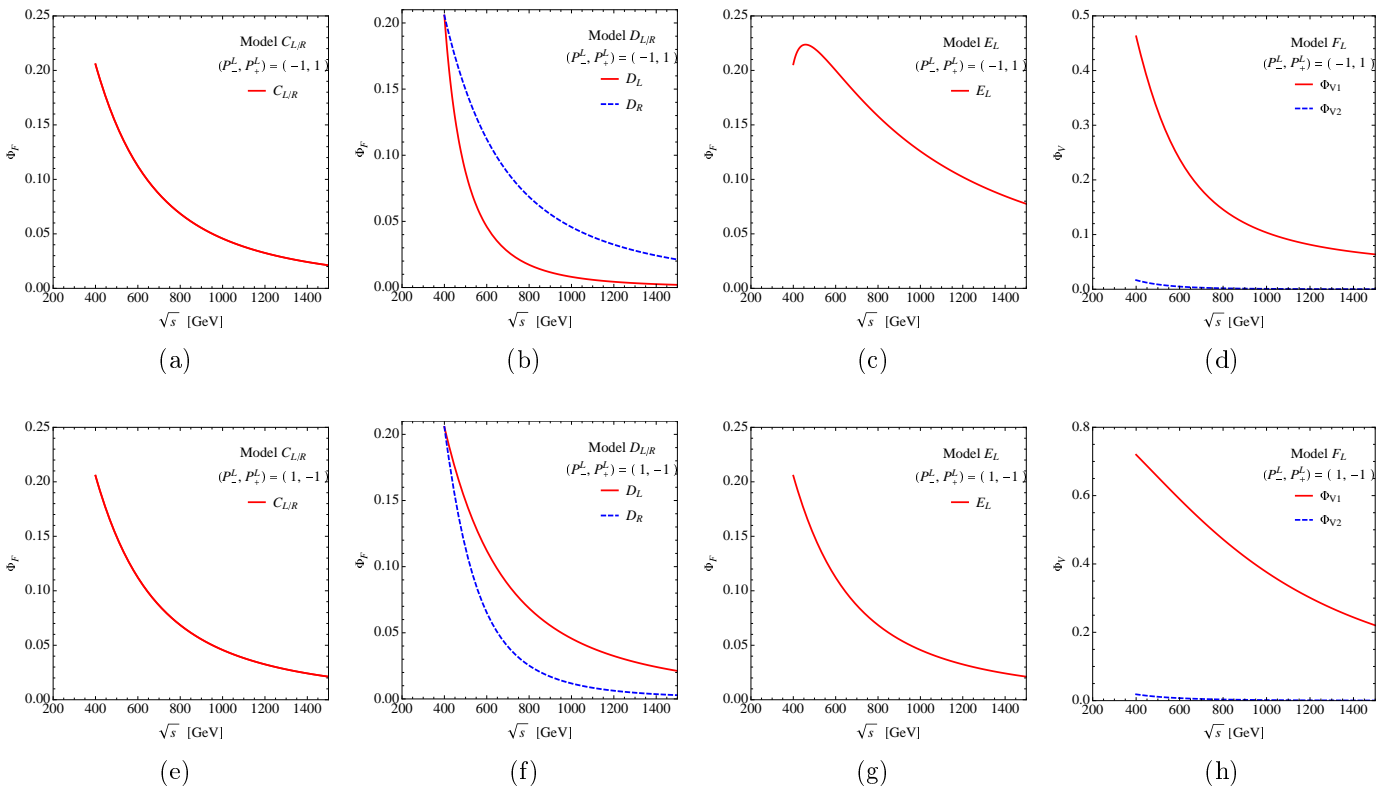}
\caption{\label{fig:decay_azimuthal_angle_correlation_coefficients}
        {Energy dependence of the coefficients $\Phi_{F}$ and $\Phi_{V1,V2}$ for
             correlated decay azimuthal-angle distributions for a spin-1/2 charged first
             KK-muon pair $\mu^\pm_{1L/1R}$ ($C_{L/R}$) and first KK-electron pair
             $e^\pm_{1L/1R}$  ($D_{L/R}$), for a spin-1/2 charged wino pair $\tilde{W}^\pm$
             ($E_L$) and for a spin-1 charged first KK-$W$ $W^\pm_1$ ($F_L$).
             The lines in the upper (lower) frames are for left-handed (right-handed)
             electron and right-handed (left-handed) positron beams, respectively.}
        }
\end{figure}

Conceptually, any azimuthal-angle correlation, which is a pure quantum-mechanical effect,
requires non-trivial interference among helicity amplitudes with different
helicity assignments as indicated by
Eqs.$\,$(\ref{eq:azimuthal_angle_correlation_coefficient_f}),
(\ref{eq:azimuthal_angle_correlation_coefficient_v1}) and
(\ref{eq:azimuthal_angle_correlation_coefficient_v2}) and so they tend to diminish
as the c.m. energy increases, as  demonstrated numerically in
Figs.$\,$\ref{fig:decay_azimuthal_angle_correlation_coefficients}.
\begin{itemize}
\item Numerically, the quantity $\Phi_F$ takes a value roughly between 0.1 and
      0.2 at $\sqrt{s}=500\, {\rm GeV}$ and sensitive to initial
      beam polarization for the processes with chiral $t$-channel contributions ($D_{L/R}$
      and $E_L$) as shown in
      Figs.$\,$\ref{fig:decay_azimuthal_angle_correlation_coefficients}(b), (c), (f) and (g),
      while it is independent
      of beam polarization in the production for a charged first KK muon pair
      $\mu^\pm_{1L/1R}$ ($C_{L/R}$) with no $t$-channel contributions as shown
      in Figs.$\,$\ref{fig:decay_azimuthal_angle_correlation_coefficients}(a) and (d).
\item One noteworthy aspect in the spin-1 case ($F_L$) is that the quantity $\Phi_{V2}$
      is too small (less than 2\,\%) in magnitude to distinguish the spin-1 $W^\pm_1$
      state from the spin-1/2 states, $\mu^\pm_1,\, e^\pm_1$ or $\tilde{W}^\pm$
      as shown in Figs.$\,$\ref{fig:decay_azimuthal_angle_correlation_coefficients}(d)
      and (h).
      This strong suppression in the spin-1 $W^\pm_1$ case is due to the cancellation
      of the corresponding production helicity amplitudes ($\sim M^2_{W^\pm_1}/E^2_{\rm cm}$)
      that is forced by the relations satisfied for saving the tree-level unitarity.
\item On the other hand, the coefficient $\Phi_{V1}$ in the $\cos\phi$ term is sufficiently
      large so that this correlation can be exploited for distinguishing the spin-1 case
      at least from the spin-0 case as indicated by the solid lines
      in Figs.$\,$\ref{fig:decay_azimuthal_angle_correlation_coefficients}(d) and (h).
\end{itemize}
{\it To summarize.} The fully-correlated azimuthal-angle correlations encoding quantum
interference between different helicity final states can provide a supplementary but not
complete method for spin measurements.\s

\begin{figure}[htb]
\centering
\includegraphics[width=18.cm]{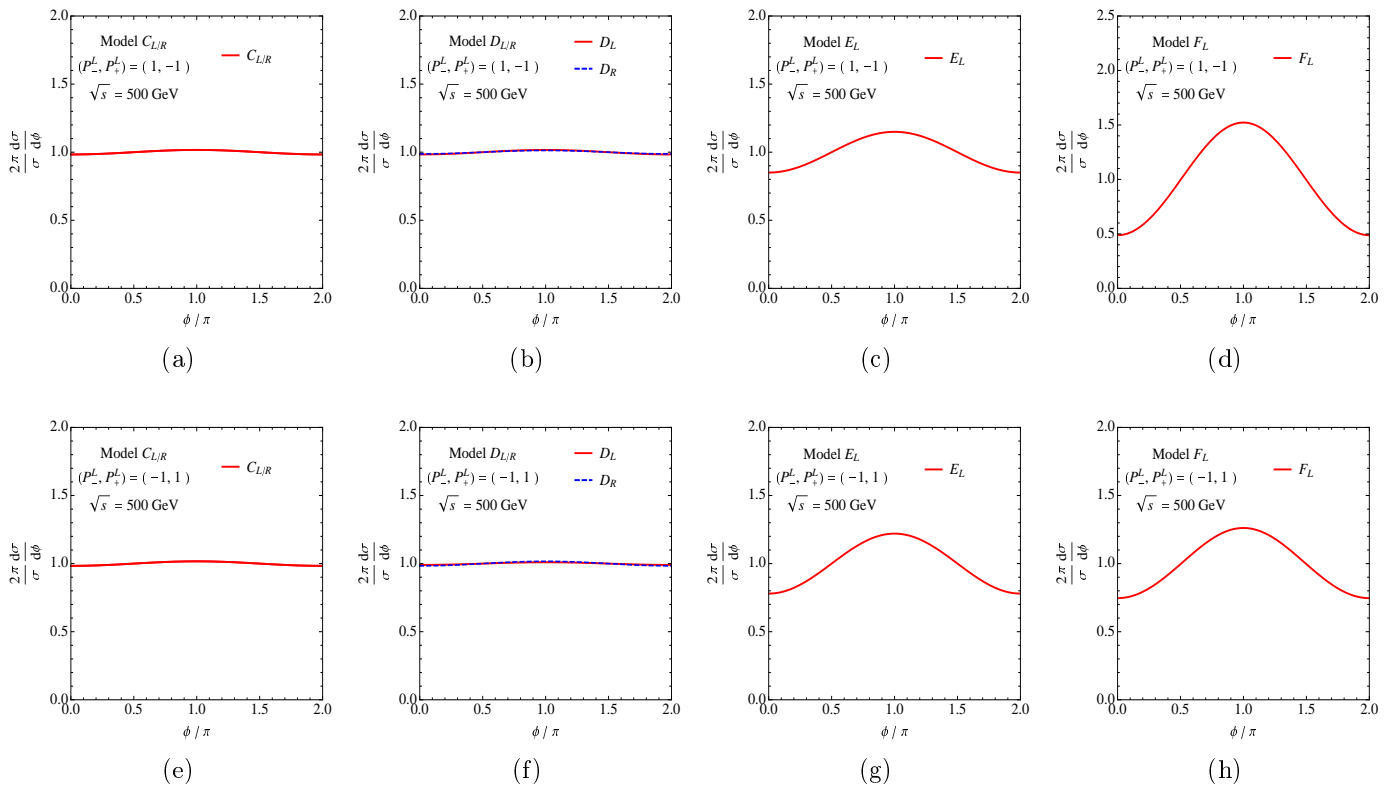}
\caption{\label{fig:decay_azimuthal_angle_correlations}
        {Correlated azimuthal-angle distributions for a spin-1/2 charged first KK-muon
             and KK-electron ($C_{L/R}$ and $D_{L/R}$), for a spin-1/2 charged wino ($E_L$)
             and for a spin-1 charged first KK-$W$ ($F_L$). The upper (lower) frames are
             for left-handed (right-handed) electron and right-handed (left-handed)
             positron beams.}
        }
\end{figure}

Based on the mass spectrum (\ref{eq:mass_spectrum}) leading to the dilution factor
$\eta_{\rm fv}=1/3$ (obtained by substituting these masses into Eq.\,(\ref{eq:eta_fv})) while $\eta_{\rm fs}=1$, we show in
Figs.$\,$\ref{fig:decay_azimuthal_angle_correlations} the fully-correlated azimuthal-angle
distributions for a spin-1/2 charged first KK muon and electron ($C_{L/R}$ and $D_{L/R}$),
for a spin-1/2 charged wino ($E_L$) and for a spin-1 charged first KK $W$-boson ($F_L$).
The plots in the upper (lower) frames are for left-handed (right-handed) electron and
right-handed (left-handed) positron beams, respectively.
\begin{itemize}
\item In the first KK muon and electron cases ($C_{L/R}$ and $D_{L/R}$), the azimuthal-angle
      correlations are too small to be distinguished from the flat distribution in the
      spin-0 case as shown in Figs.$\,$\ref{fig:decay_azimuthal_angle_correlations}(a), (b),
      (e) and (f). This tiny correlation is owing to
      the fact that we have a small coefficient $\Phi_F$ but also a small dilution factor
      $\eta^2_{\rm fv}=1/9$, which can be much larger for a small mass ratio of the parent and
      daughter particles.
\item In contrast the spin-1/2 charged wino case ($E_L$)  shows a rather distinct
      azimuthal-angle correlation as the dilution factor $\eta_{\rm fs}=1$ independently
      of particle masses, as shown in Figs.$\,$\ref{fig:decay_azimuthal_angle_correlations}(c)
      and (g).
\item In the spin-1 case the dilution factor is $\eta_{\rm vf}=8/9$ and the coefficient
      $\Phi_{V1}$ in the $\cos\phi$ term is between 0.4 and 0.6 in magnitude while
      the coefficient $\Phi_{V2}$ is extremely tiny. As a consequence, the azimuthal-angle
      correlation exhibits a distinct distribution proportional to $1+\alpha_1\cos\phi
      +\alpha_2 \cos 2\phi$ with $\alpha_1\sim 0.25$ and $\alpha_2 \sim 0$, as shown in
      Figs.$\,$\ref{fig:decay_azimuthal_angle_correlations}(d) and (h).
\end{itemize}
{\it To summarize.} We have shown for the mass spectrum (\ref{eq:mass_spectrum}) that the
spin-1/2 KK muons and electrons cannot be so easily distinguished from the spin-0 smuons and
selectrons through the azimuthal-angle correlation. In contrast, the spin-1/2 charged wino
case and the spin-1 KK $W$-boson case can be distinguished from the spin-0 cases.
However, it turned out to be difficult to establish the spin-1 nature of the KK $W$-boson
due to the strong suppression of the $\cos 2\phi$ mode, requiring other methods such as
the decay polar-angle distributions.\s

\subsection{Effects of ISR, beamstrahlung, particle widths and kinematic cuts}
\label{subsec:effects_isr_beamstrahlung_cuts}

In this subsection for a more realistic investigation we study the impact on the various
kinematic observables by initial state radiation (ISR) \cite{Skrzypek:1990qs},
beamstrahlung \cite{Chen:1991wd} and finite width of the particle $\mathcal{P}^\pm$
as well as typical kinematic cuts in an $e^+e^-$ collider
environment. We use FeynRules \cite{Christensen:2008py,Christensen:2009jx,Alloul:2013bka}
to implement all the vertices and propagators into the format of
CalcHEP \cite{Belyaev:2012qa}.
Then, we perform extensive simulations for the spin and chirality assignments
listed in Tab.$\,$\ref{tab:examples_antler_topology_process}.\s

The kinematic cuts taken in the present numerical analysis are
\begin{eqnarray}
|\cos\theta_{\ell}|\, < \, 0.9962 \quad \mbox{and}\quad
E_\ell \, > \, 10\, {\rm GeV}
\label{eq:kinematical_cut_1}
\end{eqnarray}
to ensure detection, where $\theta_\ell$ and $E_\ell$ are the polar angle and the energy
of the lepton in the laboratory frame, and
\begin{eqnarray}
\not\!{p}_T \, >\, 10\, {\rm GeV}
\label{eq:kinematical_cut_2}
\end{eqnarray}
to remove the background from $e^+e^-\to e^+e^- \ell^+\ell^-$ where the final $e^+e^-$
pair is missed. The $\mathcal{P}$ total width $\Gamma_{\mathcal{P}}$ is calculated to be
the sum of the partial widths of the two decays $\mathcal{P}^-\to \ell^-\mathcal{\bar{D}}^0$
with $\ell=e$ and/or $\mu$, for a simple analysis of the impact of the width.
ISR and Beamstrahlung effects at ILC are calculated with CalcHEP using parameters in
Tab.$\,$\ref{tab:ilc_parameters} \cite{Phinney:2007gp}.

\begin{table}[hbt]
\begin{center}
\mbox{ } \\[2mm]
\begin{tabular}{|c|c|c|c|c|c|}
\hline
Collider & $E_{\rm cm}$ ($\sqrt{s}$) [GeV]  &
           $N$ [$10^{10}$] &
           $\sigma_x$ [nm] &
           $\sigma_y$ [nm] &
           $\sigma_z$ [$\mu$m] \\
\hline\hline
ILC & 500 &
      2   & 640 &
      5.7 & 300 \\
\hline
\end{tabular}
\end{center}
\caption{\label{tab:ilc_parameters}
        {Key parameters of the initial beams at the ILC used in our numerical
             analysis. Here $N$ is the number of particles per bunch, $\sigma_{x,y}$
             are the RMS beam sizes at the interaction point, and $\sigma_z$ is the
             RMS bunch length.}
        }
\end{table}
\begin{figure}[htb]
\centering
\includegraphics[width=16.cm]{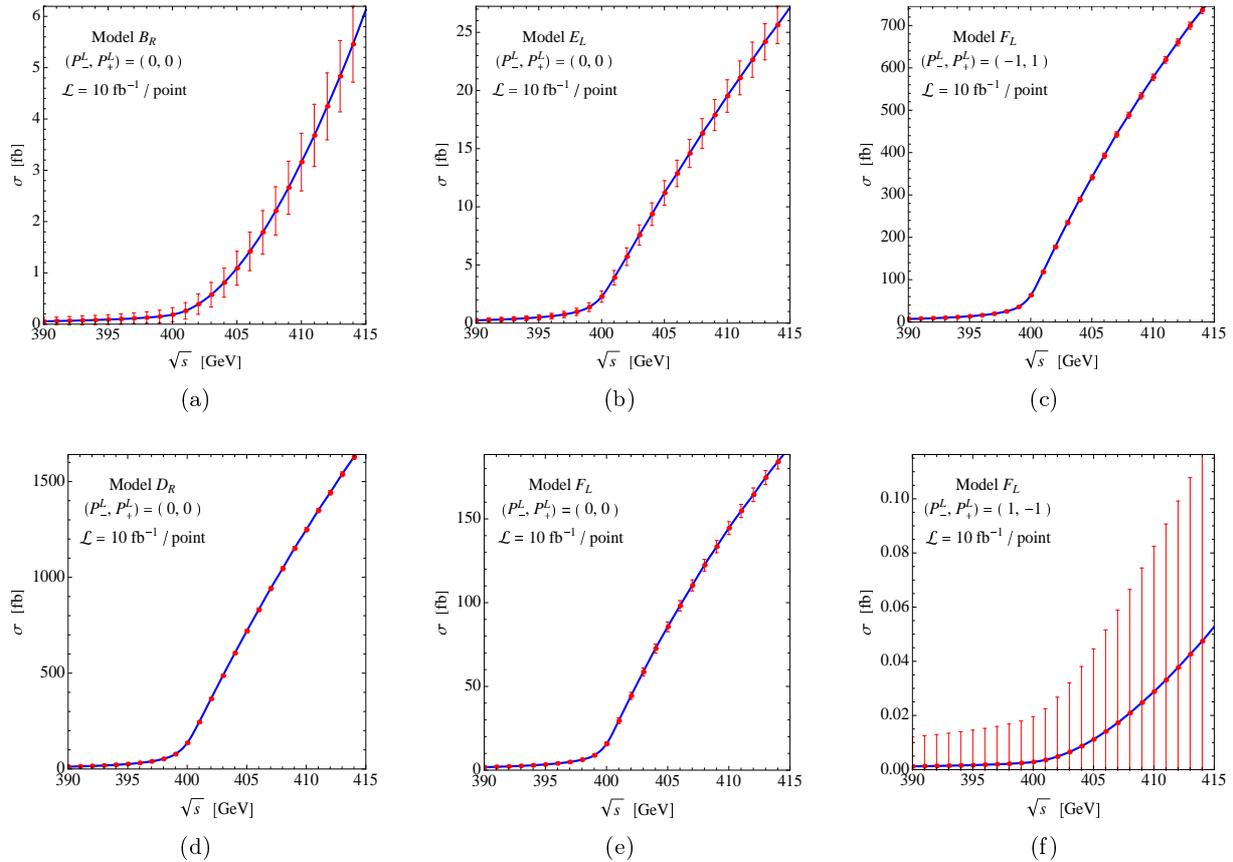}
\caption{\label{fig:threshold_excitation_MC}
        {Excitation curve of the cross section close to threshold for the production
             of a spin-0 charged $R$-selectron pair $\tilde{e}^\pm_R$ ($B_R$), a spin-1/2
             charged $R$-type first KK-electron pair $\tilde{e}^\pm_R$ ($D_R$), a spin-1/2
             charged wino pair $\tilde{W}^\pm$ ($E_L$) or a spin-1 charged first KK-$W$
             pair $W^\pm_1$ ($F_L$)  close to threshold including ISR, beamstrahlung and
             width effects as well as the kinematic cuts in
             Eqs.$\,$(\ref{eq:kinematical_cut_1}) and (\ref{eq:kinematical_cut_2});
             the statistical errors correspond to $\mathcal{L}=10\,{\rm fb}^{-1}$ per point.
             The initial beams are set to be unpolarized in the frames, (a), (b), (d) and (e),
             while the electron (positron) beams are purely left-handed (right-handed) and
             purely right-handed (left-handed) in the frames, (c) and (f).}
        }
\end{figure}

Rather than listing all the scenarios studied in the previous subsections,
we present our simulation results for only a few typical scenarios selected for
each observable. Figs.$\,$\ref{fig:threshold_excitation_MC} shows the excitation curve
of the production  of a spin-0 charged $R$-selectron pair $\tilde{e}^\pm_R$ ($B_R$),
a spin-1/2 charged $R$-type first KK-electron pair $\tilde{e}^\pm_R$ ($D_R$),
a spin-1/2 charged wino pair $\tilde{W}^\pm$ ($E_L$) or a spin-1 charged first KK $W$-boson
pair $W^\pm_1$ ($F_L$) close to threshold after the ISR, beamstrahlung and width effects as
well as the kinematic cuts in Eqs.$\,$(\ref{eq:kinematical_cut_1}) and
(\ref{eq:kinematical_cut_2}) are included. The statistical errors correspond to
$\mathcal{L}=10\,{\rm fb}^{-1}$ per point.
Except for Figs.$\,$\ref{fig:threshold_excitation_MC}(c) and (f),
the initial electron and positron beams are taken to be unpolarized. The plot of the
upper (lower) right frame is for left-handed (right-handed) electron and right-handed
(left-handed) positron beams.
\begin{itemize}
\item The production cross section can take a finite value even below threshold as
      the particle can be produced virtually with a mass smaller than its on-shell mass
      due to its non-zero width, as indicated by the tail extended toward the lower
      energy region in each frame.  Adjusting the width effect, it is evident that for
      unpolarized beams the spin-0 scalar production process ($B_R$) exhibits a slow
      $P$-wave excitation shown in Figs.$\,$\ref{fig:threshold_excitation_MC}(a) while
      the spin-1/2 fermion production processes ($D_R$ and
      $E_L$) and the spin-1 vector-boson production process ($F_L$) show a sharp
      $S$-wave excitation, as in Figs.$\,$\ref{fig:threshold_excitation_MC}(b), (c), (d)
      and (e).
\item In the spin-1 case, if the $t$-channel contribution is killed by complete right-handed
      electron and left-handed positron polarizations, the cross section rises in slow $P$
      waves near threshold in Figs.$\,$\ref{fig:threshold_excitation_MC}(f).
      Nevertheless, the number of events is very small so that it is
      expected to be quite difficult to confirm this $P$-wave pattern quantitatively.
\item The threshold behavior is not affected so much by ISR and beamstrahlung effects.
\end{itemize}
{\it To recapitulate.} The spin-0 case can be clearly distinguished
from the spin-1/2 and  spin-1 cases in the specific scenarios through the threshold scan
method, although a new method  is required for distinguishing
the spin-1 case from the spin-1/2 cases and even from the spin-0 case in the
general case, as emphasized before. \s

\begin{figure}[htb]
\centering
\includegraphics[width=18.cm]{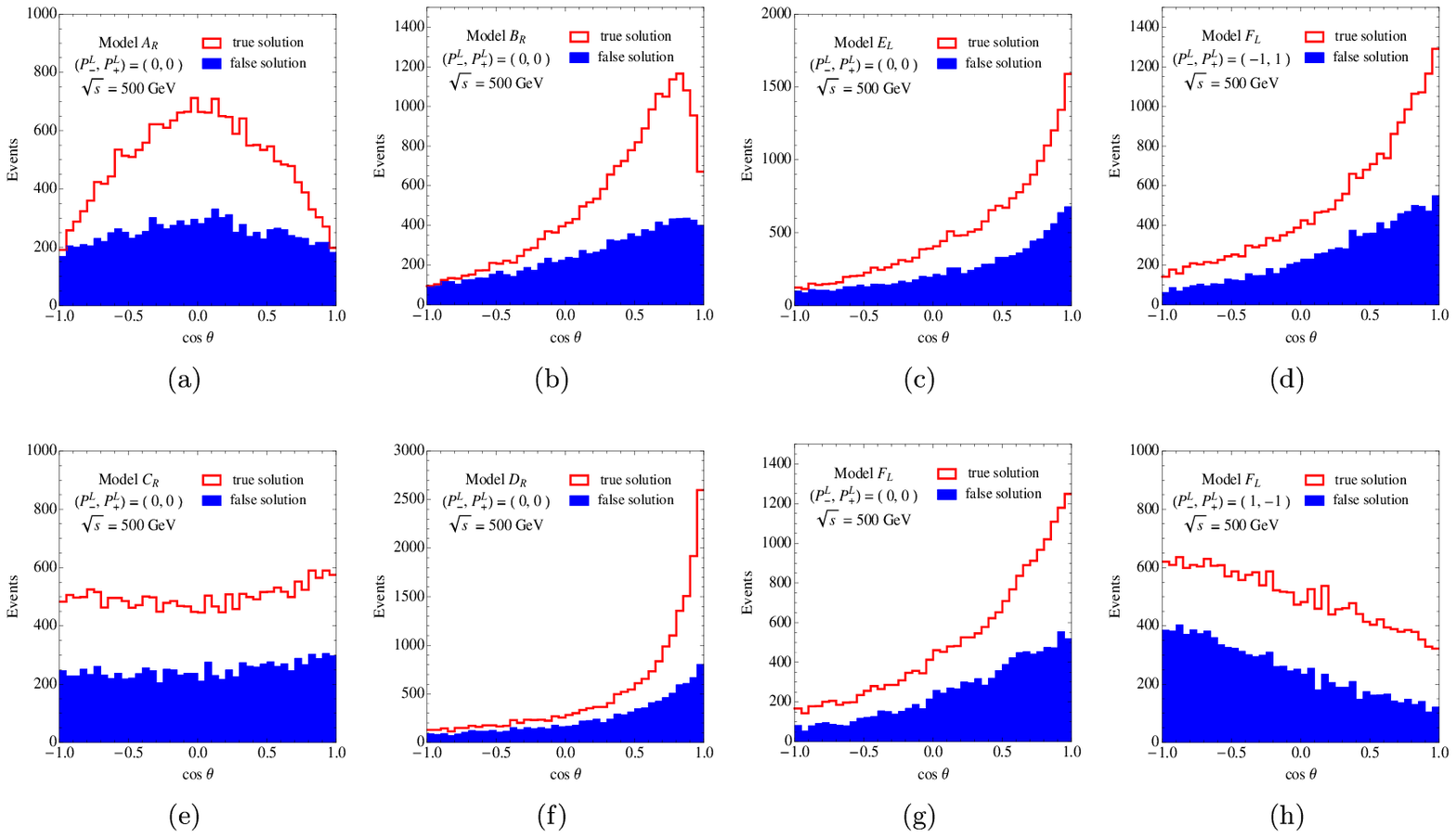}
\caption{\label{fig:production_angle_distributions_MC}
        {The polar-angle distributions with the contribution of false solution
             for the production of a spin-0 charged $R$-type smuon/selectron pair
             $\tilde{e}^\pm_R/\tilde{e}^\pm_R$ ($A_R/B_R$), a spin-1/2 charged $R$-type
             first KK-muon/KK-electron pair $\mu^\pm_1/e^\pm_1$ ($C_R/D_R$), a spin-1/2
             charged wino pair $\tilde{W}^\pm$ ($E_L$) or a spin-1 charged first KK-$W$
             pair $W^\pm_1$ ($F_L$), including ISR, beamstrahlung and width effects.
             Except for the frames, (d) and (e), the initial electron and positron beams
             are set to be unpolarized. The frames, (d) and (h),
             are for left-handed (right-handed) and right-handed (left-handed)
             electron (positron) beams, respectively. The simulation for the polar-angle
             distribution is based on a fixed number of events $N_{\rm ev}=10^4$ at
             the c.m. energy of 500 GeV.}
        }
\end{figure}

As shown before, there exists a two-fold discrete ambiguity in determining the
$\mathcal{P}^\pm$ momentum in the antler-topology event. Therefore, we show
in Figs.$\,$\ref{fig:production_angle_distributions_MC} the polar-angle distributions
with the contribution of false solution included for the production of a spin-0 charged
$R$-type smuon/selectron pair $\tilde{e}^\pm_R/\tilde{e}^\pm_R$ ($A_R/B_R$),
a spin-1/2 charged $R$-type first KK-muon/KK-electron pair $\mu^\pm_1/e^\pm_1$ ($C_R/D_R$),
a spin-1/2 charged wino pair $\tilde{W}^\pm$ ($E_L$) or a spin-1 charged first KK
$W$-boson pair $W^\pm_1$ ($F_L$), including ISR, beamstrahlung and width effects
as well as the kinematic cuts in Eqs.$\,$(\ref{eq:kinematical_cut_1}) and
(\ref{eq:kinematical_cut_2}). Except for
Figs.$\,$\ref{fig:production_angle_distributions_MC}(d) and (h), the initial electron
and positron beams are assumed to be unpolarized. The plot of the upper (lower) right-most
frame is for left-handed (right-handed) electron and right-handed (left-handed) positron
beams. For the simulation we simply take a fixed number of events $N_{\rm ev}=10^4$
at the c.m. energy of 500 GeV.
\begin{itemize}
\item The $\sin^2\theta$ law for the production of spin-0 particles (for $R$-type smuons
      ($A_R$) and $R$-type selectrons ($B_R$) close to threshold) is a unique signal for
      the spin-0 character. This feature can be confirmed in
      Figs.$\,$\ref{fig:production_angle_distributions_MC}(a) and (b)
      after the false distribution following the true distribution with a little dilution
      are extracted out from the sum of the true and false solutions.
\item However, the polar-angle distributions in the spin-1/2 and spin-1 cases
      have so much more involved patterns that it is not straightforward to
      distinguish the spin-1 case from the spin-1/2 case, unless beam polarizations
      are exploited.
\item In the spin-1 case ($F_L$) the polar-angle distribution is quite different for
      each combination of the electron and positron longitudinal polarizations as
      shown in Figs.$\,$\ref{fig:production_angle_distributions_MC}(d) and (h).
      In particular, the true polar-angle distribution in
      Figs.$\,$\ref{fig:production_angle_distributions_MC}(h)
      with right-handed electron and left-handed positron polarizations is characteristically
      different from that for the spin-1/2 first KK muon case ($C_R$)
      shown in Figs.$\,$\ref{fig:production_angle_distributions_MC}(e).
\end{itemize}
We note that the curve of the false solution is sensitively related not only to the
curve of the true solution but also to the chiral structure of the decay processes
as clearly shown by the shaded area in
Figs.$\,$\ref{fig:production_angle_distributions_MC}(h). This is
because the direction of the false solution depends not only on the the direction of the
true solution but also on the flight directions of the two leptons, whose distributions are
strongly chirality-dependent. Numerically we have confirmed that the curve with a
negative slope is due to the pure left-chiral $e W_1 \nu_1$ coupling involved in the decay
$W^-_1\to \ell^- \bar{\nu}_\ell$.
As in the excitation curves, the polar-angle distributions turn out to be not much
distorted by the ISR and beamstrahlung effects.\s

\begin{figure}[htb]
\centering
\includegraphics[width=18.cm]{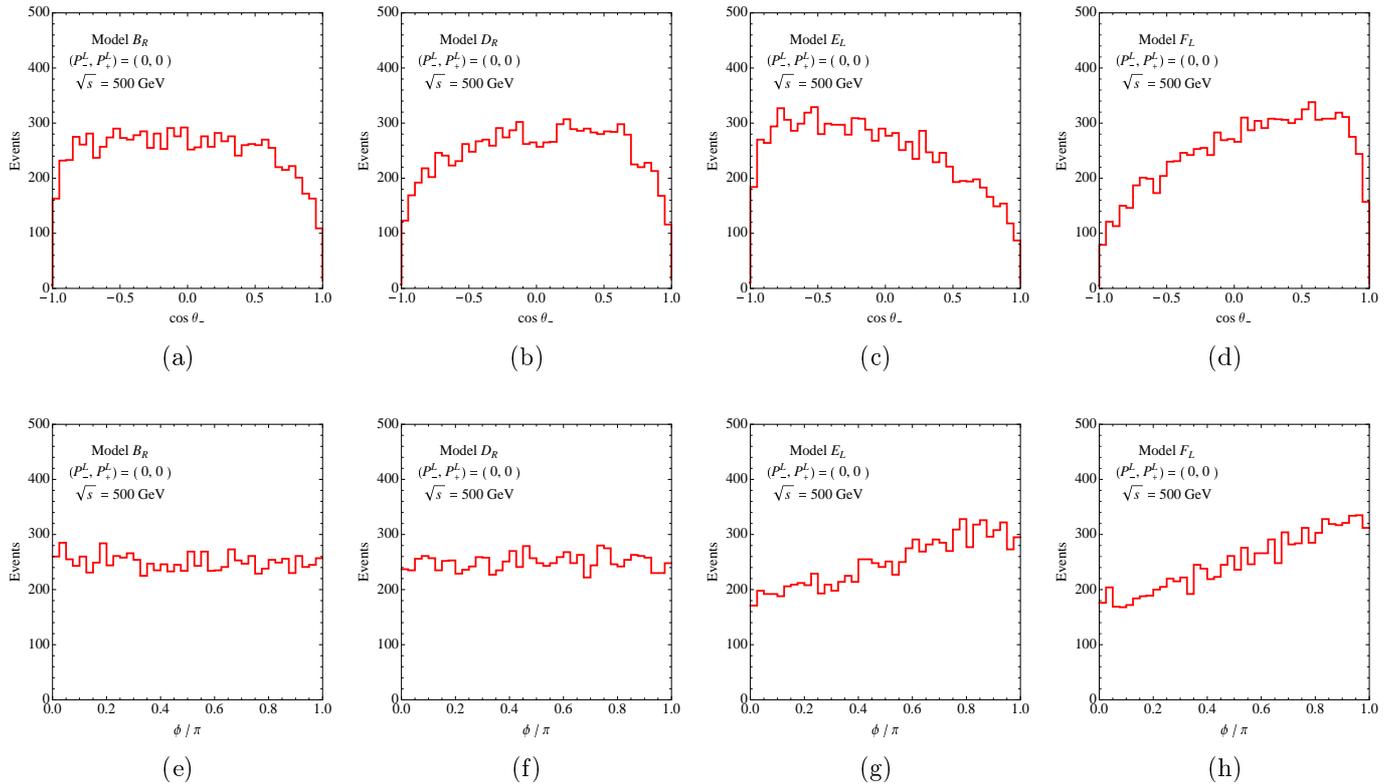}
\caption{\label{fig:decay_angle_correlations_MC}
        {The upper frames are for the single decay polar-angle distribution for the
             combination of the production of a charged pair $\mathcal{P}^-\mathcal{P}^-$
             and the sequential decay of the negatively-charged particle $\mathcal{P}^-\to
             \ell^-\mathcal{\bar{D}}^0$ and the lower frames are for the full azimuthal-angle
             correlations in the antler-topology process. For these distributions,
             we consider the production of a spin-0 charged $R$-type selectron pair
             $\tilde{e}^\pm_R$ ($B_R$), a spin-1/2 charged $R$-type first KK-electron pair
             $e^\pm_{1R}$ ($D_R$), a spin-1/2 charged wino pair $\tilde{W}^\pm$ ($E_L$) or
             a spin-1 charged first KK-$W$ pair $W^\pm_1$ ($F_L$), including ISR, beamstrahlung
             and width effects as well as the kinematic cuts in
             Eqs.$\,$(\ref{eq:kinematical_cut_1}) and (\ref{eq:kinematical_cut_2}).
             The initial electron and positron beams are assumed to be unpolarized.
             The simulation for each of the decay polar-angle distributions and azimuthal-angle
             correlations is based on a fixed number of events $N_{\rm ev}=10^4$
             at the c.m. energy of $\sqrt{s}=500\, {\rm GeV}$.}
        }
\end{figure}

Then, let us consider the single decay polar-angle distributions in the
combined process of the $\mathcal{P}^\pm$ pair production followed by the decay of a
negatively-charged particle $\mathcal{P}^- \to \ell^-\mathcal{\bar{D}}^0$ and the decay
azimuthal-angle correlations in the antler-topology process in the four scenarios, $B_R$
for a spin-0 charged $R$-type selectron pair $\tilde{e}^\pm_R$, $D_R$ for a spin-1/2
charged $R$-type first KK-electron pair $e^\pm_{1R}$, $E_L$ for a spin-1/2 chargino wino
pair $\tilde{W}^\pm$ and $F_L$ for a spin-1 first KK $W$-boson pair $W^\pm_1$. The initial
electron and positron beams are taken to be unpolarized.\s

The upper frames of Figs.$\,$\ref{fig:decay_angle_correlations_MC} show the single
decay polar-angle distributions after including the ISR, beamstrahlung and width
effects as well as the kinematic cuts in Eqs.$\,$(\ref{eq:kinematical_cut_1})
and (\ref{eq:kinematical_cut_2}) in the four different scenarios ($B_R$, $D_R$, $E_L$
and $F_L$). For the hypothetical case that no QED radiation degrades the nominal
production energy, since the ratios of the right-handed cross section with $(P^L_-, P^L_+)=(1,-1)$ over the left-handed one with $(P^L_-, P^L_+)=(-1,+1)$ for scenario $B_R$, $D_R$, $E_L$ and $F_L$ are $27.4$, $99.8$, $4.51\times 10^{-4}$ and $2.89\times 10^{-4}$ respectively, the distribution should be flat in the scenario $B_R$, linear with
a positive slope in the scenario $D_R$, linear with a negative slope in the scenario
$E_L$ and quadratic in a downward curved shape in the scenario $F_L$ for unpolarized beams, as worked out
from the plots in Figs.$\,$\ref{fig:decay_polar-angle_distributions}. However, in the
more realistic situation that ISR and beamstrahlung decrease the $\mathcal{P}^+\mathcal{P}^-$
production energy, the angular distribution is no longer linear in the spin-0 and
spin-1/2 cases and no longer curved downward in the spin-1 case, as shown in the
upper frames of Figs.$\,$\ref{fig:decay_angle_correlations_MC}. Considerable depletions
are observed at $\cos\theta_-\to \pm 1$ when the constraint on $E_{\mathcal{P}^\pm}$
tends to be most largely invalidated. However, we note that, since both the
ISR spectrum \cite{Skrzypek:1990qs} and the beamstrahlung spectrum \cite{Chen:1991wd}
can be calculated theoretically and measured precisely, the ISR and beamstrahlung effects
can be unfolded from the data, for instance, by applying a bin-by-bin correction
or a matrix inversion procedure, although we will not perform the unfolding
procedure in the present, rather simple numerical analysis. \s

The lower frames of Figs.$\,$\ref{fig:decay_angle_correlations_MC} show the full
azimuthal-angle correlations in the same set of four scenarios as in the single
decay polar-angle distributions. Firstly, it turns out that the ISR, beamstrahlung and
width as well as the kinematic cuts do not spoil the azimuthal-angle correlation
patterns. The distribution is indeed flat in the angle $\phi$ in the scenario $B_R$
for a spin-0 selectron pair and nearly flat in the scenario $D_R$ as expected from
Figs.$\,$\ref{fig:decay_angle_correlations_MC} (e) and (f).
The curves in the spin-1/2 charged wino case ($E_L$) and the spin-1 first KK $W$-boson case
($F_L$) are consistent with the simple $\cos\phi$ distribution, as shown in
Figs.$\,$\ref{fig:decay_angle_correlations_MC}(g) and (h). (We fold the range
of the angle $\phi$ into $[0, \pi]$ with respect to the line with $\phi=\pi$ for the
numerical simulation, as $\phi$ is calculated from $\cos \phi$ which can be determined
unambiguously.)  We note once more that the coefficient of the $\cos 2\phi$ mode is
strongly suppressed due to the relations for restoring tree-level unitarity. \s

\subsection{Influence from ECV interactions}
\label{subsec:influence_ecv_interactions}

Every numerical analysis performed so far is based on the assumption that electron
chirality is preserved (to very good approximation). Although the approximation is
valid in the SM with its EW gauge symmetry spontaneously breaking through the
BEH mechanism, it may be invalid in the models with additional scalar bosons and
with mass generation by different mechanisms. Evidently the ECV terms are already
constrained by various low-energy physical quantities. However, the complete
analysis of those constraints is very much involved and beyond the scope of the
present work. Therefore, before closing this lengthy section, we comment briefly
on the possible influence of the ECV terms on the spin determination rather
qualitatively, if they are not so strongly suppressed.\s

As the analytic expressions of production and two-body helicity amplitudes listed
in Sects.$\,$\ref{sec:production} and \ref{sec:two_body_decays} indicate, the ECV
effects in the production process $e^+e^-\to\mathcal{P}^+\mathcal{P}^-$ are generated
when a spin-0 scalar $S^0_s$ couples both to the $e^+e^-$ pair and also when the
$\mathcal{P}^+\mathcal{P}^-$ pair and/or the $t/u$-channel trilinear vertices
involving an electron are non-chiral, i.e. have both left-handed and right-handed
couplings while the ECV effects in the decay processes are generated by non-chiral
decay trilinear $\mathcal{P}\ell\mathcal{D}$ vertices.\s

In the ECV case, the electron and positron helicities are identical, $\sigma_-=\sigma_+$
and they form a $J_0=0$ initial two-body state. Therefore, the production polar-angle
distribution is {\rm isotropic at threshold} and the threshold excitation is in
a sharp $S$ wave except when a spin-1/2 $F^+_pF^-_p$ pair is produced only through
an $s$-channel pseudoscalar exchange ($s^{FF}_{S+}=- s^{FF}_{S-}$).
As a result, the characteristic patterns of threshold excitation and polar-angle distribution
in the ECC case can be spoiled in the presence of the ECV contributions.\s

If the decay vertices are non-chiral, the relative chirality is not maximal in
magnitude any more, i.e. $|\xi| < 1$, reducing the production-decay correlation
even for the spin-1/2 and spin-1 cases. In the extreme situation with zero
relative chirality, there exist no non-trivial decay angular distributions
for determining whether the spin of the parent particle $\mathcal{P}$ is 0 or
1/2. In contrast to the spin-1/2 case, there exists a decay-angle distribution
independent of the relative chirality in the spin-1 case. Even though the
effectiveness of this distribution is reduced by a dilution factor, the spin-1
case can be distinguished from the spin-0 and spin-1/2 cases through
the chirality-independent decay-angle correlations \cite{Christensen:2013sea}.\s

\section{Summary and conclusions}
\label{sec:summary_conclusion}

In this paper, we have made a systematic study of kinematic observables connected with
the antler-topology process $e^+e^-\to\mathcal{P}^+\mathcal{P}^-\to\ell^+\ell^-\mathcal{D}^0
\mathcal{\bar{D}}^0$ which could serve as model-independent tests for determining the spins
of the charged particles $\mathcal{P}^\pm$ and the invisible neutral particles
$\mathcal{D}^0$ and $\mathcal{\bar{D}}^0$ as well as the intermediate virtual particles
participating in the production process.\s

It is evident from our analysis that the model-independent determination of the spin quantum
numbers of new particles is a complex task even at $e^+e^-$ colliders with clean and fixed
initial-state environments and beam polarizations. The degree of complexity depend crucially
on the chiral structures as well as kinematic features of the particles. Not only threshold
excitation and angular distributions controlled through initial beam polarizations in pair
production but also angular correlations in particle decays provide powerful methods for
experimental spin measurements.\s

The predictions for the threshold excitation and the polar-angle distributions in the
production processes, separated into the ECC and ECV parts when the electron is assumed to
be massless, are summarized in Tab.$\,$\ref{tab:summary_production}. \s
\begin{table}[hbt]
\begin{center}
\begin{tabular}{|c|c|c|c|c|}
\hline
\ \ $e^+e^-$ chiralities             \ \  &
\ \ Spin $J_{\mathcal{P}}$           \ \  &
\ \  $t$ or $u$ contributions        \ \  &
\ \ Threshold excitation             \ \  &
\ \ Production polar-angle distribution         \ \        \\
\hline\hline
\ \                                  \ \  &
\ \      \raisebox{-1.5ex}{$0$}                         \ \  &
\ \      N                           \ \  &
\ \     $\beta^3$                    \ \  &
\ \     $\sin^2\theta$               \ \        \\[-1mm]
\ \                                  \ \  &
\ \                                  \ \  &
\ \     Y                            \ \  &
\ \     $\beta^3$                    \ \  &
\ \     $\sin^2\theta$ at threshold         \ \        \\
\cline{2-5}
\ \      chirality                   \ \  &
\ \      \raisebox{-1.5ex}{$1/2$}                       \ \  &
\ \      N                           \ \  &
\ \     $\beta$                      \ \  &
\ \     $1+\kappa_{1/2}\cos^2\theta$     \ \        \\[-1mm]
\ \     conserving                   \ \  &
\ \                                  \ \  &
\ \      Y                           \ \  &
\ \     $\beta$                      \ \  &
\ \     isotropic at threshold              \ \        \\
\cline{2-5}
\ \                                  \ \  &
\ \     \raisebox{-1.5ex}{$1$}                          \ \  &
\ \      N                           \ \  &
\ \     $\beta^3$                    \ \  &
\ \   $1-\kappa_1 \cos^2\theta$  \ \        \\[-1mm]
\ \                                  \ \  &
\ \                                  \ \  &
\ \      Y                           \ \  &
\ \     $\beta$                      \ \  &
\ \     isotropic at threshold              \ \        \\
\hline\hline
\ \                                  \ \  &
\ \     \raisebox{-1.5ex}{$0$}                          \ \  &
\ \      N                           \ \  &
\ \     $\beta$                      \ \  &
\ \     isotropic                    \ \        \\[-1mm]
\ \                                  \ \  &
\ \                                  \ \  &
\ \      Y                           \ \  &
\ \     $\beta$                      \ \  &
\ \     isotropic at threshold               \ \        \\
\cline{2-5}
\ \     chirality                    \ \  &
\ \     \raisebox{-1.5ex}{$1/2$}                        \ \  &
\ \      N                           \ \  &
\ \ $\beta^3/\beta\,[\,{\rm S/P}\,]$ \ \  &
\ \     isotropic                    \ \        \\[-1mm]
\ \     violating                    \ \  &
\ \                                  \ \  &
\ \      Y                           \ \  &
\ \     $\beta$                      \ \  &
\ \     isotropic at threshold               \ \        \\
\cline{2-5}
\ \                                  \ \  &
\ \      \raisebox{-1.5ex}{$1$}                         \ \  &
\ \      N                           \ \  &
\ \     $\beta$                      \ \  &
\ \     isotropic                     \ \        \\[-1mm]
\ \                                  \ \  &
\ \                                  \ \  &
\ \      Y                           \ \  &
\ \     $\beta$                      \ \  &
\ \     isotropic at threshold               \ \        \\
\hline
\end{tabular}
\end{center}
\caption{\label{tab:summary_production}
        {The threshold behavior and the polar-angle distribution of the ECC and
             ECV parts of the production process
             $e^+e^-\to \mathcal{P}^+\mathcal{P}^-$, with [S/P] standing
             for pure scalar-type or pseudoscalar-type couplings, respectively}.
             Here, the energy-dependent coefficients, $\kappa_{1/2}$ and
             $\kappa_1$, take 0 and 3/19 at threshold and they
             approach 1 asymptotically at high energies, respectively.
        }
\end{table}

In any theory with conserved chiral symmetry guaranteeing the electron mass to be zero
before EWSB such as MSSM and MUED, the ECV parts are connected with the extremely
tiny electron mass so that their contributions are negligible for high energy processes.
In this ECC case, as shown in the chirality conserving part of
Tab.$\,$\ref{tab:summary_production}, the $\sin^2\theta$ law for the production of a
spin-0 scalar pair (close to threshold) is a unique signal of the spin-0 character.
While the observation of the $\sin^2\theta$ polar-angle distribution is
sufficient for scalar particles, the $P$-wave $\beta^3$ onset of the excitation curve
is necessary but not a sufficient condition for the spin-0 character. Nevertheless,
we have found that combining the two distributions and using initial beam polarizations
to separately diagnose four $e^+e^-$ helicity combinations enable us not only to determine
the $\mathcal{P}$ spin unambiguously but also to get crucial information on the spins
of intermediate particles and the chiral structure of the couplings in the ECC case.\s

If there exist any non-negligible ECV contributions, then the
patterns of both threshold excitation and production angle distribution may be qualitatively
different from those in the ECC case, as shown in the chirality violating part of
Tab.$\,$\ref{tab:summary_production}. However, one can always use
Eq.$\,$(\ref{eq:ecc_part_extraction}) and Eq.$\,$(\ref{eq:p-odd_lr_asymmetry_c})
to extract the ECC part of pair production, as there exists at least a contribution from
an $s$-channel photon, to get spin information from threshold excitation and production angle
distribution of the ECC part. \s

Combining the production and decay processes in the antler-topology process, it is possible
to construct several correlated observables for a spin-1/2 or spin-1 particle pair
$\mathcal{P}^\pm$. Evidently there is no production-decay correlation
for a spin-0 $\mathcal{P}$, which is a characteristic feature for the spin-0 case.
The sensitivities to the $\mathcal{P}$ spin depend strongly on the chiral structure
reflected in the production and decay helicity amplitudes and the degrees of initial
and final beam polarizations. If the couplings for the decays are pure chiral and the
high degree of beam polarizations are available, then the decay polar-angle distributions
are very powerful for determining the $\mathcal{P}$ spin as the relative chiralities
serving as the polarization analysis powers are maximal. The azimuthal-angle
distribution for the difference $\phi$ between the azimuthal angles of two decay planes
also provides a supplementary method for determining the $\mathcal{P}$ spin, although
this quantum-interference effect diminishes as the c.m. energy increases.\s

If the decay vertices are not pure chiral, the sensitivities of the kinematic observables
to the particle spins are reduced. In the extreme cases of $P$-conserving pure vector or
axial vector couplings, we do not have any production-decay correlations in the spin-1/2 case.
Even in this case, there exists a non-trivial $P$-even observable in the $\cos 2\phi$
mode in the spin-1 case. However, we have found that the coefficient determined by
the production process $e^+e^-\to V^+_p V^-_p$ is strongly suppressed when the
specific relations among couplings are satisfied for saving tree-level unitarity at high
energies. As a result, it may be very difficult in those extreme cases to determine
the spins of the particle involved in the antler-topology process.\s

{\it To conclude.} It is a very complex task to determine the spins
of new particles in a model-independent way in a general theory beyond the SM. Nevertheless,
we have found that, if electron chirality invariance is valid to very good approximation,
the spin of the new particles taking part in the antler-topology process can be determined
in a model-independent way through various energy- and angle-dependent observables at
$e^+e^-$ colliders with polarized beams. Any non-chiral contributions, which
are expected to be insignificant as in many popular models beyond the SM,
render the model-independent spin determination more difficult. However we can still use
beam polarizations to extract the chirality-conserving pieces to get useful
information on the spins of new particles based on various approaches described in the
present work. After all, a high energy $e^+e^-$ collider with polarized
beams is a powerful machine for diagnosing not only the spin but also the
chirality structure of new particles, if they are kinematically available. \s

\vskip 0.5cm

\noindent
{\bf Acknowledgements.} The work of SYC was supported in part by Basic Science Research
Program through the National Research Foundation (NRF) funded by the Ministry
of Education, Science and Technology (NRF-2011-0010835) and in part by research funds of
Chonbuk National University in 2013. NDC was supported in part by PITT PACC and the U.S.
Department of Energy under grant No. DE-FG02-95ER40896.

\vskip 0.5cm

\appendix

\section{Feynman rules for interaction vertices}
\label{sec:appendix_a_feynman_rules}

The initial $e^+e^-\mathcal{S}^0_s$ and final $\mathcal{S}^0_s \mathcal{P}^+\mathcal{P}^-$ currents
for the $s$-channel $\mathcal{S}^0_s$ exchange diagram contributing to the process
$e^+e^-\to \mathcal{P}^+\mathcal{P}^-$ with $\mathcal{S}^0_s= S^0_s$ or $V^0_s$ and
$\mathcal{P}^- = S^-_p, F^-_p$ or $V^-_p$ can be parameterized in the following generic form:
\begin{eqnarray}
  J^S_{ee} &\equiv& \langle S^0_s \| e^-(p_-) e^+(p_+) \rangle
= e\, \bar{v}(p_+)\, [ s^{S}_{ee+} P_+ + s^{S}_{ee-} P_- ]\, u(p_-) \\
  J^{V\mu}_{ee} &\equiv& \langle V^0_s \| e^-(p_-) e^+(p_+) \rangle^\mu
= e\, \bar{v}(p_+)\, [\gamma^\mu\, ( s^V_{ee+} P_+
                                        + s^{V}_{ee-} P_- )]\, u(p_-) \\
  J^{SS}_S &\equiv&  \langle S^-_p(q_-) S^+_p(q_+) \|S^0_s \rangle
= 2 e M_{S_p}\, s^{SS}_S  \\
  J^{SS}_{V\mu} &\equiv&  \langle S^-_p(q_-) S^+_p(q_+) \|V^0_s \rangle_\mu
= e\, s^{SS}_V\, (q_- - q_+)_\mu \\
  J^{FF}_S  &\equiv&  \langle F^-_p(q_-) F^+_p(q_+) \| S^0_s \rangle
= e\, \bar{u}(q_-)\, [ s^{FF}_{S+} P_+ + s^{FF}_{S-} P_- ]\, v(q_+) \\
  J^{FF}_{V\mu} &\equiv& \langle F^-_p(q_-) F^+_p(q_+) \| V^0_s \rangle_\mu
= e\, \bar{u}(q_-)\, [\gamma^\mu ( s^{FF}_{V+} P_+
                               + s^{FF}_{V-} P_- )]\, v(q_+) \\
  J^{VV}_S &\equiv& \langle V^-_p(q_-) V^+_p(q_+) \| S^0_s \rangle
= 2 e M_{V_p}\, s^{VV}_S \epsilon^*_-(q_-)\cdot \epsilon^*_+(q_+) \\
  J^{VV}_{V\mu}  &\equiv&  \langle V^-_p(q_-) V^+_p(q_+) \| V^0_s \rangle_\mu
= - e\, s^{VV}_V \left[ (q_- -q_+)^\mu \epsilon^*_-(q_-) \cdot\epsilon^*_+(q_+)\right.\nonumber\\
    &&\left. \hskip 3.8cm + 2q_+\cdot\epsilon^*_-(q_-) \epsilon^{*\mu}_+(q_+)
                           - 2q_-\cdot\epsilon^*_+(q_+) \epsilon^{*\mu}_-(q_- )\right]
\end{eqnarray}
with the chiral projection operators $P_\pm = \tfrac{1}{2} (1 \pm \gamma_5)$. In the last
expression for the triple-vector vertex, the on-shell conditions $q_-\cdot\epsilon^*_-(q_-)
=q_+\cdot\epsilon^*_+(q_+)=0$ are imposed. \s

The $e\mathcal{P}\mathcal{T}$ interaction vertices $T^{e\mathcal{P}}_\mathcal{T}$ for
the $t$-channel neutral $\mathcal{T}^0$-exchange diagrams in the production
process $e^+e^-\, \to\, \mathcal{P}^+\mathcal{P}^-$ with $\mathcal{P}^\pm=S^\pm_p, F^\pm_p$
or $V^\pm_p$ and $\mathcal{T}^0= S^0_t, F^0_t, V^0_t$ can be parameterized as follows:
\begin{eqnarray}
  T^{eS}_F &\equiv&  \langle S^-_p | F^0_t  | e^- \rangle
= e\, (t^{eS}_{F+}\, P_+ + t^{eS}_{F-}\, P_-) \\
  T^{eF}_S &\equiv&  \langle F^-_p | S^0_t | e^- \rangle
= e\, (t^{eF}_{S+}\, P_+ + t^{eF}_{S-}\, P_-) \\
  T^{eF}_{V\mu} &\equiv& \langle F^-_p | V^0_t | e^-\rangle_\mu
= e\, \gamma_\mu\, (t^{eF}_{V+}\, P_+ + t^{eF}_{V-}\, P_-) \\
  T^{eV}_{F\mu} &\equiv& \langle V^-_p | F^0_t | e^-\rangle_\mu
= e\, \gamma_\mu\, (t^{eV}_{F+}\, P_+ + t^{eV}_{F-}\, P_-)
\end{eqnarray}
and the $e\mathcal{P} \mathcal{U}$ interaction vertices $U^{e\mathcal{P}}_\mathcal{U}$
for the $u$-channel doubly-charged\, $\mathcal{U}^{--}$-exchange diagrams can be
parameterized as follows:
\begin{eqnarray}
  U^{eS}_F &\equiv&  \langle S^+_p | F^{--}_u | e^- \rangle
= e\, (u^{eS}_{F+}\, P_+ + u^{eS}_{F-}\, P_-) \\
  U^{eF}_S &\equiv&  \langle F^+_p | S^{--}_t | e^- \rangle
= e\, (u^{eF}_{S+}\, P_+ + u^{eF}_{S-}\, P_-) \\
  U^{eF}_{V\mu} &\equiv& \langle F^+_p | V^{--}_t | e^-\rangle_\mu
= e\, \gamma_\mu\, (u^{eF}_{V+}\, P_+ + u^{eF}_{V-}\, P_-) \\
  U^{eV}_{F\mu} &\equiv& \langle V^+_p | F^{--}_t | e^-\rangle_\mu
= e\, \gamma_\mu\, (u^{eV}_{F+}\, P_+ + u^{eV}_{F-}\, P_-)
\end{eqnarray}
We note that in the present work, negatively-charged (positively-charged) states are
treated as particles (anti-particles), respectively.\s

The amplitudes for the two-body decay $\mathcal{P}^-(q_-)\to\ell^- (p_1)\mathcal{\bar{D}}^0(p_2)$
and its charge-conjugated process with $\mathcal{P}^- = S^-_p , F^-_p $ or $V^-_p$ and
$\mathcal{\bar{D}}^0 = \bar{S}^0_d, \bar{F}^0_d$  or $\bar{V}^0_d$ can be parameterized
in general as follows:
\begin{eqnarray}
 D^{\ell F}_S  &\equiv&  \langle \ell^- \bar{F}^0_d \| S^-_p \rangle
= e\, \bar{u}(p_1) [d^{\ell F}_{S+} P_+ + d^{\ell F}_{S-} P_-]\, v(p_2) \\
 D^{\ell S}_F  &\equiv&  \langle \ell^- \bar{S}^0_d \| F^-_p \rangle
= e\, \bar{u}(p_1) [d^{\ell S}_{F+} P_+ + d^{\ell S}_{F-} P_-]\, u(q_-) \\
 D^{\ell V}_F &\equiv&  \langle \ell^- \bar{V}^0_d \| F^-_p \rangle
= e\, \epsilon^{\mu *} (p_2) \bar{u}(p_1) \gamma_\mu\,
                   [d^{\ell V}_{F+} P_+ + d^{\ell V}_{F-} P_-]\, u(q_-) \\
 D^{\ell F}_V &\equiv& \langle \ell^- \bar{F}^0_d \| V^-_p \rangle
= e\, \bar{u}(p_1) \gamma_\mu\,
                   [d^{\ell F}_{V+} P_+ + d^{\ell F}_{V-} P_-]\, v(p_2)\,
                   \epsilon_{-}^\mu (q_-)
\end{eqnarray}
where $\ell^-$ stands for $e^-$ or $\mu^-$, which are treated as massless particles in
in our phenomenological spin and chirality analysis at high energy $e^+e^-$ colliders.\s

\section{Explicit form of the $d$ functions}
\label{sec:appendix_b_d_functions}

The explicit form of the Wigner $d$-functions, $d^{J_0}_{\Delta\sigma,\Delta\lambda}$ with
$J_0={\rm max}(|\Delta\sigma|,|\Delta\lambda|)$, needed in the present work is reproduced
below \cite{d functions:1957rs}. \s

The single $d$ function with $J_0=0$ is constant with $d^0_{0,0}=1$. The $d$ functions
with $J_0=1$ appearing both in the production and decay processes are given by
\begin{eqnarray}
&& d^1_{1,1}= d^1_{-1,-1} = \frac{1}{2}(1+\cos\theta) \nonumber \\
&& d^1_{1,-1} = d^1_{-1,1} = \frac{1}{2}(1-\cos\theta) \nonumber\\
&& d^1_{1,0} = -d^1_{-1,0} = -\sqrt{\frac{1}{2}}\sin\theta \nonumber\\
&& d^1_{0,1} = -d^1_{0,-1} = \sqrt{\frac{1}{2}}\sin\theta \nonumber\\
&& d^1_{0,0} = \cos\theta;
\end{eqnarray}
and those with $J_0=2$ appearing in the amplitudes for the production of a vector-boson
pair due to $t$-channel fermion exchange in $e^+e^-$ collisions read
\begin{eqnarray}
&& d^2_{1,2} = - d^2_{-1,-2} = \frac{1}{2}(1+\cos\theta)\sin\theta \nonumber\\
&& d^2_{1,-2} = - d^2_{-1,2} = -\frac{1}{2}(1-\cos\theta)\sin\theta
\end{eqnarray}
The $d$ functions with $J_0=1/2$ appear only in the decay processes and they are given by
\begin{eqnarray}
&& d^{1/2}_{1/2,1/2}  = d^{1/2}_{-1/2,-1/2} = \cos\frac{\theta}{2} \nonumber \\
&& d^{1/2}_{1/2,-1/2} = - d^{1/2}_{-1/2,1/2} = -\sin\frac{\theta}{2};
\end{eqnarray}
We note that the convention of Rose is adopted for the $d$ function. \s

\section{Arbitrary polarized beams}
\label{sec:appendix_c_arbitrary_polarized_beams}

The expression for the matrix element-squared for arbitrary polarized beams is
obtained as follows \cite{Hikasa:1985qi,Hagiwara:1985yu}. We denote the transverse
polarization directions $\hat{s}_\pm$ of the $e^\pm$ beams as
\begin{eqnarray}
\hat{s}_\pm = \left(\cos\varphi_\pm, \sin\varphi_\pm, 0\right)
\end{eqnarray}
where the azimuthal angles in the $x$-$y$ plane are measured from the $x$-axis
defined by the outgoing $\mathcal{P}^-$ transverse momentum in the production
process $e^+e^-\to\mathcal{P}^+\mathcal{P}^-$. We can then express the $e^\pm$
spin vectors as
\begin{eqnarray}
  s^\mu_\pm
= P^T_\pm\, (0, \hat{s}_\pm ) + P^L_\pm\, (|\vec{p}_\pm|,\, E_\pm \hat{p}_\pm)/m_e
\end{eqnarray}
The beam polarizations are limited by $0\leq P^T_\pm \leq \sqrt{1-(P^L_\pm)^2}$
with $-1 \leq P^L_\pm \leq 1$. Purely left-handed $e^\pm$ beams give $P^L_\pm =-1$
and purely right-handed $e^\pm$ beams give $P^L_\pm=+1$. While natural transverse
polarization of the $e^+e^-$ circular storage ring colliders gives $\varphi_+ =
\varphi_- + \pi$, arbitrary polarized beams are expected to be available at
$e^+e^-$ linear colliders.\s

We can now obtain the matrix element-squared for the production process
$e^+e^-\to\mathcal{P}^+\mathcal{P}^-$ combined with the subsequential $\mathcal{P}^\pm$
decays with arbitrary polarized $e^+e^-$ beams summed over the
$\mathcal{P}^\pm$ polarizations and final-state polarizations,
by choosing the transverse spin directions as
\begin{eqnarray}
\varphi_- = - \varphi \quad \mbox{and}\quad \varphi_+ = - \varphi+\delta
\end{eqnarray}
where $\varphi$ is the azimuthal angle of the $\mathcal{P}^-$ as measured from the
electron transverse momentum direction, and $\delta$ is the relative opening
angle of the electron and positron transverse polarizations. Introducing the abbreviated
notation ${\cal T}(\sigma_-,\sigma_+)$ for the correlated production-decay helicity
amplitude with the implicit assumption that the (averaged) summation  over the intermediate-
and final-state polarizations will be done, we find for the polarization-weighted distribution
\begin{eqnarray}
  \Sigma^{\mathcal{P}}_{\rm pol}
&=& \sum_{\sigma_-,\sigma'_-}\sum_{\sigma_+,\sigma'_+}
    P^-_{\sigma_-\sigma'_-} P^+_{\sigma_+\sigma'_+}\,
    {\cal T}(\sigma_-,\sigma_+)\, {\cal T}^*(\sigma'_-,\sigma'_+) \nonumber\\
&=& \tfrac{1}{4}\, [ (1 - P^L_- P^L_+) ( Q^{+-}_{+-} + Q^{-+}_{-+})
                       +(P^L_- - P^L_+) (Q^{+-}_{+-} - Q^{-+}_{-+})     \nonumber\\
  && { }\hskip 0.1cm   +(1 + P^L_- P^L_+) (Q^{++}_{++} + Q^{--}_{--})
                       +(P^L_- + P^L_+) (Q^{++}_{++} - Q^{--}_{--})       \nonumber\\
  && { }\hskip 0.1cm   +2 P^T_- P^T_+ \cos(2\varphi-\delta)\, {\rm Re}(Q^{+-}_{-+})
                       +2 P^T_- P^T_+ \sin(2\varphi-\delta)\, {\rm Im}(Q^{+-}_{-+}) \nonumber\\
  && { }\hskip 0.1cm   +2 P^T_- P^T_+ \cos\delta\, {\rm Re}(Q^{++}_{--})
                       -2 P^T_- P^T_+ \sin\delta\, {\rm Im}(Q^{++}_{--}) \nonumber\\
  && { }\hskip 0.1cm   +2 P^T_- (1-P^L_+) \cos\varphi\, {\rm Re}(Q^{+-}_{--})
                       +2 P^T_- (1+P^L_+) \cos\varphi\, {\rm Re}(Q^{++}_{-+}) \nonumber\\
  && { }\hskip 0.1cm   +2 (1+P^L_-) P^T_+ \cos(\varphi-\delta)\, {\rm Re}(Q^{+-}_{++})
                       +2 (1-P^L_-) P^T_+ \cos(\varphi-\delta)\, {\rm Re}(Q^{--}_{-+})\nonumber\\
  && { }\hskip 0.1cm   +2 P^T_- (1-P^L_+) \sin\varphi\, {\rm Im}(Q^{+-}_{--})
                       +2 P^T_- (1+P^L_+) \sin\varphi\, {\rm Im}(Q^{++}_{-+}) \nonumber\\
  && { }\hskip 0.1cm   +2 (1+P^L_-) P^T_+ \sin(\varphi-\delta)\, {\rm Im}(Q^{+-}_{++})
                       +2 (1-P^L_-) P^T_+ \sin(\varphi-\delta)\, {\rm Im}(Q^{--}_{-+}) ]
\label{eq:polarization_weighted_distribution}
\end{eqnarray}
where the electron and positron polarization matrices $P^\mp$ and the tensor
$Q^{\sigma'_- \sigma'_+}_{\sigma_- \sigma_+}$ are given by
\begin{eqnarray}
  P^\mp_{\sigma_\mp \sigma'_\mp}
&=& \frac{1}{2}
    \left(\begin{array}{cc}
      1+P^L_\mp                &  P^T_\mp {\rm e}^{-i\phi_\mp}  \\
      P^T_\mp {\rm e}^{i\phi_\mp}  &  1-P^L_\mp
          \end{array}\right) \\
   Q^{\sigma'_- \sigma'_+}_{\sigma_- \sigma_+}
&=& {\cal T}(\sigma_-,\sigma_+) {\cal T}^*(\sigma'_-,\sigma'_+)
\end{eqnarray}
with the summation over the intermediate and final-state polarizations implicitly assumed
when the elements of the tensor $Q$ are evaluated. Taking the average of the
polarization-weighted distribution (\ref{eq:polarization_weighted_distribution}) over
the azimuthal angle $\varphi$, we obtain
\begin{eqnarray}
  \overline{\Sigma}^{\mathcal{P}}_{\rm pol}
&\equiv & \int^{2\pi}_0 \frac{d\varphi}{2\pi}\, \Sigma_{\rm pol} \nonumber\\
&=& \tfrac{1}{4}\, [ (1 - P^L_- P^L_+) ( Q^{+-}_{+-} + Q^{-+}_{-+})
                       +(P^L_- - P^L_+) (Q^{+-}_{+-} - Q^{-+}_{-+})     \nonumber\\
  && { }\hskip 0.1cm   +(1 + P^L_- P^L_+) (Q^{++}_{++} + Q^{--}_{--})
                       +(P^L_- + P^L_+) (Q^{++}_{++} - Q^{--}_{--})       \nonumber\\
  && { }\hskip 0.1cm   +2 P^T_- P^T_+ \cos\delta\, {\rm Re}(Q^{++}_{--})
                       -2 P^T_- P^T_+ \sin\delta\, {\rm Im}(Q^{++}_{--}) ]
\end{eqnarray}
The last two terms are the only effect of transverse polarization to the azimuthally
integrated cross section.\s

\section{Kinematics of the Antler-topology process}
\label{sec:appendix_d_kinematics_of_the_antler_process}

When the particles, $\mathcal{D}^0$ and $\mathcal{\bar{D}}^0$, escape detection in the
correlated production-decay antler-topology process $e^+e^-\rightarrow
\mathcal{P}^+\mathcal{P}^-\rightarrow (\ell^+\mathcal{D}^0)(\ell^-\mathcal{\bar{D}}^0)$
this process is observed experimentally as
\begin{eqnarray}
e^-\, +\, e^+ \, \to\, \ell^-\, +\,\ell^+\, +\, \mbox{missing energy-momentum}
\end{eqnarray}
where the final lepton pair $\ell^-\ell^+$ can be either one of $e^-e^+$ or $\mu^-\mu^+$,
if each lepton number is strictly preserved in the underlying theory.\s

As will be explicitly shown below, if the masses, $M_{\mathcal{P}}$ and $M_{\mathcal{D}}$,
of the on-shell particles, $\mathcal{P}^\pm$ and $\mathcal{D}^0$ are a priori known,
the unobserved $\mathcal{D}^0$ and $\mathcal{\bar{D}}^0$ momenta can be determined from the
observed lepton momenta up to a twofold discrete ambiguity, in the limit where the
$\mathcal{P}$ width and photon radiation are neglected. In general, the kinematics of the
process is determined by six angles, two for the scattering, and two each for the $\mathcal{P}$
decays. Since we observe the two three-momenta of two leptons, we have in general sufficient
kinematic relations for fixing the whole configuration. A twofold discrete ambiguity occurs,
however, because the solution involves a quadratic equation. \s

As the $\mathcal{P}^\mp$ energy is fixed to be half of the beam energy, i.e.
$E_{\mathcal{P}}= \sqrt{s}/2$ in the $e^+e^-$ c.m. frame, the boost factor
$\gamma$ linking the c.m. frame to each of the $\mathcal{P}^\pm$ rest
frames is
\begin{eqnarray}
\gamma = \frac{\sqrt{s}}{2 M_{\mathcal{P}}} \quad \mbox{and}\quad
\beta  = \sqrt{1-\frac{4 M^2_{\mathcal{P}}}{s}}
\end{eqnarray}
with the boost speed $\beta=\sqrt{1-1/\gamma^2}$. In addition, the energies of the
invisible particles in the two-body decays, $\mathcal{P}^- \to \ell^- \mathcal{\bar{D}}^0$
and $\mathcal{P}^+\to\ell^+\mathcal{D}^0$, are uniquely determined by measuring the lepton
energies due to energy conservation:
\begin{eqnarray}
E_2 = \frac{\sqrt{s}}{2} - E_1 \ \ \mbox{and} \ \
E_4 = \frac{\sqrt{s}}{2} - E_3
\end{eqnarray}
with $E_1=E_{\ell^-}$ and $E_3=E_{\ell^+}$ in the laboratory frame.\s

As the particles, $\mathcal{D}^0$ and $\mathcal{\bar{D}}^0$, with an identical mass
$M_{\mathcal{D}}$, are involved in the two charge-conjugate two-body decays,
the energies of the two leptons $\ell^\pm$ are identical in the $\mathcal{P}^\pm$
rest frame:
\begin{eqnarray}
   E^*_\ell = \frac{M^2_{\mathcal{P}}-M^2_{\mathcal{D}}}{2M_{\mathcal{P}}}
\end{eqnarray}
Then, we can determine the decay lepton polar-angle $\theta_\pm$  in each of the
$\mathcal{P}$ rest frame with respect to the $\mathcal{P}^\pm$ momentum direction
depicted in Fig.$\,$\ref{fig:antler_diagram} uniquely event by event by measuring the
lepton energy $E_{\ell^\pm}$ in the laboratory frame through the relation
\begin{eqnarray}
  \cos\theta_\pm
= \frac{1}{\beta}
  \left(\frac{4 M^2_{\mathcal{P}}}{M^2_{\mathcal{P}}-M^2_{\mathcal{D}}}
        \frac{E_{\ell^\pm}}{\sqrt{s}}-1\right)
\end{eqnarray}
when the lepton mass is ignored. Furthermore, the relative orientation of the momentum
vector of $\ell^\pm$ and $\mathcal{P}^\pm$ is fixed by the two-body decay kinematics:
\begin{eqnarray}
  \cos\alpha_\pm
= \frac{1}{\beta}
  \left(1-\frac{M^2_{\mathcal{P}}-M^2_{\mathcal{D}}}{\sqrt{s} E_{\ell^\pm}}\right)
\label{eq:lab_lepton_opening_angles}
\end{eqnarray}
where the angles $\alpha_\pm$ are the opening angles between the visible $\ell^\pm$ tracks
and the parent $\mathcal{P}^\pm$ momentum directions.\s

In order to prove the existence of a twofold discrete ambiguity in determining the production
angle $\theta$, it is sufficient to solve for the $\mathcal{P}^-$ momentum direction denoted
by a unit vector $\hat{n}_{\mathcal{P}} \equiv \hat{q}_- = -\hat{q}_+$. Let us assume, for
the moment, that the two lepton three-momentum directions, denoted by $\hat{n}_\pm$, are not
parallel. Then we can expand the unit vector $\hat{n}_{\mathcal{P}}$ in terms of the
unit vectors $\hat{n}_\pm$
\begin{eqnarray}
  \hat{n}_{\mathcal{P}}
= a \hat{n}_- + b \hat{n}_+ + c (\hat{n}_-\times\hat{n}_+)
\end{eqnarray}
As shown in Eq.$\,$(\ref{eq:lab_lepton_opening_angles}), the projections of the unit vector
along the lepton momentum directions $\hat{n}_\pm$ satisfy
\begin{eqnarray}
&& \hat{n}_-\cdot \hat{n}_{\mathcal{P}} = \cos\alpha_- \\
&& \hat{n}_+\cdot \hat{n}_{\mathcal{P}} = -\cos\alpha_+
\end{eqnarray}
These two relations constrain $\hat{n}_{\mathcal{P}}$ to lie on a line
in three-dimensional space. They give
\begin{eqnarray}
&& a + b (\hat{n}_-\cdot\hat{n}_+)  = \phantom{+}\cos\alpha_- \nonumber\\
&& a (\hat{n}_-\cdot\hat{n}_+) + b  = -\cos\alpha_+
\end{eqnarray}
which can be explicitly solved:
\begin{eqnarray}
   \left( \begin{array}{l}
           a \\
           b
          \end{array}\right)
= \frac{1}{(\hat{n}_-\times\hat{n}_+)^2}
  \left( \begin{array}{cc}
           1    &  -\hat{n}_-\cdot \hat{n}_+  \\
           -\hat{n}_-\cdot \hat{n}_+  &  1
         \end{array}\right)
  \left( \begin{array}{c}
         \phantom{+}\cos\alpha_- \\
         -\cos\alpha_+
         \end{array}\right)
\end{eqnarray}
The remaining variable is determined by the condition that the vector $\hat{n}_{\mathcal{P}}$
is a unit vector, i.e. $\hat{n}^2_{\mathcal{P}} = 1$:
\begin{eqnarray}
c^2 =  \frac{(\hat{n}_-\times \hat{n}_+)^2
            -(\cos\alpha_-\hat{n}_- + \cos\alpha_+ \hat{n}_+)^2}{(\hat{n}_-\times\hat{n}_+)^4}
\end{eqnarray}
The sign of $c$ cannot be determined. This explicitly shows the twofold discrete ambiguity
mentioned before. The inequality $c^2\geq 0$ is expected to be violated only by finite
$\mathcal{P}$-width effects and by radiative corrections, and hence may serve as a test
of the $\mathcal{P}$-pair signal. Introducing the vector
$\vec{a} = \cos\alpha_- \hat{n}_- + \cos\alpha_+ \hat{n}_+$, we can rewrite the unit vector
$\hat{n}_{\mathcal{P}}$ as
\begin{eqnarray}
   \hat{n}_{\mathcal{P}}
= \frac{1}{(\hat{n}_-\times \hat{n}_+)^2}
  \left[ (\hat{n}_-\cdot \vec{a})\, \hat{n}_-
        -(\hat{n}_+\cdot \vec{a})\, \hat{n}_+
        \pm \sqrt{(\hat{n}_-\times\hat{n}_+)^2 -\vec{a}^2}\,\,
            (\hat{n}_-\times\hat{n}_+) \right]
\end{eqnarray}
determined up to a sign ambiguity. \s

In the exceptional case where the two lepton momenta are parallel, we obtain a one-parameter
family of solution for which the azimuthal angle between two decay planes is left
undetermined.\s

Let us now consider the azimuthal-angle correlations of the decay kinematics. In the
coordinate system with the $z$-axis along the $\mathcal{P}$-momentum direction,
the unit vectors denoting the $\ell^\mp$ four-momentum directions can be expressed as
follows:
\begin{eqnarray}
&& \vec{n}_-
   = E_{\ell^-}\left( \sin\alpha_- \cos\phi_-,\sin\alpha_- \sin\phi_-,
                   \cos\alpha_-\right) \\
&& \vec{n}_+
   =\left( \sin\alpha_+ \cos\phi_+, \sin\alpha_+ \cos\phi_+,
                  -\cos\alpha_+ \right).
\end{eqnarray}
Taking the scalar product between the unit vectors:
\begin{eqnarray}
  \hat{n}_-\cdot \hat{n}_+
= \sin\alpha_-\sin\alpha_+\left(\cos{\phi_-}\cos{\phi_+}+\sin{\phi_-}\sin{\phi_+}\right)
  -\cos\alpha_-\cos\alpha_+
\end{eqnarray}
and noting that $\cos\phi_+ \cos\phi_- + \sin\phi_+ \sin\phi_- = \cos(\phi_+ - \phi_-)$,
we can check that the cosine of the difference $\phi=\phi_+ - \phi_-$ of two azimuthal angles
\begin{equation}
\cos \phi  =  \frac{\hat{n}_-\cdot\hat{n}_+ + \cos\alpha_- \cos\alpha_+}{
                    \sin\alpha_- \sin\alpha_+}
\end{equation}
can be determined uniquely event by event in the correlated antler-topology process.
However, we cannot determine the sign of $\sin\phi$, of which the expression
\begin{equation}
  \sin\phi
= \frac{(\hat{n}_-\times \hat{n}_+)\cdot \hat{n}_{\mathcal{P}}}{\sin\alpha_-\sin\alpha_+}
= \pm \frac{\sqrt{(\hat{n}_-\times\hat{n}_+)^2-\vec{a}^2}}{\sin\alpha_-\sin\alpha_+}
\label{eq:sin_phi_ambiguity}
\end{equation}
has a sign ambiguity due to the twofold ambiguity in determining the momentum direction
$\hat{n}_{\mathcal{P}}$. \s

\vskip 1.5cm

% \newpage

\end{document}